\documentclass[
 pra,
 aps,
 reprint,
 superscriptaddress,
 groupedaddress,
 citeautoscript,
 floatfix,
]{revtex4-2}
\usepackage[english]{babel}
\makeatletter
\let\l@en\l@english
\makeatother
\usepackage[]{standalone}
\usepackage{anyfontsize}
\usepackage{comment}
\usepackage{import}
\usepackage{amssymb} 
\usepackage{amsmath}
\usepackage{stmaryrd}  
\allowdisplaybreaks  
\usepackage{cmap}
\usepackage[utf8]{inputenc}
\usepackage[T1]{fontenc}
\usepackage{graphicx}
\usepackage[hyperfootnotes=false,breaklinks=true]{hyperref}
\usepackage{setspace}
\usepackage{multirow, makecell}
\hypersetup{
 bookmarksopen=true,
 bookmarksopenlevel=1,
 colorlinks=true,
 linkcolor=blue,
 anchorcolor=blue,
 citecolor=blue,
 filecolor=blue,
 urlcolor=blue,
 pdfpagemode=UseOutlines,
 pdfstartview={XYZ null null 1},
}
 
\usepackage{amsmath}
\allowdisplaybreaks  
\usepackage{amssymb}
\usepackage{amstext}
\usepackage{amsfonts}
\usepackage{amsthm}
\usepackage{mathtools}

\usepackage{bm}
\usepackage{braket}
\usepackage{multirow}
\usepackage{physics}
\usepackage{enumitem}
\usepackage[
 capitalise,
]{cleveref}
\crefname{table}{Table}{Tables}%
\crefname{section}{Sec.}{Sects.}%
\Crefname{section}{Section}{Sections}%
\Crefname{table}{Table}{Tables}%

\usepackage{subcaption} 

\usepackage{algorithm}
\usepackage{algpseudocode}    
\usepackage{float}

\usepackage{graphicx}  

\def\Z{\mathbb{Z}}
\def\N{\mathbb{N}}

\def\del{\delta}

\def\M{\mathcal{M}}
\def\H{\mathcal {H}}
\def\L{\mathcal {L}}
\def\S{\mathcal {S}}
\def\T{\mathcal {T}}

\def\GSD{\operatorname{GSD}}
\def\Tr{\operatorname{Tr}}

\makeatletter
\def\blfootnote{\xdef\@thefnmark{}\@footnotetext}
\makeatother

\usepackage{amsthm}
\usepackage{braket}
\usepackage{amsmath}
\usepackage{amssymb}
\usepackage{mathtools}
\usepackage{orcidlink}
\usepackage{dsfont}

\newtheorem{Remark}{Remark}[section]  
\newtheorem{Example}{Example}[section]

\theoremstyle{definition}

\usepackage{hyperref} 
 \hypersetup{
    colorlinks=true,
    citecolor=blue,  
    linkcolor=blue,  
    filecolor=magenta,      
    urlcolor=blue,
}

\usepackage{caption}  
\usepackage{subcaption}
\begin{document}

\clearpage \newpage \twocolumngrid
\pagenumbering{arabic}

\blfootnote{This manuscript has been authored by UT-Battelle, LLC, under Contract No.~\ DE-AC0500OR22725 with the U.S.~\ Department of Energy. The United States Government retains and the publisher, by accepting the article for publication, acknowledges that the United States Government retains a non-exclusive, paid-up, irrevocable, worldwide license to publish or reproduce the published form of this manuscript, or allow others to do so, for the United States Government purposes. The Department of Energy will provide public access to these results of federally sponsored research in accordance with the DOE Public Access Plan.}

\title{ Composite-dimensional Topological Codes with Boundaries and Defects }

\author{Mohamad Mousa\,\orcidlink{0009-0000-9054-1692}}
\email{mmousa@purdue.edu}
\affiliation{\mbox{Department of Physics, Purdue University, West Lafayette, Indiana 47907, USA}}

\author{Amit Jamadagni}
\affiliation{Computational Sciences and Engineering Division, 
Oak Ridge National Laboratory, 
Oak Ridge, Tennessee 37831, USA}

\author{Eugene Dumitrescu\,\orcidlink{0000-0001-5851-9567}}
\email{dumitrescuef@ornl.gov}
\affiliation{Computational Sciences and Engineering Division, 
Oak Ridge National Laboratory, 
Oak Ridge, Tennessee 37831, USA}

\begin{abstract}
We introduce new algorithms and provide example constructions of stabilizer models for the gapped boundaries, domain walls, and $0D$ defects of Abelian composite-dimensional twisted quantum doubles. Using the physically intuitive concept of condensation, our algorithm explicitly describes how to construct the boundary and domain-wall stabilizers starting from the bulk model. This extends the utility of Pauli stabilizer models in describing non-translationally invariant topological orders with gapped boundaries. To highlight this utility, we provide a series of examples, including a new family of quantum error-correcting codes where the double of $\mathbb{Z}_4$ is coupled to instances of the double semion (DS) phase. We discuss the codes' utility in the burgeoning area of quantum error correction with an emphasis on the interplay between deconfined anyons, logical operators, error rates, and decoding. We also augment our construction, built using algorithmic tools to describe the properties of explicit stabilizer layouts at the microscopic lattice-level, with dimensional counting arguments and macroscopic-level constructions building on pants decompositions. The latter outlines how such codes' representation and design can be automated. Our results are validated by a series of error-correcting threshold calculations comparing our code's performance with standard surface codes. To do so, we introduce a composite dimensional belief propagation decoder with ordered statistics that utilizes combination sweeps. Going beyond our worked-out examples, we expect our explicit step-by-step algorithms to pave the path for new higher-dimensional codes to be discovered and implemented in near-term architectures that take advantage of various hardware's distinct strengths. 
\end{abstract}
\date{\today}

\maketitle
\tableofcontents               

\section{Introduction}
 
Quantum error correction distills many noisy physical quantum degrees of freedom into a smaller number of logical quantum degrees of freedom with exponentially suppressed error rates \cite{Shor1995,Gottesman1997,Dennis2002}. Further, topological fault-tolerant quantum computation utilizes topological order to encode and, with low error rates, manipulate logical degrees of freedom encoded in a non-local logical subspace~\cite{Kitaev2003,Nayak2008}. Recently, a number of qubit experiments have demonstrated how topological codes' logical error rates are suppressed with increasing system size. Recent examples include 31 superconducting qubits~\cite{Satzinger2021}, 49 qubits in a neutral-atom array~\cite{Bluvstein2023}, 81 neutral-atom platform~\cite{SalesRodriguez2025} as well as many others~\cite{exp1,exp2,exp3,exp4,exp5,exp6,exp7,google2024}. 

As the number of qubits increases, it is of growing importance to determine which quantum error-correcting codes are best suited to maximize performance on different quantum hardware architectures. In this work, we provide a new direction to help make this determination. 
We work within the context of the family of topological codes based on Abelian twisted quantum doubles \cite{Kitaev2003,Yuting2013}, which encompasses generalizations to the well-studied surface code. Here, we explore the interplay of three generalizations to the surface code: local Hilbert space dimension, boundaries and defects, and spatial anisotropy to provide new routes towards error correction. 

One natural generalization of the qubit-based surface code is to leverage higher-dimensional $d>2$ local Hilbert spaces as a resource (qudits) and use higher-dimensional codes. This route is promising because qudits have shown promise in accelerating the simulation of high‑dimensional quantum systems~\cite{Meth2025,Sawaya2020}, compiling algorithms~\cite{Bocharov2017,Gokhale2019,Chu2023}, magic-state distillation~\cite{Campbell2012,Campbell2014}, and quantum control~\cite{Lindon2023,Kiktenko2020}. As a result, experiments have been developed to realize qudit systems using architectures based on donor spins in silicon~\cite{FernandezDeFuentes2024}, ultracold atoms and molecules~\cite{Vilas2024,Chaudhury2007}, optical photons~\cite{Kues2017,Chi2022}, superconducting circuits~\cite{Brock2025,Nguyen2024,Roy2025,Wang2025}, trapped ions~\cite{Leupold2018,Ringbauer2022}, and vacancy centers~\cite{Adambukulam2024,Soltamov2019}.

As we will explore in this work, higher-dimensional composite local Hilbert spaces provide a route to simultaneously utilize multiple topological phases in a single quantum error-correcting code. This direction is less explored, since translationally invariant systems are simpler to study \cite{Haah2013}. Translationally invariant (generalized) Pauli stabilizer codes in prime dimensions were shown to be equivalent to copies of higher-dimension toric codes $D(\Z_d)$ \cite{Haah2021Classification}. Additionally, for composite-dimension qudits, translationally invariant Pauli stabilizer models can realize the bulk of Abelian twisted quantum doubles~\cite{Tyler2022}. These works of Refs.~\citenum{Bombin2012Universal, Bombin2014Structure, Haah2021Classification, Tyler2022}, while covering all qudit dimensions, did not address the case of translationally non-invariant models. We here explore spatially anisotropic codes with an additional emphasis on the richness that composite-dimension qudits offer. This spatial anisotropy is known to enable new properties like universality with non-Abelian islands \cite{Laubscher2019}.

However, constructing spatially anisotropic codes requires a careful treatment of the boundaries that separate distinct topological phases~\cite{Bombin2008family,Beigi2011,Kitaev2012,Cong2016,Cong2017,Kong2017,tqdbdry,operatoralg2024,sptsewing2024,bdry2025}. This was absent from previous works that considered Pauli stabilizer models of Abelian twisted quantum doubles for example \cite{Tyler2022}. In particular, the existing abstract constructions~\cite{Beigi2011,Cong2017} of these boundaries and defects may not be experimentally efficient in certain cases. For example, while the bulk of each phase separately has a representation in terms of familiar Pauli stabilizer models, it is unclear if the gapped boundaries and defects can also have a Pauli stabilizer description. Answering this question has practical consequences for implementation as well as for assessing the computing power of these codes. A simple systematic method to construct these boundaries and $0$D defects within the Pauli Stabilizers is still lacking. 

In this work, we show that boundaries, domain walls and point-like defects of Abelian twisted quantum doubles can be written in terms of generalized Pauli stabilizers, thus generalizing results of Ref.~\citenum{Tyler2022}. By providing an algorithm to construct them starting from the bulk theory, we prove this \textit{constructively}. Our construction relies on local condensation, so it does not introduce new degrees of freedom~\cite{sptsewing2024,bdry2025}, non-Pauli stabilizers \cite{Beigi2011,Cong2016,Cong2017}, nor rely on lattice defects \cite{Bombintwist2010}. It also stays within the framework of topological phases, without reverting to gauging symmetries of their underlying symmetry-protected phases~\cite{sptsewing2024}. It also automatically produces topologically complete models, meaning that all logical operators are macroscopic~\cite{bravyi2010,operatoralg2024}. Even other choices for the initial quantum double stabilizers, like spatially disconnected stabilizers \cite{operatoralg2024}, will automatically give models with topological order. As such, the resulting Hamiltonian for the boundary will also be described by local stabilizers. This construction offers both a simple, intuitive understanding of the boundaries as well as an efficient method that can automate boundary construction for various phases and lattices. Importantly, by relying on anyon condensation as opposed to consistency relations \cite{Hu2017,Hu2018,Beigi2011,Cong2017}, it extends the systematic construction of Pauli stabilizers to also include the $0D$ defects \cite{Barkeshli2013} of Abelian twisted quantum doubles, which is, to our knowledge, unique to this work.

Our construction further facilitates the exploration of a new family of codes that comprises generalized holes, boundaries, and defects of higher-dimensional codes. For simplicity, we exemplify this new family of codes by studying the smallest composite number $d = p \times q$ where $p=q=2$. This could be realized by four-dimensional qudits or, alternatively, in the near term, by grouping qubits in pairs. We find new qualitative properties for higher-dimensional codes, such as tunable error biases, which can be exploited to simplify noise models~\cite{Tuckett2019}. Their syndrome extraction process can also exhibit advantages compared to the qubit case.

\section{Summary of main results}

This section summarizes the main results of and outlines our work. Readers primarily interested in explicit code constructions, decoding algorithms, and physical examples may focus on the background formalism in Sec.~\ref{sec:codes} and new results in Sec.~\ref{sec:composite}. Readers seeking a general and systematic framework for constructing gapped boundaries, domain walls, and defects within a Pauli stabilizer formalism may consult Sec.~\ref{sec:stabilizer_boundaries}. The global properties of the resulting codes, including ground-state degeneracy and logical operator structure, are analyzed in Sec.~\ref{sec:GSD} using both microscopic and macroscopic approaches.

Sec.~\ref{sec:codes} describes the background and quantum double and twisted quantum double formalisms used throughout. Specifically, we work with the double of $\Z_4$, $D(\mathbb{Z}_4)$, as a base model. We discuss periodic and open boundary conditions. In addition to the smooth and rough boundaries, we introduce an even boundary that is distinct and leads to different types of planar codes. Further, we use our boundary construction algorithm to derive a set of domain walls which generalize Bombin's~\cite{Bombintwist2010} $\Z_2$ twist. Sec.~\ref{sec:DS} then moves on to describe the bulk and boundaries of the Double Semion phase model, which may be obtained by boson condensing the parent $\Z_4$ model. Throughout, we provide examples of how different combinations of boundaries and domain walls (co-dimension 1 defects) realize logical codes. For example, codes of logical dimension $D = 4^M$ may be constructed via planar or spherical ($M=0$), open boundary condition planar codes ($M=1$), surface or toric ($M=2$), or higher $2M$-boundary surface or $M$-torus or other combinations involving twists ($M>2$). Interestingly, even boundaries lead to qubit (rather than logical $d=4$-qudits) codes. Taken together, we have codes with logical dimensions $D = 4^M 2^{M'}$.

Building on these constructions and intuition, in Sec.~\ref{sec:composite} we go beyond codes with a single bulk phase. The main tool to do so, generalizing the notion of punctures, is condensing \textit{local} DS patches (co-dimension $0$ defects as opposed to global condensation~\cite{Tyler2022}), from the parent $\Z_4$ phase. In this composite and non-translationally invariant code, we find that $N+1$ DS patches realize an $N$ logical qubit code. Combining this result with deformations of the parent $\Z_4$ model, we get a total composite logical dimension of $D = 4^M \times 2^{N+M'}$. We provide the logical Pauli operators that act on the patch codes' qubit topological codespace.

After a brief review of stabilizer codes and the noise models for qudit-based codes, we then detail how such codes can be decoded. This involves a variety of modified minimum-weight-perfect-matching decoder (MWPM), an integer linear programming decoder (ILP), and a belief propagation with ordered statistics and combination sweep decoder (BP-OSD-CS). We then use these decoders to find code capacity thresholds for the five codes we considered.
 
After describing our new codes and their decoding strategies, Sec.~\ref{sec:stabilizer_boundaries} presents rigorous constructions and the main systematic methodologythath led to our codes' discovery. Specifically, we provide an algorithm (\ref{alg:Bdry}) that concretely illustrates how to gauge from a parent code---meaning to insert boundaries, domain walls, and $0$D defects into Abelian twisted quantum doubles---to the new composite codes. We emphasize the concept of anyon condensation, which is used to conceptually understand high-level code properties, such as their interfaces. In addition,  condensation results in certain code gauging strategies. These tools can be used to alter codes from the bottom up (i.e., microscopic $\rightarrow$ macroscopic) within a generalized Pauli stabilizer formalism. Importantly, we carefully examine how condensation depends on both pre-existing bulk and boundary degrees of freedom. 

To provide intuition, and to see the algorithm in action, a series of well-known codes and their boundaries are re-derived within this formalism in Sec.~\ref{sec:bdryalg}. To rigorously describe our DS patch codes, we then systematically constructed boundaries and domain walls in Secs.~\ref{sec:bound_algs} and ~\ref{sec:DWalg} respectively. As an additional illustrative example, highlighting further possible modifications, the generalization from the $\Z_2$ twist to $\Z_4$ is exposed in Ex.~\ref{Example:emdw}. An important example of the domain wall between $\Z_4$ and DS is treated heuristically in Ex.~\ref{ex:z4dsdwsimple} and, using the full machinery of our construction, in App.~\ref{sec:dsz4dwnotsimple}. Further examples, such as the construction of $0$D defects, are detailed in App.~\ref{sec:appinvdw}.

To insert these topological components into different topological codes under a unified framework, Sec.~\ref{sec:GSD} computes the ground state degeneracies in the presence of \textit{multiple} boundaries, domain walls, and defects. Based on microscopic commuting projector stabilizer models, these calculations rely on counting arguments that take into account the order of old and new constraints and the changing dependencies on one another. They are central to deriving the logical dimension of our family of new quantum error correcting codes. Ultimately, the properties of the topological codes do not depend on the exact layout, which enables us to then analyze codes at scale. 

In Sec.~\ref{sec:lego_code}, a generic code description that is agnostic to concrete stabilizer models is used to compute the codespace dimension and provide intuition about the logical operators of different codes. This is carried out using the pants decomposition of any g-genus 2D orientable manifold with the topological components of codes. This confirms the results of Sec.~\ref{sec:GSD} and illustrates how interested readers may further modify or generalize our codes as desired. Finally, Sec.~\ref{sec:fin} concludes by discussing this work's implications, open questions, and future research directions. 

\section{Background}
\label{sec:codes}
Kitaev's quantum double models~\cite{deWild1995, Kitaev2003} are both physical realizations of the mathematical quantum double construction by Drinfeld~\cite{Drinfeld1988} as well as concrete lattice instances of Dijkgraaf-Witten Topological Quantum Field Theories (TQFTs)~\cite{DijkgraafWitten1990}. In addition, they underpin a family of Quantum Error-Correcting Codes (QECCs)~\cite{Kitaev2003}. 

The bulk of a TQFT $A$ is described by a Hamiltonian, which we denote as $H_A$. Generically, the TQFT is spatially supported on a surface $\M_A$ which may have a boundary $\partial \M_A$. In addition to the bulk interactions, the boundary will be described by $H_{\partial A}$ living on $\partial \M_A$. Further, two distinct TQFTs $A,B$ supported on surfaces $\partial \M_A$ and $\partial \M_B$ must have a domain wall between them if  $\partial \M_A \cap \partial \M_B \neq \varnothing$. This domain wall will be described by a Hamiltonian $H_{A \cap B}$ which lives on the common boundary of the two TQFTs $\partial \M_A \cap \partial \M_B$. Altogether, the composite system will be described by a Hamiltonian. 

\begin{equation}
\begin{aligned}
 H_{\text{tot}} &= \sum_{\M_A} H_A + \sum_{\M_B} H_B + \sum_{\partial {\M_A}} H_{\partial A} + \sum_{\partial {\M_B}} H_{\partial B} \\
 &\quad + \sum_{\partial \M_A \cap \M_B} H_{A \cap B}
\end{aligned}
\end{equation}

It is evident that the boundaries of a TQFT are trivial specific examples of the domain walls between this TQFT and the vacuum TQFT. Conversely, gapped domain walls between two topologically non-trivial phases are closely related to their boundaries with the vacuum. In fact, as is discussed in Sec.~\ref{sec:DWalg}, a gapped domain wall between two topological phases $A$ and $B$ is precisely the gapped boundary of the topological order $A \otimes B$~\cite{Beigi2011,Kong2014}. 

In the lattice picture, the models are defined on a 2D orientable surface $\M$, possibly containing a spatial boundary $\partial \M$. The surface is then triangulated into a lattice $\Gamma$ consisting of vertices $V$, edges $E$, and plaquettes $P$ (faces). Each spatial boundary $\partial \M$ has to be associated with a consistent boundary theory that is compatible with the bulk theory \cite{Kitaev2003,Kitaev2012,Cong2016,Cong2017,Kong2017}. 

In (twisted) quantum double models \cite{Kitaev2003,Yuting2013}, a qudit, taking values in a finite group $G$, is associated with each edge of the lattice $e\in E$ \cite{quditedge}. These qudits live in a Hilbert space $\mathcal{H}_e$ spanned by the orthonormal basis generated by the group elements $\{|g\rangle_e: g \in G\}$. The total Hilbert space is then the tensor product of all the edge Hilbert spaces $\mathcal{H}_{tot}=\bigotimes_{e\in E}\mathcal{H}_e$. The case where the group is $\Z_2$ corresponds to the toric code, where a qubit is associated with each edge~\cite{Kitaev2003}. Since the basic results in this paper can also be shown in Abelian groups, we focused on the case of the cyclic group $\Z_4$, which has a simple implementation using two qubits or $d=4$ qudits (hereafter referred to as qudits). In addition, the Hamiltonian terms are built from the generalized Pauli stabilizers \cite{Gheorghiu2014}, defined below. 

\begin{Remark}
With appropriate definitions, all physical properties of the model do not depend on the microscopic details of the lattice. \textit{Any} triangulation of the surface $\M$ will give rise to the same topological phase \cite{Kitaev2003,Yuting2013}. For simplicity, we follow the most prominent conventions, as they appear in the literature, and work with a square lattice.
\end{Remark}

In Sec.~\ref{sec:Z4} we review the $\Z_4$ surface code, which is the generalization of the surface code for the case of qudits~\cite{Kitaev2003}. In addition, by deriving the gapped boundaries and domain walls, we use $\Z_4$ as a tool to review defects and edges. Afterward, in Sec.~\ref{sec:DS} we review a Doubled Semion code construction which is a twisted quantum double version of the $\Z_2$ toric code~\cite{Yuting2013}, and is also closely related to the $\Z_4$ code~\cite{Tyler2022}.

\subsection{The $D(\Z_4)$ Quantum Double}
\label{sec:Z4}
In higher-dimensional Hilbert spaces, the stabilizers of codes will be built out of the generalized Pauli $X$ and $Z$ operators \cite{Gheorghiu2014}, which are also called the shift and clock operators. In the rest of the paper, we simply refer to them as Pauli operators. For the case of $D(\Z_N)$, they are defined as:
\begin{equation}\label{eq:genpauli}
X=\sum_{0\leq j \leq N-1}|j+1\rangle\langle j|, \quad Z=\sum_{0 \leq j \leq N-1} e^{2\pi ij/N}|j\rangle\langle j|
\end{equation}
For $N=4$, the explicit matrix representations at a given site are:
\begin{equation}\label{eq:genpauli}
X \;=\;
\begin{pmatrix}
0 & 1 & 0 & 0 \\[4pt]
0 & 0 & 1 & 0 \\[4pt]
0 & 0 & 0 & 1 \\[4pt]
1 & 0 & 0 & 0 
\end{pmatrix},
\qquad
Z \;=\;
\begin{pmatrix}
1 & 0 & 0 & 0 \\[4pt]
0 & i & 0 & 0 \\[4pt]
0 & 0 & -1 & 0 \\[4pt]
0 & 0 & 0 & -i
\end{pmatrix}.
\end{equation}
They obey the following relations,
\begin{equation}
X^4 = Z^4 =1,    \quad ZX=i XZ, \quad ZX^{\dagger}=-iX^{\dagger}Z.
\end{equation}

Using these operators, now define for every vertex $v$ and plaquette $p$ the star operators $A(v)$ and the plaquette operators $B(p)$. We also introduce shorthand pictorial symbols for later use.

\subsubsection{$D(\Z_4)$ Bulk}
\label{sec:Z4_bulk}
The bulk Hamiltonian of $\Z_4$ is then,
\begin{equation} \label{eq:Hz4}
H_{\Z_4} = -\sum_{v \in V}(A(v) + H.C.) -\sum_{p \in F}(B(p) + H.C.)
\end{equation}
where H.C. denotes the hermitian conjugate and the sum is over vertices and plaquettes belonging to the bulk region.

\begin{equation}
  \text{\includegraphics[width=0.48\textwidth]{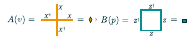}}
  \label{eq:z4}
\end{equation}

Since all the terms of the Hamiltonian commute, they define a stabilizer code generated by all the star and plaquette operators:
\begin{equation}
    \mathcal{S}_{\Z_4}= \langle A,B \rangle
\end{equation}

\textit{Excitations of the bulk}--There are $16$ local excitations (including the vacuum) which are generated by the different fusions of the pure electric charges $\{1,e,e^2,e^3\}$ and the pure magnetic fluxes $\{1,m,m^2,m^3\}$ satisfying $e^4=m^4=1$. These are gapped excitations from the Hamiltonian perspective, as they cost finite energy to create. Alternatively, in the stabilizer picture, they are either errors or logical operators, depending on whether they violate stabilizers or not, respectively. Lastly, from the TQFT perspective, they can be thought of as localized particles. In particular, they are called anyons as they can have exotic statistics which are neither fermionic nor bosonic~\cite{Wilczek82}. We switch between all three pictures depending on the context. Here, we will treat them as anyons for the moment, and denote the anyons of the quantum double $D(\Z_4)$ by $\mathcal{A}(\Z_4)$. 

When two anyons come close together, they can be regarded as one emergent anyon; this process is called fusion \cite{Kong2022}. In Abelian theories, fusion outcomes are deterministic, which means the fusion of any two anyons results in only one new anyon \cite{Kitaev2003,Gould_1993,Wang2020}. Explicitly, we have:
\begin{equation}
  e^{p} m^{q} \otimes e^{r} m^{s} \;=\; e^{\,p+r} m^{\,q+s} \quad p,q,r,s \in \{0,1,2,3\}
\end{equation}
 Equivalently, we can also write it using the fusion symbols:
\begin{equation}\label{eq:fusion}
a \otimes b = \sum_{m} N_{ab}^{m} \, m \quad a,b,m \in \mathcal{A}(\Z_4)
\end{equation}
 where the fusion symbol $N_{ab}^m = \delta_{a+b,m}$. The fusion of anyons in $\Z_4$ thus forms a group $\mathcal{F} = \Z_4 \times \Z_4$.
 
 In addition to fusion rules, anyons carry an intrinsic topological spin given by the diagonal $\mathcal{T}$-matrix:
 \begin{equation}\label{eq:spin}
\mathcal{T}({e^{p}m^{q}}) =  i^{pq}
\end{equation}
where we used the shorthand $\mathcal{T}(a) = \mathcal{T}_{aa}$ since it is a diagonal matrix. We can then see that all the pure electric charges, $\{1,e,e^2,e^3\}$, the pure magnetic fluxes, $\{1,m,m^2,m^3\}$, and the dyon $e^2m^2$ are bosons. 

\begin{figure}[htbp]
 \includegraphics[width=0.50\textwidth]{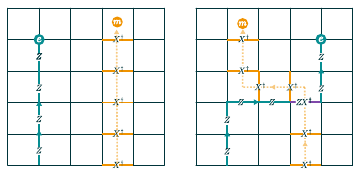}\caption{\label{fig:embraid} Strings for creating e and m anyons in $\Z_4$. (b) Braiding and $e$ and an $m$ anyons, the two strings overlap at the purple edge.}
\end{figure}

As is common in topological phases, bosonic excitations (anyons with topological spin $=1$) can still braid non-trivially with other excitations. The braiding of two anyons $a$ and $b$ around each other produces a $U(1)$ phase denoted here by $B(a,b)$. In Abelian theories, this phase can be directly computed from the $\mathcal{T}$-matrix as \cite{Kitaev2003}:
\begin{equation}\label{eq:braid}
B(a,b) = \frac{\mathcal{T}(a \otimes b)}{\mathcal{T}(a)\, \mathcal{T}(b)}
\end{equation}
 In Abelian theories, braiding is thus completely determined by the topological spin of the generators $e$ and $m$, which is $i$ in this case. Applying Eqs.~\eqref{eq:fusion}, \eqref{eq:spin} and \eqref{eq:braid} we have 
\begin{equation}\label{eq:braidz4}
B(e^{p}m^{q},e^{r}m^{s}) = i^{ps+qr}.
\end{equation}
The braiding statistics can also be read from the microscopic string operators creating and transporting the anyons, Fig.~\ref{fig:embraid}. Note that strings with opposite direction will have the inverse operators. This means that if the $m$ string goes down or left, it will act with $X$, and if the $e$ string goes left or down, it will act with $Z^\dagger$. This follows from the implicit orientation of the lattice fixed through the definition of the Hamiltonian terms \eqref{eq:Hz4}. All other anyonic braiding statistics can be deduced using these two generators. 
The braiding statistics of anyons play a crucial role in defining the $\Z_4$ models as quantum error-correcting surface, toric, or other related codes. Closed Wilson loops of anyons commute with all the Hamiltonian terms and thus define/implement logical operators acting within the code space \cite{bombin2013introduction}. The caveat is that if two loops of anyons braid non-trivially, then their actions are dependent. In addition, loops that are deformed into one another, via the action of the terms in the Hamiltonian, are also equivalent.

\subsubsection{$D(\Z_4)$ Boundaries}
\label{sec:bdryz4}
The original model for fault-tolerant quantum computing by anyons was defined on a Torus~\cite{Kitaev2003}, which is a closed manifold, so there was no need to discuss what occurs at boundaries. Introducing smooth and rough boundaries, a seminal follow-on work~\cite{Bravyi1998} defined $D(\mathbb{Z}_2)$ surface (planar) codes which have become ubiquitous QECCs.

In this section, we focus on the generalization of the $\Z_2$ case to the gapped boundaries of TQFTs described by (twisted) quantum doubles based on the $D(\mathbb{Z}_4)$. Note that our notion of boundaries, thus far, corresponds to a boundary between a non-trivial TQFT and the topologically trivial vacuum, where no anyons exist. This means that at the bulk-to-boundary map, a bulk deconfined (an anyon that does not need extra energy to propagate) anyon may only condense (be identified with the vacuum) or become confined depending on the details of the boundary. The confined anyons will still be gapped, and they will live on the $(1+1)$D boundary. In fact, the $(1+1)$D gapped boundaries of topological phases are completely determined by the anyons they condense~\cite{Kapustin2011,Kong2014}. These anyons for a certain boundary form a subgroup of the fusion group, which is called a Lagrangian subgroup. This subgroup $\mathcal{L}$ satisfies three properties \cite{Etingof2009,Wang2020,Kapustin2011}:
\begin{enumerate}
\label{enu:Lagrange}
    \item All anyons $a \in \L$ are bosons, $\T(a)=1$.
    \item All anyons $a,b\in \L$ braid trivially with each other $B(a,b)=1$.
    \item For any anyon $c \notin \L$, there exists an anyon $a' \in \L$ such that $B(a',c)\neq 1$. 
\end{enumerate}

Intuitively, the Lagrangian subgroup describes a group of anyons that can be consistently identified with the vacuum. In addition, identifying this group with a vacuum will confine all other anyons outside the subgroup \cite{Levin2013}. From the spin and braiding statistics of the $\Z_4$ model, we can find three boundaries:
\begin{equation}\label{eq:z4Ls}
\begin{aligned}
    \L_{\text{smooth}} &=\{1,m,m^2,m^3\}, \quad \L_{\text{rough}}=\{1,e,e^2,e^3\},\\
    \L_{\text{even}}&=\{1,e^2,m^2,e^2m^2\}.
\end{aligned}
\end{equation} 
The first two boundaries generalize the smooth and rough boundaries of the $\Z_2$ surface code to the $\Z_4$ case. The last boundary is a new boundary that condenses all anyons with even exponents.

\begin{figure}[htbp]
\includegraphics[width=0.48\textwidth]{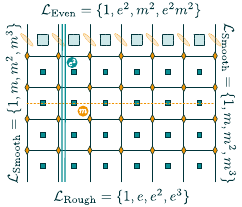}
\caption{\label{fig:z4surface1} A logical-qubit $\Z_4$ surface code composed of three different boundaries. The logical $\widetilde{Z}$ operator is a vertical string operator, transporting an $e^2$ particle between the rough and even boundaries. The logical $\widetilde{X}$ operator is a horizontal string operator transporting an $m$ particle between the two smooth boundaries. } 
\end{figure}

Suppose the model lives on a square surface $\M$ with a boundary, $\partial \M$ Fig.~\ref{fig:z4surface1}. Then the Hamiltonian in Eq.~\eqref{eq:Hz4} has to be changed to include boundary terms in order to realize these three boundaries:

\begin{equation}\label{eq:hz4bdry}
\begin{aligned}
    H_{\Z_4,\text{Smooth}} &= H_{\Z_4}-\sum_{v\in \partial \M} (A(v)) +H.C.~\\
        H_{\Z_4,\text{Rough}} &=  H_{\Z_4} -\sum_{p\in \partial \M} (B(p)) +H.C.~\\
        H_{\Z_4,\text{Even}} &= H_{\Z_4} -\sum_{p,e\in \partial \M} (B^2(p)+ X^2(e))+H.C.
\end{aligned}
    \end{equation}
Explicitly, the three boundaries can be constructed using the stabilizers in Eq.~\ref{eq:z4bdrystab}. 

\begin{equation}
  \text{\includegraphics[width=0.48\textwidth]{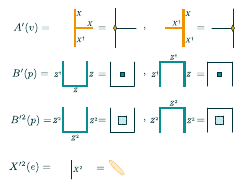}}
  \label{eq:z4bdrystab}
\end{equation}

We see that the boundary terms are derived from the bulk terms with possibly a higher power or restricted support. We will formalize this notion in Sec.\ref{sec:stabilizer_boundaries}, where a general algorithm is given that constructs all boundaries of Abelian twisted quantum doubles in terms of Pauli stabilizers. A $\Z_4$ surface code will then be a choice of these boundaries. We provide one example here, while other examples can be treated analogously using methods of \ref{sec:GSD}.
\begin{Example}[$\Z_4$ on a Surface]
Consider a $\Z_4$ code on a square surface with two opposing (left/right) smooth boundaries as well as an even (top) and a rough (bottom) boundary as in Fig.~\ref{fig:z4surface1}. We used the usual notation where a solid line represents an $e$ particle (pure electric charge) living on the direct lattice, while dashed lines represent an $m$ particle (pure magnetic flux) living on the dual lattice.

The ground state is two-fold degenerate. We discuss methods for computing the ground state degeneracy ($\GSD$) for different surfaces and boundaries in Sec.~\ref{sec:GSD}. The logical operators can be chosen to be the vertical world-line of a $e^2$ that condenses on the even and the rough boundaries, in addition to the horizontal $m$ world-line string that condenses on the smooth boundaries. From the braiding equation Eq.~\eqref{eq:braidz4}, we can read $B(e^2,m)=-1$, and from the fusion Eq.~\eqref{eq:fusion} we have the relations $(e^2)^2=1$. The square of the \textit{entire} $m$ string condenses on the top even boundary, and it is therefore identified with the identity. Since the two logical operators anticommute and square to identity, they generate the one-qubit Pauli group, and we define the $e^2$ and $m$ worldline strings as the $\bar{Z}$ and $\bar{X}$ logical operators, respectively. Note that, had the top boundary been rough, rather than even, the code would have defined a logical $d=4$ qudit. 
\end{Example}

\subsubsection{$D(\Z_4)$ Domain walls and defects} 
In addition to boundaries with vacuum, topological phases can have boundaries with other compatible phases or, in particular, with the same phase. Those boundaries are called domain walls for distinction. They are also called defects, as they usually correspond to defects in the lattice \cite{Bombintwist2010}. We reserve the term defects for the $0$D ends of a finite domain wall \cite{,Barkeshli2013,Barkeshli2023,Barkeshli2024}. Here, we focus on invertible domain walls within $D(\Z_4)$. An invertible domain wall is a wall that can be followed by another wall, producing the transparent (trivial) domain wall that acts as identity \cite{Kong2022}. In $D(\Z_4)$, we have four invertible walls. Labeled by their action on the generators $e$ and $m$, they are: 
\begin{equation}\label{eq:twistsz4}
\begin{aligned}
W^{1} &: (e,m) \mapsto (e,m),\\
W^{e\leftrightarrow e^{-1}} &: (e,m) \mapsto (e^{-1},m^{-1}),\\
W^{e\leftrightarrow m} &: (e,m) \mapsto (m,e),\\
W^{e\leftrightarrow m^{-1}} &: (e,m) \mapsto (m^{-1},e^{-1}).
\end{aligned}
\end{equation}

The duality domain wall $W^{e \leftrightarrow m}$ is also known in the literature as a twist \cite{Bombintwist2010}. The inverse domain wall $W^{e \leftrightarrow e^{-1}}$ was trivial in the $D(\Z_2)$ case. Domain walls can be brought together to form new domain walls. The new product is called stable if no local interaction near the domain wall can change the wall~\cite{Tan2015}. This is denoted by $W^i \otimes  W^j = W^k$. In the $\Z_4$ case, the identity wall (transparent) naturally satisfies $W^1 \otimes W^i = W^i \otimes W^1=W^i$. All walls square to the transparent wall: $W^i \otimes W^i = W^{1}$. Further, $W^{e \leftrightarrow e^{-1}} \otimes W^{e \leftrightarrow m}  = W^{e \leftrightarrow m^{-1}}$ and $W^{e \leftrightarrow m} \otimes W^{e \leftrightarrow m^{-1}}  = W^{e \leftrightarrow e^{-1}}$. It is then clear that the domain walls of $\Z_4$ form a $\Z_2 \times \Z_2$ Abelian group under fusion. The Pauli Stabilizers of these domain walls are given in Eqs.~\eqref{eq:defectlegos1} and \eqref{eq:defectlegos2}. 
 
\begin{equation}
  \text{\includegraphics[width=0.48\textwidth]{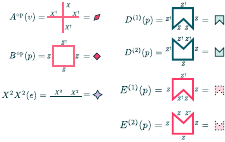}}
  \label{eq:defectlegos1}
\end{equation}
\begin{equation}
  \text{\includegraphics[width=0.50\textwidth]{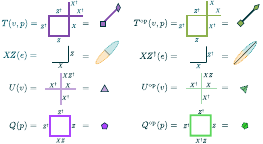}}
  \label{eq:defectlegos2}
\end{equation}

\begin{figure}[htbp]
  \centering
  \includegraphics[width=0.50\textwidth]{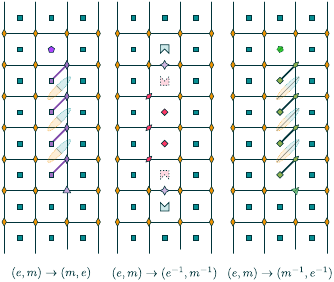}
  \caption{Stabilizers layout for the $3$ non-trivial invertible $\Z_4$ domain walls ending with $0$D defects. Definitions of symbols are given in Eqs.~\eqref{eq:defectlegos1} and \eqref{eq:defectlegos2}.   }
  \label{fig:z4dws}
\end{figure}

The exact layout of the stabilizers implementing the four domain walls is shown in Fig.~\ref{fig:z4dws}. At the end of a finite domain wall, a $0D$ defect lives. These defects can be used for quantum computing on their own \cite{Bombintwist2010,bombin2011clifford,yoder2017surface,kesselring2018boundaries,scruby2020hierarchy,Petiziol2024,Cong2016,Cong2017,Kobayashi2023,Kobayashi2024}. Sec.~\ref{sec:stabilizer_boundaries} discusses how domain walls, in addition to $0D$ defects, can be constructed systematically from bulk stabilizers. Note that the domain walls in Fig.~\ref{fig:z4dws} realize their mappings in any $\Z_n$ for appropriate generalized Pauli matrices, Eq.\eqref{eq:genpauli} (but they do not necessarily exhaust all the invertible domain walls for $\Z_n$ with $n> 4$).

\subsection{Double Semion Twisted Quantum Double}
\label{sec:DS}
 
We are interested in the error-correcting properties of the $\Z_4$ phase and its closely related phases. In this section, we review the doubled semion (DS) phase, which is closely related to $\Z_4$ as we will discuss later in Sec.~\ref{sec:stabilizer_boundaries}. The doubled semion (DS) phase is an Abelian topological order first realized as a string-net model~\cite{levinwen2005}. It is a twisted $\Z_2$ gauge theory. It is also the simplest twisted quantum double, which is a generalization of the Kitaev quantum double~\cite{Yuting2013}. All twisted quantum doubles with Abelian orders can be described in terms of Pauli stabilizers~\cite{Tyler2022}.

\subsubsection{DS Bulk}
To begin, let us consider the bulk stabilizer Hamiltonian describing the DS phase as described in Ref.~\citenum{Tyler2022}. As in the $D(\Z_4)$ case, the model will live on a lattice consisting of qudits. As illustrated in Eq.~\eqref{eq:Z4DSops0}, the operator $F(v,p)=A(v)B(p)$ denotes the product of adjacent $\Z_4$ vertex and plaquette operators, while the $C$ operators are a generalization of the $\Z_2$ twist~\cite{Ani2020}. However, in the context of $\Z_4$, since $[X^2,Z^2]=0$ these operators are now mutually compatible and they may overlap.
\begin{equation}
  \text{\includegraphics[width=0.50\textwidth]{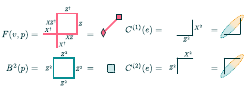}}
  \label{eq:Z4DSops0}
\end{equation}

The bulk DS Hamiltonian is
\begin{eqnarray}
\label{eq:HDS}
H_\text{DS} &=& -\sum_{v \in V}(F(v,p) + H.C.) - \sum_{p \in F}B^2(p) \nonumber \\
            &\; & - \sum_{e\in E} (C^{(1)}(e) + C^{(2)}(e)).   
\end{eqnarray} 
Since all the stabilizers commute, this model likewise constitutes a stabilizer code with a stabilizer group 
\begin{equation}
\mathcal{S}_{\mathrm{DS}} \;=\; \langle\, F,\, B^{2},\, C \rangle.
\end{equation}

The $4$ anyons in the DS model are $\{1,s,\overline{s},b\}$. The $s$ ($\overline{s}$) is called a (anti-)semion, while the $b$ anyon is the boson. These particles obey the fusion rules
\begin{equation}\label{eq:DSfusion}
\begin{aligned}
    s \otimes s &= 1, \quad \overline{s} \otimes \overline{s} = 1, \quad b \otimes b =1 \\
    s\otimes \overline{s} &= b, \quad s \otimes b = \overline{s}, \quad \overline{s}\otimes b = {s},
    \end{aligned}
\end{equation}
which form a $\Z_2 \times \Z_2$ group just like the $\Z_2$ toric code. However, their spin statistics are different. The semion has a spin $i$, which is the square root of the spin of the fermion, whence the name~\cite{Canright1989}.

\begin{equation}\label{eq:DSspin}
\begin{aligned}
    \T(s) = i , \quad \T(\overline{s})=-i, \T(b) =1
    \end{aligned}
\end{equation}

The braiding statistics can again be read from Eq.~\eqref{eq:braid}
\begin{equation}\label{eq:DSbraid}
\begin{aligned}
B(s,s)              &= -1, &\qquad 
B(\overline{s},\overline{s}) &= -1, &\qquad 
B(b,b)              &=  1, \\[6pt]
B(s,\overline{s})   &=  1, &\qquad 
B(s,b)              &= -1, &\qquad 
B(\overline{s},b)   &= -1 .
\end{aligned}
\end{equation}

\begin{figure}[htbp]
  \centering
  \includegraphics[width=0.48\textwidth]{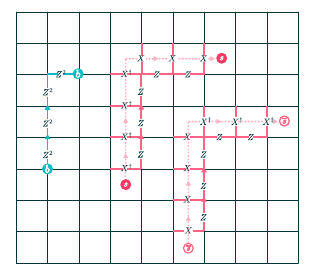}
  \caption{Ribbon operators for DS anyons. As discussed in Eq.~\eqref{eq:dsribbs}, they are inherited from $\Z_4$ ribbon operators.   }
  \label{fig:dsribbs}
\end{figure}

The ribbon operators for the DS anyons are inherited from the $\Z_4$ anyons, Fig.~\ref{fig:dsribbs}. The mapping is as follows:
\begin{equation}\label{eq:dsribbs}
\begin{aligned}
      1& \mapsto  \{1,e^2m^2\}, \quad s \mapsto \{em, e^3m^3\}, \\ \overline{s}& \mapsto \{em^3,e^3m\}, \quad b\mapsto \{e^2,m^2\}
\end{aligned}
\end{equation}  

This redundancy in the ribbon operators stems from the stabilizer embedding of DS in $\Z_4$ as discussed in \ref{sec:stabilizer_boundaries}. In this embedding, the ribbon operator for the $e^2m^2$ anyons in $\Z_4$ commutes with all the stabilizers of the Hamiltonian. One choice for the ribbon operators for the different anyons is shown in Fig.~\ref{fig:dsribbs}. Multiplying the ribbon operators by the one for $e^2m^2$ results in another equivalent ribbon. In this code, the $F$ operators act as syndromes detecting the boson while the $B$ and $C$ operators, having eigenvalue $-1$ denote the presence of a (anti-)semion. Reference~\citenum{Tyler2022} showed how this phase may be put on a torus to construct a pair of logical qubits. In the next sections, we will generalize this result.

\subsubsection{DS Boundaries}
As we briefly described in Sec.~\ref{sec:bdryz4} and will expand on in Sec.~\ref{sec:stabilizer_boundaries}, each boundary is completely determined by its Lagrangian subgroup~\ref{enu:Lagrange}. In the case of DS, we only have one non-trivial boson $b$ and thus only one Lagrangian subgroup:
\begin{equation}
    \L_{\text{DS}}=\{1,b\},
\end{equation}
From the DS's anyonic braiding statistics Eq.~\eqref{eq:DSbraid}, we see that condensing the $b$ boson confines all other non-trivial anyons at the boundary. From the correspondence of Lagrangian subgroups and boundaries, DS has only one boundary. 
\begin{equation}\label{hz4bdry}
\begin{aligned}
    H_{\text{DS},\text{Boundary}} &= H_{\text{DS}}-\sum_{v,e\in \partial \M} (A^2(v)+Z^2(e))   
\end{aligned}
    \end{equation}
   
    As before, the boundary terms are inherited from the bulk terms. The stabilizers of the boundaries are:
\begin{equation}
  \text{\includegraphics[width=0.48\textwidth]{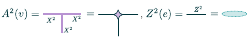}}
  \label{eq:DSBdrystabs}
\end{equation}
    
\begin{figure}[htbp]
  \centering
  \includegraphics[width=0.40\textwidth]{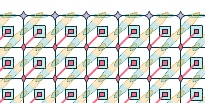}
  \caption{Stabilizers for the unique boundary of the DS phase. At this boundary, only the boson $b$ condenses. }
  \label{fig:dsbdry}
\end{figure}

 Fig.~\ref{fig:dsbdry} shows the exact layout of stabilizers at this unique boundary of DS.
 
This means that, in the absence of any punctures, any DS surface code has $\GSD= 1$. On a torus, however, the model will have $\GSD=4$ just like the toric code. In addition, the only invertible domain wall between two DS phases is the transparent one.

\section{Composite Topological Codes} 
\label{sec:composite}

In this section, we present an investigation of the utility of using composite-dimensional qudit codes exemplified by the $D(\Z_4)$ surface code, the related even code introduced in Sec.~\ref{sec:Z4}, and the DS code introduced in Ref.~\citenum{Tyler2022} and reviewed in Sec.~\ref{sec:DS}. In addition, we also discuss spatially anisotropic topological codes (non-translationally invariant codes). These are exemplified by a hybrid DS-$\Z_4$ patch code introduced below in Sec.~\ref{sec:patch_code}. We discuss the construction of the codes from Pauli stabilizers, their boundary conditions, and their logical operators. Finally, we present numerical simulations to determine their thresholds and logical error rates subject to different noise models.

The $\Z_2$ surface code had only two types of boundaries: smooth and rough \cite{Bravyi1998} and only one type of nontrivial invertible domain walls, namely the one implementing the $e\leftrightarrow m$ duality. With these topological properties, many constructions that are variants of the toric and surface codes have been proposed. These constructions include, but are not limited to, adding punctures to increase the number of qubits in $\Z_2$ surface code~\cite{Dennis2002,brown2017,Benhemou2022}, and adding twists that affect the $\GSD$~\cite{Bombintwist2010,Ani2020,landahl2021}. Although these constructions had diverse properties, they were limited by the possibilities afforded by the dimensionality and group-theoretic properties of $\Z_2$. A similar structure is also common for the slightly higher-dimensional qutrits \cite{iqbalqutrit,fernHowSU2$_4$Anyons2017,Gokhale2019,qcqutrit} and, in fact, all prime-dimensional qudits. 

As reviewed in Sec.~\ref{sec:Z4} and Sec.~\ref{sec:DS}, two new properties are unlocked for the $\Z_4$ surface code: the existence of a new boundary (the even boundary), and the existence of child phases accessible through condensation (the DS phase). This is in addition to new domain walls Eq.\eqref{eq:twistsz4}. The details of the methodological construction of these topological components are deferred to Sec.~\ref{sec:bound_algs}. In this section, we will focus on four main codes: i) the $\Z_4$ planar code, ii) the $\Z_4$ even code, iii) the DS code, and iv) DS-$\Z_4$ hybrid patch code. We begin with an introduction to stabilizer code metrics.

\subsection{Stabilizer Codes}

 Assume we have a certain Hilbert space $\H$ with an associated group of linear operators $\mathcal{G}$. A stabilizer code $\S$ is any Abelian subgroup of operators $\S \subset \mathcal{G}$. For example, $\S= <A_v, B_p>$ for the toric code Eq.\eqref{eq:z4}~\cite{Gottesman1997}. The code subspace $\mathcal{C}$ is a subset of vectors (codewords) in the Hilbert space that are fixed (stabilized) by elements of $\S$: $$\mathcal{C} = \{\ket{c} \ \in \H \ | \ s\ket{c} = \ket{c} \forall s \in \S \}$$
 
 Usually, the stabilizer code is implemented as commuting Hamiltonian terms (with local terms in the case of topological codes). Such a Hamiltonian will have the code subspace as elements of the ground subspace. The dimension of the code space $\dim(\mathcal{C})$ is the logical subspace used for quantum computing. Assuming the code is constructed from logical qubits (which is the case for qubit- and dim-4 qudits), the number $k \in \N$ is defined as: $$k= \log_2\left(\dim(\mathcal{C})\right)$$ The number of physical qubits used to implement the system is usually denoted by $n$. For commensurability, we will continue to count qudit-based codes in terms of qubits, including necessary ancilla qubits. This offers a conservative estimate of the new codes, assuming only native qubits are available. We quickly note that dim-4 qudits are actually easier to manufacture from qubits than dim-3 qutrits since they do not need an extra ancilla~\cite{qcqutrit}. However, the total overhead requires more careful treatment. 

 Logical operators are a subset of operators $L  \in \mathcal{G} $ that commute with the stabilizers $[l,s] = 0, \forall l\in L, s\in \S$. Their action on the codespace is quotiented by stabilizers, since the stabilizers' action is trivial on any codeword. The set of distinct logical operators is then $\widetilde{L} = L/\S$. For any operator, its weight is the cardinality of its support set. That is simply the number of qudits it acts on. Since in topological codes any operator can be multiplied by stabilizers that act trivially, the weight usually refers to the minimum weight within a coset. For a stabilizer code, the code distance is the minimum weight of all non-trivial logical operators and is denoted by $d$ \cite{Gottesman1997}.
 $$d = \min_{l\in L,\ l \notin \S}(||l||) $$
The three numbers $n,k,$ and $d$ are important metrics for any code; the code is usually denoted by the triplet $[[n,k,d]]$ \cite{Gottesman1997}.

Many bounds exist for quantum error-correcting codes; most of them are generalizations of known classical bounds \cite{ knillTheoryQuantumErrorcorrecting1997, calderbankQuantumErrorCorrection1997}. Importantly, the following bound applies for topological stabilizer codes on a Euclidean surface \cite{bravyiTradeoffsReliableQuantum2010}: $$k d^2 \leq cn$$
for some constant $c$ that depends on the weight of the stabilizers and the dimension of qudits constituting the system. This bound shows that the family of topological codes cannot simultaneously satisfy the criteria for a good quantum code \cite{Calderbank1996Goodcodes}: \begin{equation}
k / n \geq c_1 \text { and } d / n \geq c_2
\end{equation}
for constants $c_1$ and $c_2$. Different lattice geometries, such as hyperbolic surface codes or more general qLDPC codes, were shown to evade this bound \cite{breuckmannHyperbolicSemiHyperbolicSurface2017,panteleevAsymptoticallyGoodQuantum2022,bravyiHighthresholdLowoverheadFaulttolerant2024}. Even though Abelian topological codes were shown to not satisfy a good encoding rate, and cannot implement transversal universal quantum computing without either magic-state distillation or non-Abelian ingredients \cite{jochym-oconnorDisjointnessStabilizerCodes2018,Laubscher2019, bravyi_universal_2005,eastin_restrictions_2009,bravyi_classification_2013,litinski_magic_2019,bao_magic_2022,gidney_magic_2024,chen_efficient_2026}, their relative accessibility both theoretically and experimently continue to lead efforts to implement quantum computing and quantum error-correction \cite{exp1,exp3,exp4,exp5,exp6,google2024}.

\subsection{Noise Model}\label{sec:noise_model} Modeling noise is an important procedure to better design quantum error-correction codes with realistic noise resilience. It further affects the decoding strategies and the thresholds obtained from simulations. For the qubit case, two standard error models are widely used. First, the pure bit-flip (or phase flip) Pauli error channel. For a given physical error rate $p$, this channel applies an $X$ operator with probability $p$ independently on each qubit. The pure $X$ channel is often used as a zeroth-order test for codes obtaining the code capacity threshold. Second, the depolarizing noise model applies either $X,$ $Y,$ or $Z$ with probability $p/3$ each~\cite{steaneErrorCorrectingCodes1996,bennettMixedstateEntanglementQuantum1996,shorQuantumErrorCorrectingCodes1996}.

For the case of qudits, we consider widely used generalizations to these two channels \cite{duclos-cianciKitaevsZ_dCodesThreshold2013, andristErrorThresholdsAbelian2015, watsonFastFaulttolerantDecoder2015}. For simplicity, we express these channels in terms of $d=4$ qudits. The generalization for a general dimension is straightforward. For a given physical error rate $p$, we have a pure $X$ noise channel that acts as identity with a probability $(1-p)$ and acts with either $X, X^2$, or $X^3$ with probability $p/3$ each. A pure $Z$ noise channel can be defined in a similar manner. For the depolarizing noise, the channel again acts with identity with probability $(1-p)$, and acts as $X^aZ^b$ for $\{a,b\} \subset \{0,1,2,3\}$ and $(a,b) \neq (0,0)$ with probability $p/15$ each~\cite{duclos-cianciKitaevsZ_dCodesThreshold2013, andristErrorThresholdsAbelian2015, watsonFastFaulttolerantDecoder2015}.

We note that this is a conservative noise model that should work well if the qudits are actually made out of qubits, since it treats all the powers of the generalized Pauli operators equally. If the experimental system has native qudits, one can expect that errors will be further suppressed by their powers. For example, an excitation to the second excited state will have a quadratically suppressed probability.

\subsection{Decoding} \label{sec:decoding} 
There is another way of thinking about codes, which is more amenable to classical error-correction. In a Pauli stabilizer code, the check matrix $H$ contains the code's stabilizers in its row space. The symplectic form of $H$ separates $Z$ and $X$ terms as: $H = [Hz| Hx]$ and thus has a size $ ||\S||\times 2n$, where $||\S||$ is the cardinality of the generators of the stabilizer group. A valid logical ground state $g$ will be a $2n$ vector satisfying $H.g = 0 $.

Generically, the system will be initialized in one of the ground states (or a superposition of them), and the check matrix will then be repeatedly measured to detect errors. If a detectable error occurs, the erroneous state $x$ will fail to satisfy some stabilizers $H.x = s$. Here, $s$ is called the syndrome. The job of a decoder for a stabilizer code is essentially to estimate $x$ given $s$. This is done by finding a correction $c$ such that $H.(x+c) = 0$. The weight of an error is defined as the number of qudits it non-trivially acts on. For example, $e = \begin{bmatrix} 1 & 0 & 2 & 0 & \dots & 0 \end{bmatrix}^{\top}$ means there is a $Z$ error on qudit $1$ and a $Z^2$ error on qudit $3$. Such an error has a weight equal to 2: $wt(e) = 2$. This is in line with the definition for qubits \cite{Gottesman1997,nielsen_chuang_2010}. Ideally, all errors with a Hamming weight less than $t= \lfloor\frac{d-1}{2}\rfloor$ are correctable \cite{knillTheoryQuantumErrorcorrecting1997}. However, in practice, this might not hold for approximate decoders.

There are two major kinds of decoders. First, {Quantum Maximum Likelihood (QML)} decoders estimate the most likely individual error $x$. This problem was shown to be NP-complete \cite{berlekampInherentIntractabilityCertain1978}. Their most famous algorithms include minimum-weight perfect matching (MWPM) \cite{edmondsMaximumMatchingPolyhedron1965,edmondsPathsTreesFlowers1965,Dennis2002,higgottSparseBlossomCorrecting2025}, union find \cite{delfosseAlmostlinearTimeDecoding2021}, renormalization group \cite{bravyiQuantumSelfCorrection3D2013,poulinIterativeDecodingSparse2008,watsonFastFaulttolerantDecoder2015}, Belief propagation \cite{roffeDecodingQuantumLowdensity2020,panteleevDegenerateQuantumLDPC2021}, neural networks and reinforcement learning~\cite{krastanovDeepNeuralNetwork2017,torlaiNeuralDecoderTopological2017,chamberlandDeepNeuralDecoders2018,seniorScalableRealtimeNeural2025,huEfficientUniversalNeuralNetwork2025,zhangLearningNeuralDecoding2025}. These decoders work to best estimate the single error that occurred, usually by assuming that the lower weight errors are more probable. These are highly parallelizable algorithms that are standard for simulations and hardware architecture implementation~\cite{cauneDemonstratingRealtimeLowlatency2024,liyanageFPGAbasedDistributedUnionFind2024,ziadLocalClusteringDecoder2025,wolanskiAmbiguityClusteringAccurate2025,barberRealtimeScalableFast2025,seniorScalableRealtimeNeural2025}.

Second, {Degenerate Quantum Maximum Likelihood (DQML)} decoders attempt to solve the more useful question: what was the logical ground state before the error? That is, instead of estimating only the most likely error $x$, they estimate the most likely logical equivalence class of errors. This problem is harder than finding the most likely error, and it is $\#$P-complete \cite{iyerHardnessDecodingQuantum2013}. One prominent algorithm for these decoders is the BSV decoder, which is based on tensor networks \cite{bravyiEfficientAlgorithmsMaximum2014, ferrisTensorNetworksQuantum2014, chubbGeneralTensorNetwork2021}.

There exist efficient packages for sampling qubit- and qudit-based circuits \cite{gidney2021stim, kabirSdimQuditStabilizer2025}. In addition, many implementations for different types of decoders are available \cite{higgott2022pymatching, wu2023fusion, tuckett2020tailoring, UnionFindCPP, roffe_2022_ldpc, qiskit_qec_2022, roffe_2021_bposd}. However, all of these decoders either assume qubits, prime-dimensional qudits (or treat composite dimensions as Galois fields), or CSS codes. The case with dimension-4 qudits and non-CSS codes is less developed. In the following, we report three prominent types of decoders that we developed.

\subsubsection{MWPM-4 decoder} \label{sec:mwpm_decoder}

The most straightforward way to decode $\Z_4$ is to adapt MWPM \cite{edmondsMaximumMatchingPolyhedron1965,edmondsPathsTreesFlowers1965,Dennis2002,higgottSparseBlossomCorrecting2025} to qudits by splitting the problem into even and odd errors. The intuition is simply that even errors can be paired together $2+2 \equiv 0 \pmod{4}$ while odd errors need to be added similarly $1+3 \equiv 0 \pmod{4}$. We will call this the MWPM-4 decoder. Let us illustrate with the case of $\Z_4$ surface code with pure $X$ noise as an example. Even syndromes will arise from the $X^2$ errors, while odd syndromes will arise from the $X$ and $X^3$ errors. One can then split the syndrome vector $s$ into the odd part: $s_{\text{odd}} \equiv s \pmod{2}$, and the even part: $s_{\text{even}} \equiv \frac{s(s-1)(3-s)}{2}  \pmod{2}$. We can then decode the odd syndromes together. Note that after the modulus operation, the decoder only sees a binary string of $1$s and $0$s, so it will try to match two $X$ errors together ($1+1 \not\equiv 0 \pmod{4} $), for example. After decoding the odd part, there will be even residuals. These even residuals, along with originally even syndromes, can be decoded together in one last step. Of course, this lost information degrades the decoder's performance and does not fully exploit the Hilbert space of the qudits.

 \begin{figure}[htbp]
    \centering
        \begin{subfigure}[b]{0.47\textwidth}
        \includegraphics[width=\textwidth]{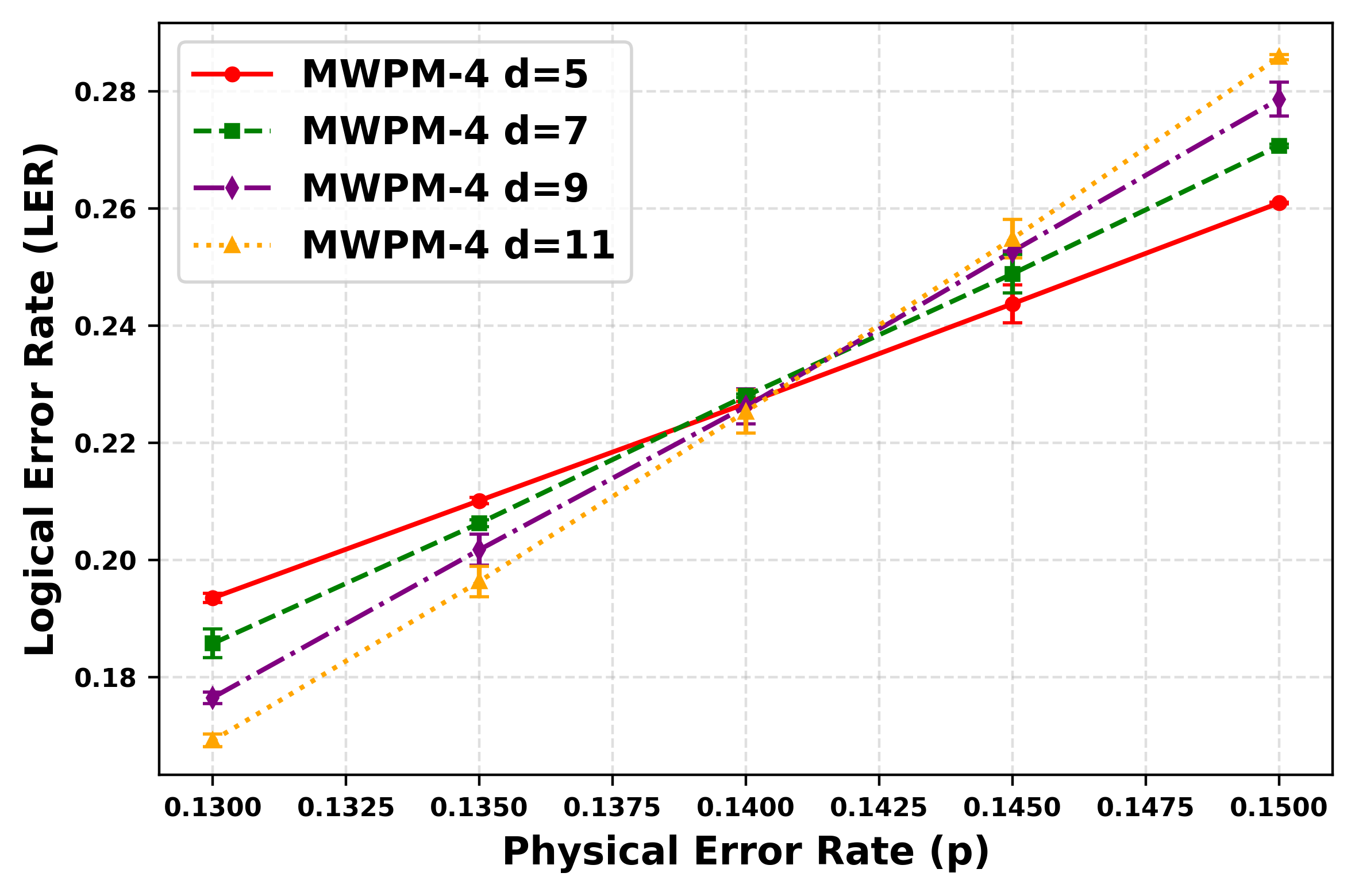}
 
    \end{subfigure}
    \hfill
    \begin{subfigure}[b]{0.47\textwidth}
      \includegraphics[width=\textwidth]{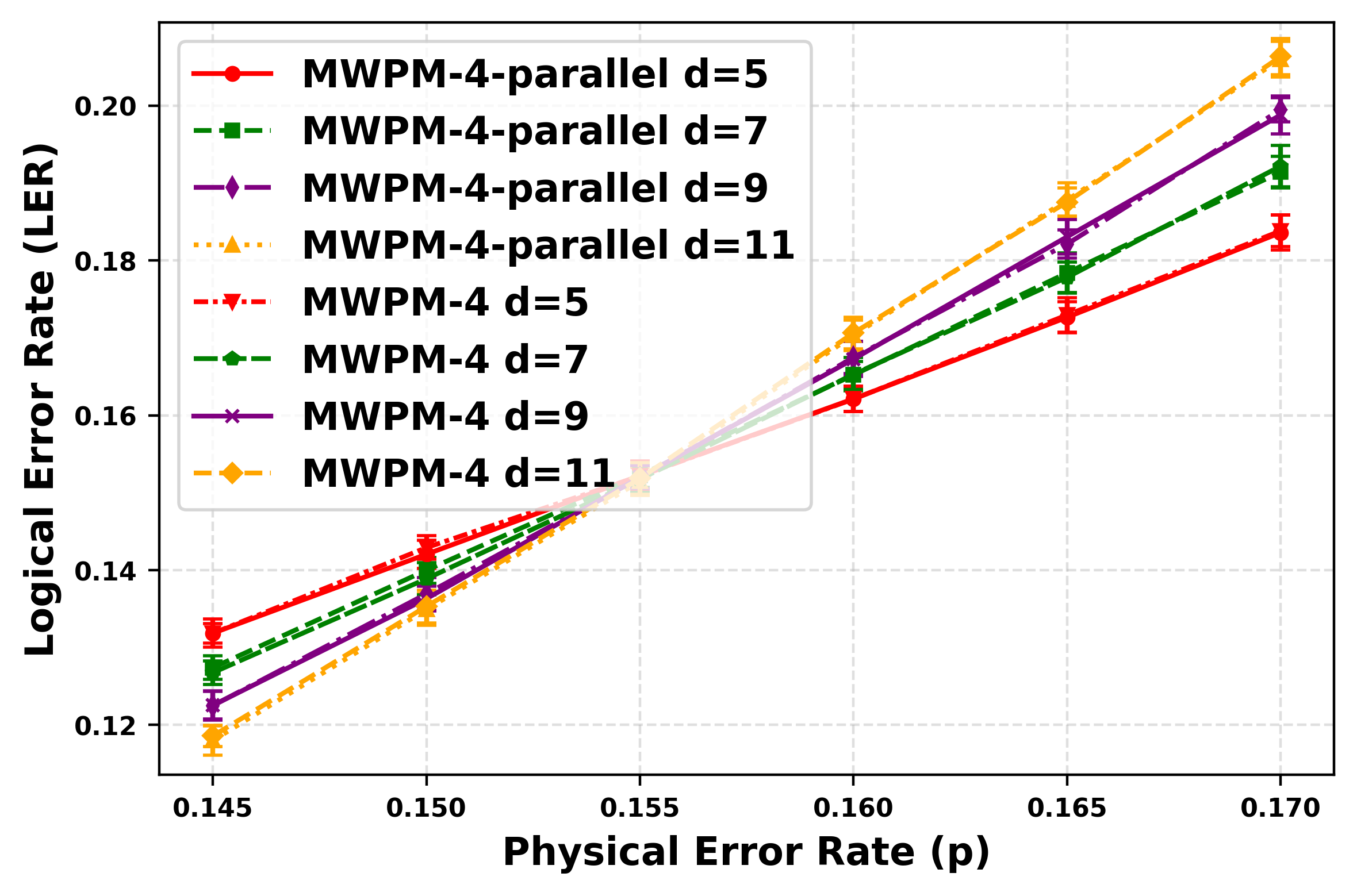}
 
    \end{subfigure}
     \caption{\label{fig:z4_mwpm4} Top: code capacity threshold of the $\Z_4$ surface code, Eq.~\eqref{eq:z4} subject to pure $X$ noise using the MWPM-4 decoder. The threshold is around $14\%$, which is comparable to the $15.25\%$ of Ref.~\citenum{duclos-cianciKitaevsZ_dCodesThreshold2013}. Bottom: code capacity threshold of the even code of Sec.~\ref{sec:even_code} subject to the same noise. The threshold is around $15.5\%$, which is higher than the surface code. The sequential and parallel forms of the decoder behave very similarly.}
\end{figure}

Evidently, using this naive decoder, one finds a code capacity threshold of $14\%$ for the $\Z_4$ surface code as in Fig.~\ref{fig:z4_mwpm4}, which is close to the RG decoders' result of around $15.25\%$ in Ref.\citenum{duclos-cianciKitaevsZ_dCodesThreshold2013}. Both decoders are well lower than the hashing bound of $18.9\%$, which is expected to be saturated from statistical mechanical arguments \cite{duclos-cianciKitaevsZ_dCodesThreshold2013, andristErrorThresholdsAbelian2015}. Even though they are not optimal, this choice of decoders is incredibly fast, as it only scales by a constant factor over standard MWPM decoders. To tackle more codes, such as the even code introduced in Sec.~\ref{sec:even_code}, care must be taken when extracting syndromes and splitting them into their parity parts. As some stabilizers have even orders, they will give even syndromes if the error is odd. This can be easily tackled by accounting for the parity of the order of the stabilizers. Furthermore, instead of following the sequential steps described above, one can parallelize by decoding the $s_{\text{odd}}$ and $s_{\text{even}}$ parts simultaneously. The even part is guaranteed to succeed, while the odd part may leave residuals that require another round of decoding. The number of rounds is again the same; however, the second even round will be much less dense than in the sequential version.

We found that the two methods give nearly identical results near the threshold for the even code. In the surface code, two odd checks can trigger an even syndrome if they touch the same odd check, preventing the parallel decoder from reaching threshold. This problem does not affect the even decoder as the logical operators square to identity. While MWPM-4 is already fast, we note that this caveat may be useful for certain hardware implementations. We finally note that the same scheme can readily work for any dimension that is a power of two (MWPM-$2^{\log_2(d)}$). As for the depolarizing noise channel discussed in Sec.~\ref{sec:noise_model}, the decoder struggles with it, giving $17.25\%$ for the $\Z_4$ surface code and $17.75 \%$ for the even code. While this decoder can handle CSS codes, it requires significant adaptations to handle non-CSS codes. Given its relatively sub-optimal performance, we chose to pursue other decoding strategies.

\subsubsection{ILP decoder} \label{sec:ilp_decoder}

Certainly, MWPM-4 is nowhere near optimal despite its speed. To establish accurate code thresholds, we opted for Integer Linear Programming (ILP) decoders. Evidently, the decoding can be cast as a direct formulation of the most-likely-error problem \cite{landahlFaulttolerantQuantumComputing2011}. Given a check matrix $H$ and a syndrome $s$, the problem is to minimize the Hamming weight of the estimated error $x$ subject to $H.x = 0$. These decoders are widely used for benchmarking other decoders \cite{,takadaIsingModelFormulation2024,cainCorrelatedDecodingLogical2025,beniTesseractSearchBasedDecoder2025,sakashitaFastAccurateDecoder2026}. These decoders are usually very slow as they find the optimal solution. They rely on software packages like CPLEX or Gurobi that solve the problem with exact integers \cite{cplex,gurobi}. The main advantage of these decoders, aside from their accuracy, is that they can handle any type of code or noise. They also handle the $\Z_4$ logic easily by using auxiliary variables to handle the modulus relations. These decoders are not to be confused with Linear Programming (LP) decoders ~\cite{feldmanUsingLinearProgramming2005,liLPDecodingQuantum2018,fawziLinearProgrammingDecoder2021,javedLowComplexityLinearProgramming2024,guPowerLimitationsLinear2025} which involve a relaxation step to continuous variables followed by an ordered statistics decoding (OSD) phase.

Using this decoder, we can find thresholds for all the codes discussed in this work. We give representative results here while summarizing all results in Sec.~\ref{sec:numerical_results}. Using these more accurate decoders, we find a code capacity threshold of $18\%$ for the $\Z_4$ surface code and $17.5\%$ for the even code. The higher thresholds of the $\Z_4$ surface code compared to the $\Z_2$ surface code can be understood as there are different error channels, enabling the decoder to use different syndromes to better predict the correction. Alternatively, if we compare a qubit-based code to a qudit-based code such that each 4-dimensional qudit is composed of two qubits, then for the same physical error rate, one qubit constituting the qudit code experiences almost half the physical error that an individual qubit experiences inside the qubit code. While we expect both codes to have a threshold close to $18.9\%$, which is the hashing bound \cite{andristErrorThresholdsAbelian2015}, two reasons might contribute to this sub-optimality. First, this is a QML decoder and not a DQML decoder. Second, we have set time limits for its runs due to computational time constraints. Evidently, the logical error rate is lower for the even code as the possible logical error paths are fewer.

 \begin{figure}[htbp]
    \centering
        \begin{subfigure}[b]{0.47\textwidth}
        \includegraphics[width=\textwidth]{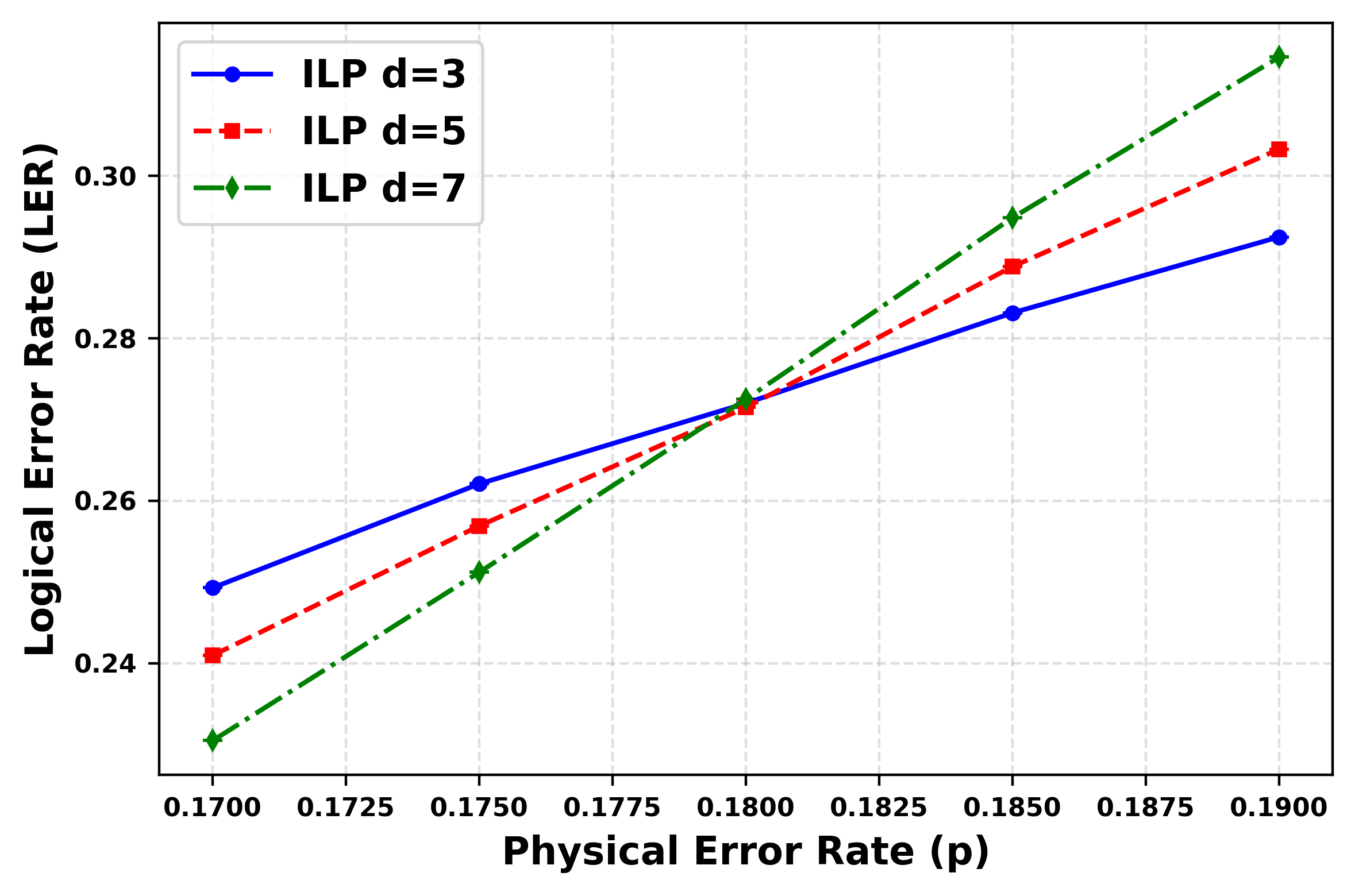}
 
    \end{subfigure}
    \hfill
    \begin{subfigure}[b]{0.47\textwidth}
      \includegraphics[width=\textwidth]{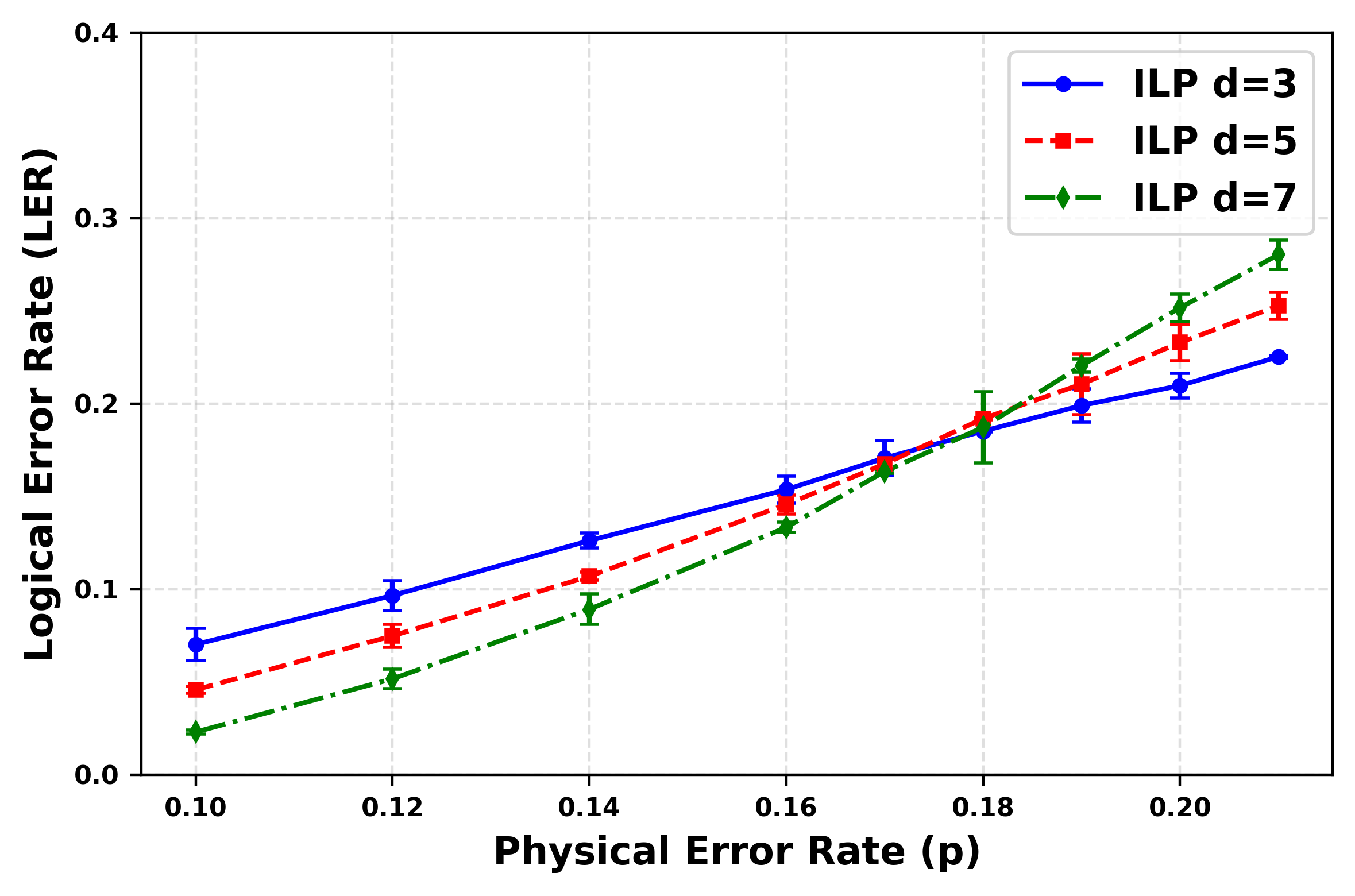}
 
    \end{subfigure}
     \caption{\label{fig:z4_mwpm4} Top: code capacity threshold of the $\Z_4$ surface code, Eq.~\eqref{eq:z4} subject to pure $X$ noise using the ILP decoder. The threshold is around $18\%$, which is comparable to the theoretical maximum $18.9\%$ of Ref.~\citenum{duclos-cianciKitaevsZ_dCodesThreshold2013, andristErrorThresholdsAbelian2015}. Bottom: code capacity threshold of the even code of Sec.~\ref{sec:even_code} subject to the same noise. The threshold is around $17.5\%$.}
\end{figure}

\subsubsection{BP-OSD-4 decoder} \label{sec:bposd_decoder}

 The two extremes of a very fast but less optimal decoder (MWPM-4) and the slow but accurate decoder (ILP) beg for a middle ground where relatively fast decoding gives adequate results. To achieve this, we turned to belief propagation decoding, which is a widely used classical decoder \cite{mackayGoodErrorcorrectingCodes1999}. This message-passing algorithm again tries to find the minimum-weight error $x$ such that $H.x=s$. It computes marginal probabilities for individual qubits based on the information from their adjacent parity and check operators. Importantly, when the Tanner graph \cite{tanner1981recursive} of the check matrix is a tree, it exploits this structure to do fewer summations and is guaranteed to converge \cite{pearl1988probabilistic,kschischangFactorGraphsSumproduct2001}. It was also shown to do well if the graph is sufficiently tree-like \cite{weiss2000correctness,richardson2001capacity,yedidia2003understanding}.
 
 The problem with loops is that they lead to split beliefs: in the quantum case, there are multiple possible solutions due to quantum degeneracy \cite{panteleevDegenerateQuantumLDPC2021,MultipathSummationDecoding}. Two solutions, $H.x_1=s$ and $H.x_2=s$, can each receive a fractional estimate. This will return the solution $x=x_1 + x_2$; however, this only guarantees $H.x \equiv s \pmod 2$, which is not equal to $s$ in general (for the qubit case, $H.x=0$, which fails for any non-zero syndrome). This problem is addressed by the ordered statistics decoder (OSD), which is designed to handle situations where the BP algorithm does not converge. Importantly, the OSD step uses the fractional results of the first BP step as soft decision inputs. This method was used in classical decoding to lower error floors, and was used relatively recently in a quantum setting \cite{roffeDecodingQuantumLowdensity2020, fossorierSoftdecisionDecodingLinear1995,panteleevDegenerateQuantumLDPC2021}.  We defer the specific implementation of the message-passing stage to App.~\ref{sec:belief_propagation_app}, while we describe the OSD stage here.
 
 Let us assume we are dealing with pure $X$ noise for simplicity, as described in Sec.~\ref{sec:noise_model}. When the belief propagation of App.~\ref{sec:belief_propagation_app} fails, it returns fractional values for qudits instead of integer values. This fractional soft information will be a $n \times 4$ matrix $q$, where the row $q_i$'s entries represent the probabilities of the $i$th qudit to have experienced an error of $(1,X,X^2,X^3)$. Intuitively, the qudits that are physically further from the error will receive a high-confidence estimate. Assume qudit $i$ is such a qudit far from the error, the estimated error vector will generically look like $q_i = (0.90,0.05,0,0.05)$, which means we are quite confident this qudit did not experience an error. Another example of reliable qudits may be a pair of qudits that experienced an error while qudits around them did not. In this case, the BP stage will provide a confident estimate of the errors of the two isolated qudits.
 
 The first step in the OSD stage is then to order the qudits in the check matrix by the BP algorithm's confidence in their values, from least reliable to most reliable. The OSD stage is based on the observation that while the check matrix is not full-rank and cannot be used to directly find the error by inversion ($x= H^{-1}.s$), we can still find a full-rank submatrix $H_A$ inside the ordered check matrix $\widetilde{H} = [ H_A| H_B]$ where inversion is possible (i.e. such that $H_A^{-1}.s= x_A$). In the following, we will refer to the two sets of qudits as $\{A\}$ (less reliable and with independent columns) and $\{B\}$ (more reliable). The full estimated error is then $x = [x_A | x_B]$, where $x_B$ is formed from the hard decision of the decoder BP stage.
 
 The second step is to actually invert the submatrix $H_A$ to find the zeroth error estimate $x_A^{[0]}= H_A^{1}.s_A$. In binary codes, the underlying Hilbert space is just the field $\mathbb{F}_2$ where standard Gaussian elimination can be used to invert any square matrix with non-zero determinant. This is the widely used implementation of this step in qubit-based codes~\cite{roffe_2021_bposd,panteleevDegenerateQuantumLDPC2021,roffe_2022_ldpc}. The important difference is that the four-dimensional qudits are a ring and not a field, as evidenced by the non-invertible element $2\times2=0 \equiv \pmod{4}$. The straightforward solution is to use Smith's normal form (SNF) \cite{norton2000structure,dummit2004abstract}, which can handle the case of rings to solve the equation $H_A.x=s$ by decomposing $H_A = UDV$ where $U$ and $V$ are invertible, and $D$ is diagonal. The equation then becomes $U\cdot D \cdot V \cdot x=s$, which can be rearranged to $D \cdot y = t$, with $y \coloneqq V\cdot x $ and $t \coloneqq U^{-1} \cdot s$.

While the SNF method is mathematically clear, the diagonalization step is performed only once for the code-check matrix. In each shot, the two matrix multiplications to revert from the auxiliary variables $y$ and $s$ dominate the computation. A faster method is to perform Gauss elimination in two steps. We reduce rows with unit pivots $\{1,3\}$. Since these units are invertible, we can also normalize so that the pivots are all $1$. Then we reduce over the even pivots without normalizing. These are only row operations, so they are cheaper than the full SNF; however, again, this is done once per code. The bottleneck is the per-shot computation, as we will discuss. After these two reducing steps, the check matrix becomes:

\begin{equation}
\overline{H_A}=\left(\begin{array}{ccc}
I_{r_1} & P_{1 } & P_{f} \\
0 & 2 I_{r_2} & 2 P'_{1 } \\
0 & 0 & 0
\end{array}\right)
\end{equation}

Here, $I_{r_1}$ and $I_{r_2}$ are identity matrices of the unit and even pivots $r_1$ and $r_2$, respectively. $P_1$ and $P'_1$ only have $\{0,1\}$ entries, and $P_f$ has free qudits $n_f$ with no restricted entries. This is Howell's form of the matrix \cite{mulders2003mathematical,sarkar2024qudit}. The crucial point is that, for the even pivots, there is ambiguity in both the SNF and Howell's form. Since an equation like $2x=0$ has two solutions ($x \in \{0,2\}$) over $\Z_4$. We have to do a combination sweep in order to see which choice is better. Since the number of possibilities grows as $2^{|r_2|}$ where $|r_2|$ is the number of even pivots, we can not explore all the possibilities. This situation is similar to the combination sweep \cite{roffeDecodingQuantumLowdensity2020}.

We devised a simple way to tackle this. First, we note that the BP soft information might have been confident that certain qudits should take one of the two choices. We then assign all the qudits of the even pivots to their preferred BP choices. This is the baseline greedy solution. We then try flipping single qudits from the least reliable $\gamma$ qudits among the even pivots. This step can be denoted by $r_2\text{-CS-1}$. Higher combination sweeps ($r_2\text{-CS-$\omega$}$) can also be considered. We empirically found that truncating at $\omega =2$ is satisfactory. It gives total trials $1+\gamma + {{\gamma}\choose{2}}$. For codes of sizes $d \in \{3,5,7,9\}$, a value of $\gamma = 16$ was used. After these trials, we select the trial with the lowest Hamming weight.

 This is a zeroth-order solution since it only solved for the least-reliable qudits $\{A\}$ while fixing the most reliable qudits $\{B\}$ to their BP beliefs. Higher-order OSD involves flipping some of the more reliable qudits $\{B\}$ and seeing if this reduces the Hamming weight of the estimated error. This is called a combination sweep (CS) \cite{fossorierIterativeReliabilitybasedDecoding2001, panteleevDegenerateQuantumLDPC2021, roffe_2021_bposd}. Since the combinations very quickly become intractable, algorithms usually focus on the lowest-weight combinations, like flipping a single or two qubits at a time \cite{fossorierIterativeReliabilitybasedDecoding2001, panteleevDegenerateQuantumLDPC2021, roffe_2021_bposd}. In our qudit case, each qudit must be randomly flipped to one of the $3$ values $\{1,2,3\}$, which adds a constant factor of $3$ for the single-qudit sweeps and a factor of $3^2$ for the pair sweeps. However, some qudits are constrained by even pivots, and thus some of the flipping choices are not solvable. This is checked for every trial, and the trial is skipped if it is unsolvable.
 
 The combination sweep step for the reliable qudits is implemented in multiple ways. For example, one can set a number of free qudits $\lambda$ and try all single and pair combinations in this set~\cite{panteleevDegenerateQuantumLDPC2021}. This will involve $3\times \lambda +9 \times {{\lambda}\choose{2}}$ trials. Alternatively, we can set a number $\Omega$ and try all combinations of Hamming weight $\Omega$ over the entire free-qudit space $n_f$ \cite{roffeDecodingQuantumLowdensity2020}. For $\Omega =2$, this will involve $3\times n_f +9 \times {{n_f}\choose{2}}$. In practice, we found it useful to set an upper bound on the number of trial qudits controlled by the parameter $\Gamma$. We opted for using single combinations up to $3 \times \Gamma$ and pairs up to $\Gamma$. This brings the total trials to $9\times \left[\Gamma + {{\Gamma}\choose{2}}\right]$. We used $n_f$-CS-2 with $\Gamma \text{ up to } 19$. The algorithm is summarized in Alg.~\ref{alg:osd_cs}.
\begin{algorithm}[H]
\caption{OSD-CS for $\mathbb{Z}_4$}\label{alg:osd_cs}
\begin{algorithmic}[1]
  \Procedure{OSD}{$H$, $s$, $q$, $\omega=2$, $\Omega =2$, $\gamma$, $\Gamma$}
    \State Order $H$'s columns by $q$-confidence to form $\widetilde{H} = [H_A | H_B]$.
    \State Reduce $\widetilde{H}$ to Howell's form.
    \State Assign hard beliefs for even pivot qudits $r_2$ and free qudits $n_f$.
    \State Initialize: $e_{\text{best}} \gets \emptyset$, $w_{\text{min}} \gets \infty$, $q_{\text{best}} \gets \infty$.
    \For{even trial sweep $q_\sigma \in \text{even pivots } r_2$ ($\sigma \leq \gamma$)}
        \For{free trial sweep $q_\tau \in \text{free qudits } n_f$ ($\tau \leq \Gamma$)}
            \If{any even pivot has an odd rhs}
                \State \textbf{continue} \Comment{The trial is unsolvable}
            \EndIf
            \State Compute estimated error $e$.
            \If{$\text{wt}(e) < w_{\text{min}}$ \textbf{or} ($\text{wt}(e) = w_{\text{min}}$ \textbf{and} $q < q_{\text{best}}$)} \Comment{Tie-breaking}
                \State $e_{\text{best}} \gets e$, $w_{\text{min}} \gets \text{wt}(e)$
            \EndIf
        \EndFor
    \EndFor
    \State \Return $e_{\text{best}}$
  \EndProcedure
\end{algorithmic}
\end{algorithm}
  
\begin{figure}[htbp] 
        \includegraphics[width=0.47\textwidth]{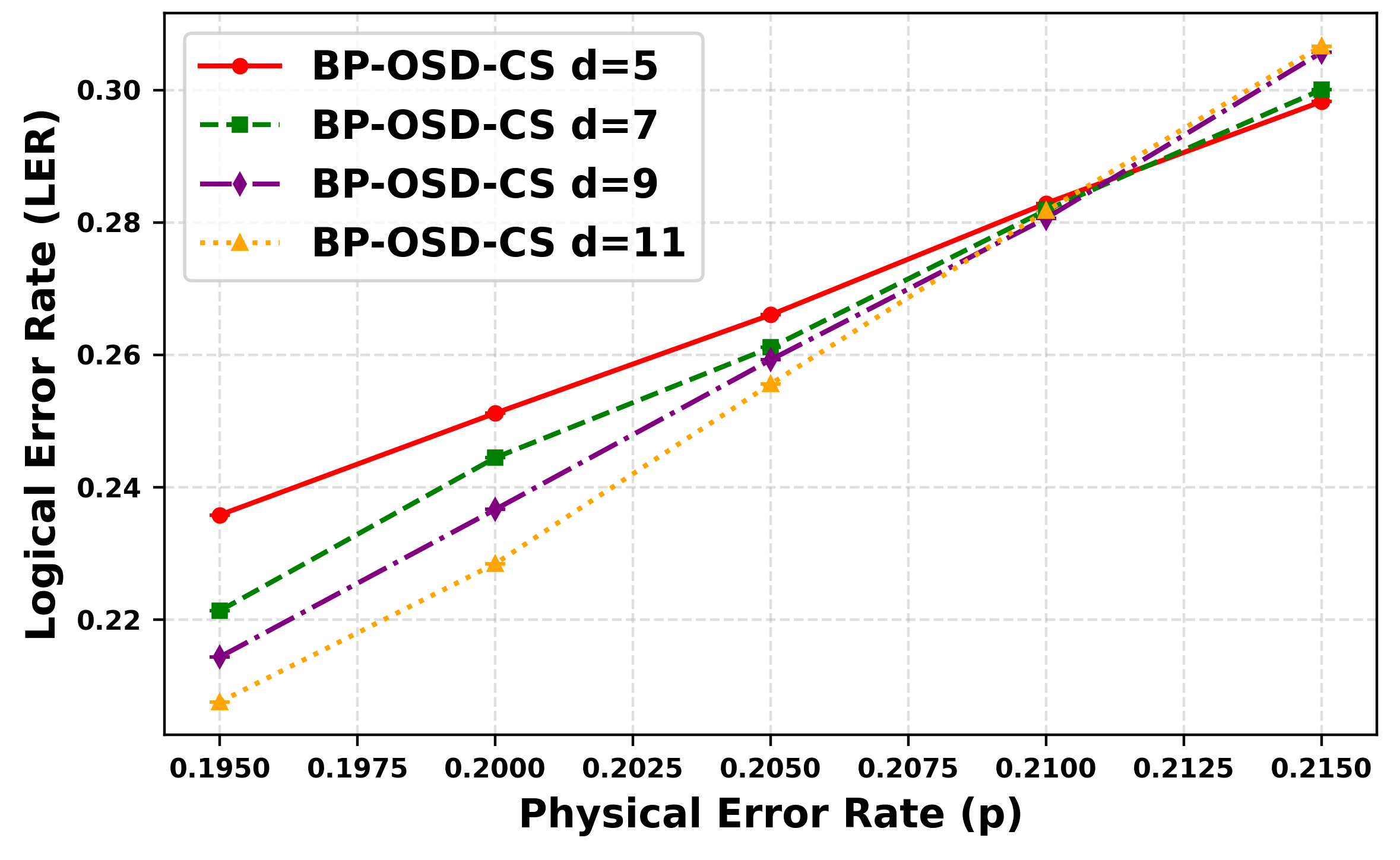}
        \caption{\label{fig:ds_threshold} Threshold calculation of the DS code on a torus of Sec.~\ref{sec:ds_code} subject to pure $X$ noise. This was done with $10^6$ shots with belief-propagation with OSD and CS as illustrated in Alg.~\ref{alg:osd_cs} and in App.~\ref{sec:belief_propagation_app}.}
 
\end{figure}

An example of a threshold calculation is shown in Fig.~\ref{fig:ds_threshold}. Here, the DS code is simulated against pure $X$ noise, achieving a threshold of $21\%$. This is a $10\%$ improvement from the same calculation with machine-learning tools using non-Pauli codes, which gave $9.5\%$ \cite{varonaDeterminationSemionCode2020}.

We finally note that the combination sweep for the free qudits $n_f$ or the even pivots can be done in parallel, avoiding explicitly running the double for loops of Alg.~\ref{alg:osd_cs}.

\subsection{Numerical Simulations of qudit codes}\label{sec:numerical_results}
In this section, we present the numerical results of decoding new composite topological codes using the noise model of Sec.~\ref{sec:noise_model} and decoders of Sec.~\ref{sec:decoding}. Let us begin by providing the explicit stabilizer layouts for these codes. These include the even code introduced in Fig.~\ref{fig:z4surface1}, the DS code explained in Sec.~\ref{sec:ds_code}, and a new hybrid DS-$\Z_4$ code. 

\subsubsection{Planar $\Z_4$ code benchmarking}
\label{sec:planar_code}
\begin{figure}[htbp]
 \includegraphics[width=0.49\textwidth]{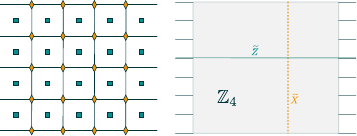}\caption{\label{fig:z4_logical} $D(\Z_4)$ surface code with opposite smooth and rough boundary conditions, and which has $\GSD = 4$. The logical operators $\widetilde{Z}$ and $\widetilde{X}$ are a string of Pauli $Z$s and a string of Pauli $X$s, respectively. }
\end{figure}

Let us begin by discussing the $\Z_4$ surface code. This is the simplest topological code using qudits, and there is extensive literature on its thresholds. The code layout is identical to the well-known surface code but with qudits instead, as shown in Fig.~\ref{fig:z4_logical}. From statistical mechanical arguments, a threshold of $18.9\%$ is expected for the pure $X$ noise \cite{andristErrorThresholdsAbelian2015} while renormalization group (RG) decoders achieved a threshold of around $15.4\%$. Our simplistic MWPM-4 described in Sec.~\ref{sec:mwpm_decoder} achieves a threshold of $14\%$, which is comparable to the RG decoders. However, both the Integer Linear Programming (ILP) solver and the Belief Propagation decoder (BP) achieve a threshold of around $18\%$, which is much closer to the theoretical maximum. For the depolarizing noise, we find thresholds of $27\%$ and $26.5\%$ for ILP and BP, respectively.

\subsubsection{Even code results}
\label{sec:even_code}

\begin{figure}[htbp]
 \includegraphics[width=0.48\textwidth]{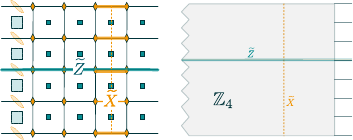}\caption{\label{fig:even_logical} $D(\Z_4)$ surface code with two opposite smooth boundaries, a rough, and an even boundary (here denoted by a zigzag boundary). It has a $\GSD = 2$. The logical operators $\widetilde{Z}$ and $\widetilde{X}$ are a string of Pauli $Z^2$s (denoted with double lines) and a string of Pauli $X$s, respectively. }
\end{figure}

The even code introduced in Fig.~\ref{fig:even_logical} is a simple variant on the $\Z_4$ surface code. It effectively restricts the code to its even sectors, thereby realizing a logical qubit within the logical qudit of the $\Z_4$ surface code. This is achieved by changing one of the rough boundaries of the surface code to a new boundary we call the even boundary; see Sec.\ref{sec:bdryz4}. This changes the logical $\widetilde{Z}$ operator to be a string of $Z^2$ Pauli operators as shown in Fig.~\ref{fig:even_logical}. This lowers the logical error rate by at least half, as the odd $Z$ operators cannot induce a logical error. We could not find any results for this code in the literature, although we expect its thresholds to be similar to those of the $\Z_4$ surface code, as they share the same bulk. Indeed, we found for the pure $X$ noise a code capacity threshold of $17.5\%$ and $18\%$ for the ILP and BP decoders, and only $15.5\%$ for the MWPM-4 decoder. For the depolarizing noise, ILP and BP gave $27\%$ and $26\%$, respectively. The MWPM-4 decoder cannot handle the complex depolarizing noise over $\Z_4$. Importantly, there is a biased suppression of pure $Z$ vs pure $X$ noise due to the asymmetric nature of this code. While they have the same thresholds, the logical error rate for the $Z$ noise was found to be significantly lower (by almost a factor of $2$ at $d=5$ and $p=10\%$).

\subsubsection{DS code results }
\label{sec:ds_code}
A DS code can be defined on a torus just like the toric code \cite{Kitaev2003, Yuting2013}. The code can be realized using qubits and non-Pauli stabilizers, as it is a twisted quantum double. It was simulated using machine learning decoders subject to pure $X$ noise, achieving a threshold of $9.5\%$ and a threshold of $10.5\%$ for depolarizing noise \cite{varonaDeterminationSemionCode2020}. The same reference presents thresholds of $7.6 \%$ (pure $X$) and $7.4\%$ (depolarizing) with MWPM decoder, and  $9.4 \%$ (pure $X$) and $10.5\%$ (depolarizing) with multi-layer perceptron decoding.

\begin{figure}[htbp]
 \includegraphics[width=0.48\textwidth]{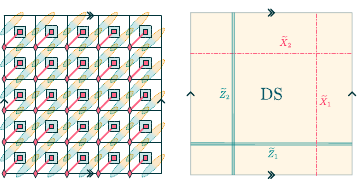}\caption{\label{fig:dslogical} Double Semion code on a torus which has $\GSD = 4$. The logical operators $\widetilde{Z}$ are strings of Pauli $Z^2$ creating the boson $b$. While the logical $\widetilde{X}$ are strings of Pauli $X$ and $Z$s creating the semion $s$; see Fig.~\ref{fig:dsribbs}. }
\end{figure}

Equivalently, the code has a description in terms of generalized Pauli stabilizers for 4-dimensional qudits \cite{Tyler2022}. Using this later description, we were able to simulate the code using an ILP decoder, obtaining thresholds of $17.5\%$ and $20.5\%$ for pure $X$ and depolarizing noise, respectively. Moreover, the BP decoder achieves thresholds of $20.5\%$ and $22\%$, respectively. These numbers are almost double the thresholds for the qubit case due to the fact that half of the anyons of the $\Z_4$ are confined in the DS phase; see Fig.~\ref{fig:bdrycond}. These confined anyons translate to error strings that the decoder can see and reliably correct since it sees the whole string and not only the two ends.

These simulations used a number of shots up to $5\times 10^6$ and a capped ILP time of $12s$ per shot. This time limit explains the less-expected result of the ILP solver getting a lower threshold than BP. We also note that BP with OSD can match the ILP accuracy with a very exhaustive OSD stage as discussed in Sec.~\ref{sec:bposd_decoder}. Finally, similar to the qubit case, the twisted nature of the code results in the depolarizing noise threshold being close to the pure noise. This can be easily seen in the Pauli stabilizer case simulated here, as condensing the anyon $e^2m^2$ induces an equivalence between $X$ and $Z$ errors.

\subsubsection{Hybrid DS-$\Z_4$ code results}
\label{sec:patch_code}

\begin{figure}[htbp]
 \includegraphics[width=0.48\textwidth]{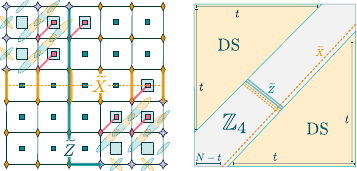}\caption{\label{fig:hybrid_logical} Layout for a hybrid DS-$\Z_4$ code on a plane which has $\GSD = 2$. The logical operators $\widetilde{Z}$ are strings of Pauli $Z^2$ creating the boson $b$. While the logical $\widetilde{X}$ are strings of Pauli $X$. Alternatively, a string of simultaneous $X^2$ and $Z^2$ can also create a logical $\widetilde{X}$ between the two DS patches. }
\end{figure}

Finally, the most interesting case is when we spatially combine DS and $\Z_4$ phases to form a spatially anisotropic composite-dimensional code. One simple layout for this code is shown in Fig.~\ref{fig:hybrid_logical}. This code combines two phases with two thresholds. As we found in Sec.~\ref{sec:planar_code} and Sec.~\ref{sec:ds_code}, the DS code has a higher pure $ X$-noise threshold, while the planar code has a higher depolarizing-noise threshold. It is hard to determine a priori the new thresholds for the new code, given the BP algorithm as a decoder.

We still expect that the logical error rate will be lower than that of the planar code, as the average weight of the logical $\widetilde{X}$ operators is now higher. The domain wall between the DS and $\Z_4$ condenses $\{1,e^2m^2\}$ of the $\Z_4$ anyons, while all anyons of DS can propagate into the $\Z_4$. In contrast to our prior examples, involving a single TQFT, the $\L_{\text{$\Z_4$-DS}}$ domain wall is non-invertible. This implies that particles do not tunnel across symmetrically. For example, the $e$ anyon cannot tunnel into the DS phase as it is confined. This is natural since the two phases have a different number of anyons. We defer the details of how this domain wall is constructed, and its relation to anyon condensation, to Ex.~\ref{ex:z4dsdwsimple}. Since the wall is non-invertible, its action cannot be read from the generators $\{e,m\}$ alone. A more complete description of domain walls and their relations to degeneracy is discussed in Sec.~\ref{sec:lego_code}. For illustration, we now give the DW's action on each anyon of $\Z_4$. 
\[
W^{\text{$\Z_4$-DS}}:
\begin{cases}
1,e^2m^2   & \mapsto 1, \\[6pt]
em,e^3m^3  & \mapsto s, \\[6pt]
em^3,e^3m  & \mapsto \overline{s}, \\[6pt]
e^2,m^2    & \mapsto b,\\[6pt]
\text{otherwise} & \mapsto \text{confined}
\end{cases}
\]

\begin{figure}[h]
\includegraphics[width=0.48\textwidth]{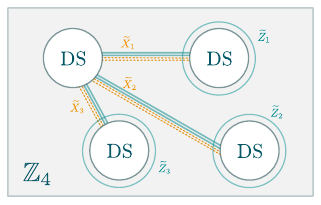}
\caption{\label{fig:ds_patches} Hybrid three-logical-qubit $\Z_4$-DS patch code comprising of four DS patches inside a $\Z_4$ bulk. The logical $\widetilde{Z}$ operators consist of $e$ loops (solid lines) around three patches. The logical $\widetilde{X}$ operators are an $e^2m^2$ ($2$ solid lines and $2$ dashed lines) string between three patches and one patch.  }
\end{figure}

One can naturally explore other layouts for these codes. For example,  we can push the patches to the bulk of the $\Z_4$ as in Fig.~\ref{fig:ds_patches}, also see Ex.~\ref{Example:z4noncontpatches} gives $\GSD = 8$.

The patches can exchange $e^2m^2$ anyons between them. These strings commute and square to one $e^4m^4=1$. We therefore identify them with the Pauli $\widetilde{X}_i$ operators on each of the three logical qubits. In addition to patches exchanging $e^2m^2$s, we can also form non-contractible loops, of $e$ or $m$ world-lines, around each patch, which are confined in the DS phase, and they cannot therefore be contracted to trivial loops inside the patch. These loops are depicted in Fig.~\ref{fig:ds_patches}. Interestingly, the loops again square to unity, this time because the $e^2$ anyons are \textit{deconfined} inside $DS$ and the, e.g., $e^2$, loop can be contracted. Further, the  $e$ or $m$ world-line loops anticommute with the Pauli $\widetilde{X}_i$ operators as $e^2m^2$ and $e$ anticommute Eq.~\eqref{eq:braidz4}. They can then be used to define the Pauli $\widetilde{Z}_i$ logical operators acting on each new logical qubit.

To summarize this section, we present the thresholds of different codes subject to different noise models and using ILP, BP-OSD-CS, and MWPM-4 decoders in Table~\ref{tab:codes_thresholds}.

\begin{table}[h]
\centering
 
\renewcommand{\arraystretch}{1.3}
\begin{tabular}{|c|c|c|c|c|c|}
\hline
\textbf{\#} & \textbf{Codes} & \textbf{Noise} & \textbf{ILP} & \textbf{BP-CS} & \textbf{MWPM-4} \\
\hline
\multirow{2}{*}{1} & \multirow{2}{*}{Planar $\mathbb{Z}_2$} & Pure X & 11\% & 10.75\% & 10.3\% \\ \cline{3-6}
 & & Depol. & 18\% & 16\% & 15.25\% \\
\hline
\multirow{2}{*}{2} & \multirow{2}{*}{Planar $\mathbb{Z}_4$} & Pure X & 18\% & 18\% & 14\% \\ \cline{3-6}
 & & Depol. & 27\% & 26.5\% & 17.25\% \\
\hline
\multirow{2}{*}{3} & \multirow{2}{*}{Even} & Pure X & 17.5\% & 18\% & 15.5\% \\ \cline{3-6}
 & & Depol. & 27\% & 26\% & 17.75\% \\
\hline
\multirow{2}{*}{4} & \multirow{2}{*}{DS} & Pure X & 17.5\% & 21\% & N/A \\ \cline{3-6}
 & & Depol. & 20.5\% & 22\% & N/A \\
\hline
\multirow{2}{*}{5} & \multirow{2}{*}{DS - $\mathbb{Z}_4$} & Pure X & 18.5\% & 19\% & N/A \\ \cline{3-6}
 & & Depol. & 26.5\% & 26.5\% & N/A \\
\hline
\end{tabular}
\caption{Code capacity thresholds across different noise models and decoders.}
\label{tab:codes_thresholds}
\end{table}
 
\begin{figure}[htbp]
        \includegraphics[width=0.50\textwidth]{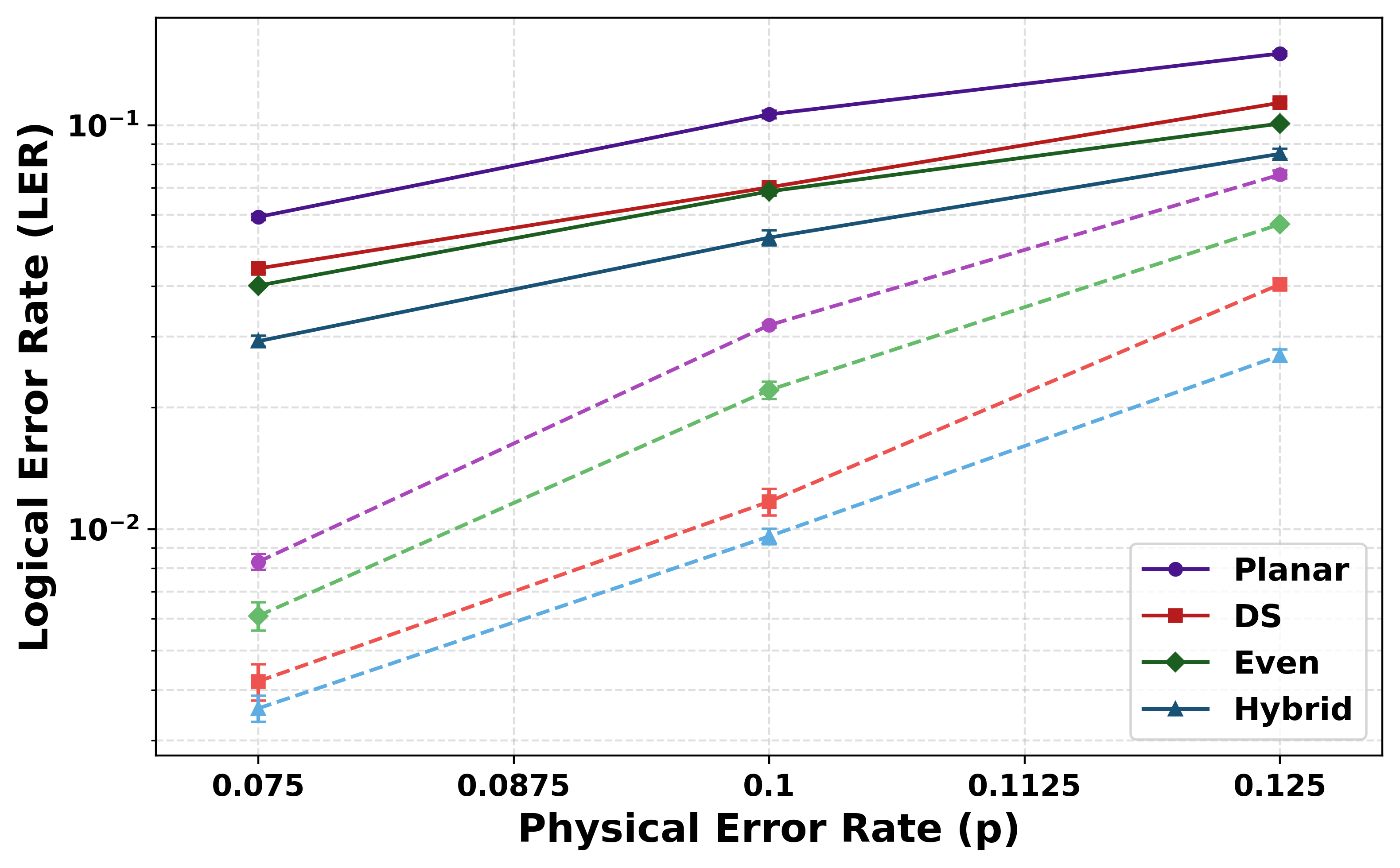}
 
     \caption{\label{fig:lers_codes} Comparison of logical error rates among different codes subject to pure $X$ noise using the belief-propagation algorithm discussed in Sec.~\ref{sec:decoding}. Solid lines represent $d=3$ while dashed lines represent $d=7$. The plot shows monotonic error suppression between the different codes.}
\end{figure}

The computing resources used to produce the table results are a single node with 64 cores (multicore processing on a single node), with a maximum job run time of 24 hours for the ILP decoder. However, for the latest belief propagation-based decoder, the run times are largely less than 10 hours. We used shots between $5\times10^5$ and $10^8$. These results show that using the same number of qudits, one can still achieve different thresholds or logical error rates by virtue of their connectivity. Indeed, as shown in Table~\ref{tab:codes_nkd}, the encoding rates and number of qubits (assuming each qudit is made out of two qubits) are comparable for these codes, while they still exhibit different behaviors as evidenced in Table~\ref{tab:codes_thresholds}. As shown in Fig.~\ref{fig:lers_codes}, the logical error rates are successively suppressed as we change the codes. For instance, both the even and the hybrid code use the same asymptotic number of qudits, however, the hybrid code has a lower logical error rate. This shows that spatial structure, composite-dimensionality and stabilizer connectivity are three valuable resources for topological codes.
\begin{table}[h]
\centering
 \renewcommand{\arraystretch}{1.3}
\begin{tabular}{|c|l|c|c|c|}
\hline
\textbf{\#} & \textbf{Codes} & \textbf{$n$} & \textbf{$k$} & \textbf{$d$} \\
\hline
1 & Planar $\mathbb{Z}_2$ & $2d^2$ & 1 & $d$ \\
\hline
2 & Planar $\mathbb{Z}_4$ & $4d^2$ & 2 & $d$ \\
\hline
3 & Even & $4d^2$ & 2 & $d$ \\
\hline
4 & DS & $4d^2$ & 4 & $d$ \\
\hline
5 & DS - $\mathbb{Z}_4$ & $4d^2$ & 2 & $d$ \\
\hline
\end{tabular}
\caption{Code parameters $n$, $k$, and $d$ for each evaluated code family.}
\label{tab:codes_nkd}
\end{table}

\section{Microscopic Boundary Formulation}\label{sec:stabilizer_boundaries}

To make the results and codes above rigorous, this section derives the lattice models presented and provides an algorithm to do so. Our algorithm takes as input a set of stabilizers and returns a new set of stabilizers describing the new Abelian twisted quantum double, which may contain boundaries and domain walls in addition to bulk terms. We present examples highlighting its use in constructing DS patches embedded in a $\mathbb{Z}_4$ parent phase at both a concrete and conceptual level.

Conceptually, we show how the understanding of the classification of boundaries can aid their construction from the bulk Hamiltonian terms. This is manifested in a concrete lattice-level algorithm. The concrete construction of the new codes in the previous section is a by-product of this algorithm. Its utility can extend well beyond that. The intuition and perspective presented here can be especially useful for the recently proposed schemes that use multiple phases to perform universal quantum computing \cite{huang2025hybrid,manjunath2026universal}.

This section is organized into five subsections. Sec.~\ref{sec:bdry_class} lays out the classification of the boundaries and discusses differences from other works, Sec.~\ref{sec:condensation} explains at a conceptual level why boundaries can be constructed from bulk terms, Sec.~\ref{sec:bdryalg} gives the concrete lattice-level algorithm for their constructions, Sections~\ref{sec:bound_algs} and ~\ref{sec:DWalg} give examples of boundaries and domain walls construction. The construction of $0D$-defects is deferred to App.~\ref{sec:appinvdw}.

\subsection{Boundaries Classification}\label{sec:bdry_class}
The classification of the boundaries and domain walls of a twisted quantum double based on the (not necessarily Abelian) group $G$ is given by the Lagrangian algebras of the topological order~\cite{Kong2014,Kong2017}. As seen in Sec.~\ref{enu:Lagrange}, the set of mutually compatible condensing anyons completely specifies a boundary. In the case of untwisted quantum doubles, these Lagrangian subgroups are in one-to-one correspondence with a subgroup $K \subseteq G$ and a function $\omega: K \times K \to U(1)$ that associates a $U(1)$ phase to each ordered pair of elements in the subgroup $K$. This function is a 2-cocycle since it obeys:
\begin{align}
\delta\omega(g_{1},g_{2},g_{3})
      &=\frac{\omega(g_{2},g_{3})\,\omega(g_{1},g_{2}g_{3})}
             {\omega(g_{1}g_{2},g_{3})\,\omega(g_{1},g_{2})}=1,
      &&\forall\,g_{1},g_{2},g_{3}\in G,\\[6pt]
\omega(e,g)&=\omega(g,e)=1,
      &&\forall\,g\in G.
\end{align}
The first line is the trivial coboundary condition, while the second line is a gauge choice that can always be made \cite{Beigi2011}. Further, the function $\omega$ is defined only up to an equivalence class $[\omega] \in H^2(K,U(1))$.

As an example, we provide a translation between the two descriptions for the case of $\Z_4$.

\begin{table}[h]
\begin{tabular}{|c|c|c|c |}
\hline
boundary       & $K$   & [$\omega]$ & $\L$    \\ \hline
rough & $\{0\} $ &  trivial   & $\{1,e,e^2,e^3\}$        \\ \hline
smooth &  $\{0,1,2,3\} $    &  trivial   &  $\{1,m,m^2,m^3\}$       \\ \hline
even & $ \{0,2\}$  &  trivial   & $\{1,e^2,m^2,e^2m^2\}$        \\ \hline
\end{tabular}
\caption{Rosetta stone with cocycle ($K\subseteq G$,$\omega$) and Lagrangian subgroup $\L$ descriptions of $\mathbb{Z}_4$ boundaries.}
\end{table}

In the case of twisted quantum doubles, extra constraints hold for the boundary \cite{tqdbdry}. For example, the DS phase is based on the group $\Z_2$, which has two subgroups, namely the trivial subgroup and the whole group. However, since it is a twisted quantum double with a non-trivial 3-cocycle, it cannot have a boundary with a non-trivial subgroup \cite{tqdbdry}.

\begin{table}[h]
\begin{tabular}{|c|c|c|c |}
\hline
 Boundary       & $K$   & [$\omega$] & $\L$    \\ \hline
 DS-Vacuum & $\{0\} $  &  trivial   & $\{1,b\}$        \\ \hline
\end{tabular}
\caption{Translation between subgroup and cocycle ($K\subseteq G$,$\omega$) and Lagrangian subgroup $\L$ descriptions of the boundaries of the DS phase.}
\end{table}

In what follows, we will not deal with the classification in terms of cocycles directly. Instead, we will only use the physically intuitive Lagrangian subgroup of anyons explicitly, while the cocycle picture will be implicit. This is contrary to Refs.~\citenum{ Beigi2011,tqdbdry,Cong2016, Cong2017} that abstractly formulated how to construct microscopic lattice models for a (twisted) quantum double's gapped boundary or domain walls using cocycles. In the case of quantum doubles, they are formed by defining subgroup variants of the star and plaquette operators \eqref{eq:z4}. The modified star operators have a modified action that associates a $U(1)$ phase depending on the qudits' values. Consequently, they are not readily implementable experimentally. This issue will arise for $1D$ domain walls and $0D$ defects that will need an extra $U(1)$ phase (those with a nontrivial 2-cocycle)~\cite{ Beigi2011,tqdbdry,Cong2016, Cong2017}. From another perspective, these abstract terms, while correct, are not derived from the bulk terms. 

Other constructions were proposed recently to obtain Pauli stabilizers for the boundaries of Abelian (twisted) quantum doubles \cite{operatoralg2024}. One important difference is that our algorithm also produces the $0$D defects. In addition, it produces topologically complete models, meaning that the only local operators that commute with the stabilizers are products of other stabilizers~\cite{bravyi2010,operatoralg2024}. Other choices for the initial (twisted) quantum double stabilizers are still valid for our algorithm. For example, spatially disconnected stabilizers introduced in Ref.~\citenum{operatoralg2024}, will not require extra steps to ensure topological completion in our construction.

\subsection{Boundaries by Local Condensation}
\label{sec:condensation}
The operators presented in the previous works~\cite{ Beigi2011,tqdbdry,Cong2016, Cong2017}, while sufficient to mathematically study the properties of each boundary, do not provide a simple prescription for experimental implementation. To this point, for \textit{Abelian} twisted quantum doubles, a description of the bulk in terms of generalized Pauli stabilizers was recently provided~\cite{Tyler2022}. However, a simple construction of Pauli stabilizers for the boundaries and domain walls was still lacking. Condensation of anyons is a suitable tool to study boundaries \cite{Kong2014,Burnell2018,colorcond2024}. Inspired by the correspondence between the classification of boundaries and Lagrangian subgroups of the bulk, we present an algorithm to construct different boundaries, domain walls, and $0D$ defects of Abelian topological orders described by (twisted) quantum doubles using local condensation.

Our construction of the boundaries is similar to the procedure for condensing the bulk of the untwisted quantum double to produce twisted quantum doubles~\cite{Tyler2022}, but with two modifications. First, we condense a Lagrangian group instead of a partial condensation that leaves non-trivial topological order afterward. Second, we only condense a region in space that leaves the boundary terms we are interested in.

 \begin{figure}[htbp]
\includegraphics[width=0.48\textwidth]{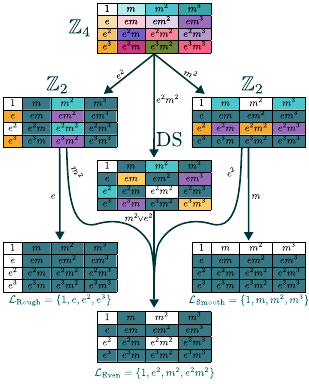}
\caption{\label{fig:bdrycond}Different condensation paths starting from a $D(\Z_4)$ bulk. The top three paths $\{e^2,m^2, e^2m^2\}$ produce $\Z_2$ toric codes with different domain walls with the $\Z_4$ and DS phase, respectively. Further condensing $\{m^2,e^2,e^2 \sim m^2 \}$ produces the even boundary of the $\Z_4$. Anyon labels with the same color are identified, while those with a dark background are confined. Seen differently, the final paths to $\Z_4$ boundaries are also different boundaries for the $\Z_2$ and DS embedded in the $\Z_4$. }
\end{figure}

\textit{Partial condensation}-- In $\Z_p$ where $p$ is a prime, only rough and smooth boundaries exist. Thus, all punctures in $\Z_p$ are either rough or smooth or a mixture of them. Anyon condensation is much richer in composite dimension qudits $\Z_{pq}$. This can be illustrated with the example of $\Z_4$ ($p=q=2$), as summarized in Fig.~\ref{fig:bdrycond}. Starting with the bulk theory ($\Z_4$), one can condense any set of bosons that mutually braid trivially, relaxing the last requirement of~\ref{enu:Lagrange} (that anyons outside the condensed subgroup braid non-trivially with an anyon in the group). For example, condensing only $e^2$ in $\Z_4$ is allowed and results in the $\Z_2$ Toric code \cite{iqbal2018}, which can be thought of as a new kind of puncture inside the $\Z_4$. This condensation is partial in the algebraic sense since, after condensation, non-trivial topological order remains. For example, condensing $e^2$ leaves the anyon classes $\{[1],[e],[m^2],[em^2]\}$. These are the remaining classes because condensing $e^2$ identifies $1 \sim e^2 $, $e\sim e^3$, $m^2\sim e^2m^2$ and $em^2 \sim e^3m^2$. The rest of the anyons (e.g. $m$) are now confined, as they braid non-trivially with one of the condensed anyons (e.g. $e^2$). Confined anyons require energy to propagate in the phase, and they are no longer elementary excitations. This process is shown in Fig.\ref{fig:bdrycond}, where confined anyons have a darker background and anyons of the same color belong to the same class. Thus, condensing $e^2$ yields the $\Z_2$ toric code, where the spins and braiding relations are inherited from the $\Z_4$ operator algebra. In the same way, as discussed in Sec.~\ref{sec:DS}, condensing $e^2m^2$ gives the DS phase. In general, all twisted quantum doubles with Abelian orders can be constructed similarly from condensing certain quantum doubles~\cite{Tyler2022}.

\textit{Maximal condensation}--On the other hand, condensing a Lagrangian subgroup~\ref{enu:Lagrange} will leave only the trivial topological order. One such Lagrangian subgroup is $\L_{rough}=\{1,e,e^2,e^3\}$. This subgroup is generated by condensing either $e$ or $e^3$. If we condense $e$ after condensing $e^2$, we end up with the trivial topological order where all deconfined anyons are identified with identity. This is the bottom left of Fig.\ref{fig:bdrycond}.

\textit{Local condensation}--Condensing a set of anyons globally, whether a Lagrangian group or one of its subgroups, will switch the topological phase into another one. However, if the condensation is carried over locally, we end up with two phases with a domain wall at their interface. If one of the phases is vacuum, then the interface will be a boundary (maximal condensation); otherwise, it will be a domain wall between two phases (partial condensation). In the $\Z_4$ example, condensing $\L_{rough}=\{1,e,e^2,e^3\}$ over a region $R$ of space will result in vacuum in that region. Additionally, the boundary $\partial R$ will correspond to the rough boundary of $\Z_4$ Eq.~\eqref{eq:z4Ls}. Similarly, condensing $e^2$ over a region $R$ will result in $\Z_2$ over that region. The boundary $\partial R$ will be a domain wall between $\Z_4$ and $\Z_2$. Condensing $m^2$ over $R$ would have resulted in the $\Z_2$ phase over that region. However, it will correspond to a different domain wall between $\Z_4$ and $\Z_2$. The correspondence between anyon condensation and boundaries ensures that all domain walls can be constructed this way.

The takeaway from the $\Z_4$ example is that condensing anyons over a region $R$ should be enough to construct all possible boundaries at $\partial R$. Since any domain wall between two phases is a boundary of a related phase, we will focus on boundaries without loss of generality. We will treat concrete examples of domain walls in Sec.\ref{sec:DWalg} and finite domain walls ending with $0$D defects in App.~\ref{sec:appinvdw}.

\subsection{Algorithm to Construct Boundary Stabilizers}\label{sec:bdryalg}
 
We now describe the algorithm for constructing the boundary stabilizers. Suppose we have the (twisted) quantum double, which has the bulk Hamiltonian: 
\begin{equation}
    {H}=-\sum_vA(v)-\sum_p B(p)-\sum_e C(e) + H.C.
\end{equation}

where the $C(e)$ are edge terms required for the case of twisted quantum doubles (e.g. Eq.\eqref{eq:HDS}). Here, the summation is over all vertices $V$, edges $E$, and plaquettes $P$ as the model is assumed to live on a closed surface $\M$. And, as before, the star and plaquette operators are the standard tensor products of generalized Pauli operators; Eq.~\eqref{eq:Hz4}. The terms of this Hamiltonian generate a stabilizer group, which we denote by $\mathcal{S}_H$.
\begin{equation}
    \mathcal{S}_H = \langle A,B,C \rangle
\end{equation}

We want to modify both the Hamiltonian and the stabilizer group to have a spatial boundary $\partial R$ bounding a region $R \subset \M$ of the lattice. In addition, as boundaries are in $1-1$ correspondence with Lagrangian subgroups (the anyons they condense), we have to choose the kind of boundary which corresponds to a Lagrangian subgroup $\mathcal{L}$. We do this in six steps: 

\textit{First}, we choose a $2D$ region $R \subset \M$ in the lattice with a $1D$ boundary $\partial R$ and choose a Lagrangian subgroup $\L$ for the boundary. The type of the boundary is independent of its microscopic spatial geometry (one does not need to fix a specific shape of the boundary beforehand). The boundary can live on the direct or dual lattice, or a mixture of them. One choice for the boundary is shown by the pink line in Fig.~\ref{fig:toricbdry1}.

\textit{Second}, we add all ribbon operators $F^a(t)$ that create the condensed anyons in $\mathcal{L}$ inside $R$ or on $\partial R$ to the Hamiltonian. In the case of dyons, for example, condensing $em \in \L$, the shortest ribbon will act on two edges. We call these the new stabilizer terms $\mathcal{S_L}$
\begin{equation}
    \mathcal{S}_{\mathcal{L}}=\{F^{a}(t) \ | \ a \in \mathcal{L} \}
\end{equation} 

\textit{Third}, we remove old stabilizers that do not commute with new ones. The physical interpretation is that the removed operators correspond to the smallest contractible loops of anyons that will become confined and cannot tunnel through the boundary to the vacuum. The set of now confined stabilizers is: 
\begin{equation}
    \mathcal{S}_C=\{h\in \mathcal{S}_H \ | \exists l\in \mathcal{S}_{\mathcal{L}}, [h,l] \neq 0  \}.
\end{equation}

\textit{Fourth}, we construct the new stabilizer group $\mathcal{S}_{H'}$:
\begin{equation}
    \mathcal{S}_{H'}=\mathcal{S}_{\mathcal{L}} \cup (\mathcal{S}_H \backslash \mathcal{S}_C)
\end{equation}
The generators of this stabilizer group are denoted by $J(v,p)$ and they fall into one of three categories: (a) old stabilizer group $S_H$ generators $A(v),B(p),C(e)$ that continue to commute with the new ones in $\S_{\L}$, (b) the new stabilizers of the condensed anyons $\S_\L$, (c) local products of removed old stabilizer terms $J(v,p) = \prod_{S_C} A(v)B(p)C(e)$ that commute with the new terms $\S_\L$. The terms in (c) are promoted from being generated in the old code to being generators in the new code. For example, it maybe be the case that $A(v_1)$ and $B(p_1)$ do not commute with the new stabilizer group $\S_\L$ but that their generat\textit{ed} product $J(v_1,p_1)=A(v_1)B(p_1)$ does. In this case, we add $J(v_1,p_1)$ as a generat\textit{or} of the new stabilizer group $\mathcal{S}_{H'}$. 

\textit{Fifth,} we erase any qudits that are trivially condensed. These will include the qudits inside $R$ or on the boundary $\partial R$. As they are acted on by the stabilizers in $\S_\L$.  

\textit{Sixth,} restrict the stabilizers at the boundary to the  non-trivial qudits only. If, after this step, new qudits are condensed, then repeat step five until no more qudits are condensed  \footnote{In this case, it is optional to remove these qudits or keep them.}. Keeping them will result in a boundary that strictly follows the boundary $\partial R$. We choose to erase all trivialized qudits for clarity. The new Hamiltonian will be the summation of the generators of $\mathcal{S}_{H'}$.
\begin{equation}
\begin{aligned}
  H' &= \sum_{v\in \text{Bulk}} A(v) - \sum_{p\in \text{Bulk}} B(p)\\
     & - \sum_{e\in \text{Bulk}} C(e)
           - \sum_{(v,p)\in \partial R} J(v,p)
\end{aligned}
\end{equation}
Where $A,B$, and $C$ are the bulk stabilizers that do not overlap with $\partial R$. $J(v,p)$ represents a generic boundary term that can depend on a vertex and an adjacent plaquette at the same time. The algorithm is summarized in Alg.\ref{alg:Bdry}.

\begin{algorithm}[H]
\caption{Boundary Stabilizers for $H$}\label{alg:Bdry}
\begin{algorithmic}[1]
  \Procedure{Bdry}{$ H, R, \L_R$}
    \Require A region $R$ with boundary $\partial R$ and a Lagrangian subgroup $\mathcal L_R$.
    \State Measure the stabilizers $\mathcal S_{\mathcal L}$ that create anyons in $\mathcal L_R$ inside $R$ or on $\partial R$.
    \State Remove stabilizers $\mathcal S_C\in\mathcal S_H$ that do not commute with $\mathcal S_{\mathcal L}$.
    \State Construct $\mathcal S_{H'}=\mathcal S_{\mathcal L}\cup\bigl(\mathcal S_H\setminus\mathcal S_C\bigr)$.
    \While{condensed qudits exist}
      \State Erase trivialized qudits.
      \State Restrict boundary stabilizers to their non-trivialized support.
    \EndWhile
    \State\Return $ {H_{\partial R}}$
  \EndProcedure
\end{algorithmic}
\end{algorithm}

\subsection{Examples of Boundary Construction}
\label{sec:bound_algs}

To illustrate this algorithm, we consider, in order of increasing complexity, the explicit construction of a few example boundaries.  First up is the well-known smooth boundary of the surface code. The star and plaquette operators are Eq.\eqref{eq:z4}:

\begin{equation}
  \text{\includegraphics[width=0.48\textwidth]{Figures/z4.pdf}}
  \label{eq:z42}
\end{equation}

\begin{Example}[Smooth boundary for the $\Z_4$ surface code]
\label{Example:smoothtoric}
\begin{figure}[!htbp]
\includegraphics[width=0.40\textwidth]{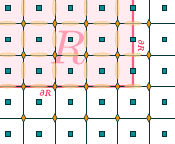}
\caption{\label{fig:toricbdry1} After the first two steps of the algorithm, the edge operators $X$ (yellow ellipses) are measured inside the selected region $R$ and also on the boundary $\partial R$.}
\end{figure}

First, we choose a connected region $R$. As illustrated in Fig.~\ref{fig:toricbdry1}, $R$ is bounded by $\partial R$, which is denoted by the pink path. We wish for this smooth boundary to condense $m$ particles, so the Lagrangian subgroup is $\L=\{1,m,m^2,m^3\}$. Second, to condense the Lagrangian subgroup anyons, we will measure the shortest ribbon operator that creates the generator ($X$) at each qubit inside $R$ and at the pink path $\partial R$. Fig.~\ref{fig:toricbdry1} shows the resulting picture after the first two steps. Third, we need to remove the old $B(p)$ stabilizers that do not commute with the measured $X$ operators. This includes all the plaquettes inside $R$ as well as those that share any edge with $\partial R$. The star operators will commute, and they will not be removed. Fourth, since the only products of the old operators that commute with the new stabilizers are the star operators, there is nothing to do in this step. Importantly, those star operators right at the boundary $\partial R$ will still commute with the new Hamiltonian, so they remain. After the first four steps, we reach the configuration in Fig.~\ref{fig:toricbdry2}. 

\begin{figure}[!htbp]
 \includegraphics[width=0.40\textwidth]{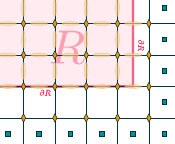}\caption{\label{fig:toricbdry2} After the fourth step of the algorithm, the stabilizers that do not commute with the new $X$ operators are removed. These include any plaquette that shares any edge with an $X$ operator.}
\end{figure}

In the fifth step, for clarity, we remove the edges that have condensed to vacuum as they are decoupled from the rest of the system. Any edge measured with the short string local-$X$ operator condenses into a trivial vacuum. After condensing these edges, we arrive at Fig.~\ref{fig:toricbdry3}. The star operators on the boundary $\partial R$ will now have a support of one edge. This is written in a consistent notation in Fig.~\ref{fig:toricbdry3b}.
\begin{figure}[htbp]
 
        \includegraphics[width=0.30\textwidth]{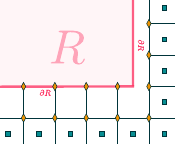}
        \caption{\label{fig:toricbdry3} After removing the qudits measured by the $X$ operator, star operators have one edge support at the boundary.}
   
\end{figure}

This is a perfectly valid boundary for the surface code. The seemingly rough boundary is actually a smooth boundary because it condenses $m$ anyons. Had we chosen the pink path to lie in the dual lattice only, we would not have needed to deal with this, as exemplified by the other vertical pink line. Removing the edges that are now touched by only one $X$ operator gives us the (geometrically) smooth boundary~\cite{Bravyi1998} illustrated in Fig.~\ref{fig:toricbdry4}. Sixth, we restrict the new stabilizers to their support on the non-trivial qudits. Some of the vertices now have a star that is supported on only 3 edges.
 \begin{figure}[htbp]
    \centering
        \begin{subfigure}[b]{0.23\textwidth}
        \includegraphics[width=\textwidth]{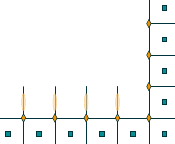}
        \caption{\label{fig:toricbdry3b} Smooth boundary for the $\Z_4$ surface code that strictly follows spatial boundary $\partial R$.}
    \end{subfigure}
    \hfill
    \begin{subfigure}[b]{0.23\textwidth}
      \includegraphics[width=\textwidth]{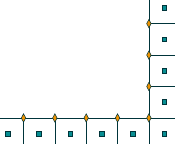}\caption{\label{fig:toricbdry4} After the last step of the algorithm, we recover the familiar smooth boundary for $\Z_4$.}
    \end{subfigure}
\end{figure}
\end{Example}
 
In example~\ref{Example:smoothtoric}, the fourth step was trivial because of the simplicity of the surface code. In the next example, we treat the slightly more complicated case of a new boundary in $\Z_4$.

\begin{Example}[Even boundary for the $\Z_4$ surface code]
\label{ex:Z_4even}

\begin{figure}[htbp]
  \centering
  \subfloat[\label{fig:z4e2m21} Region $R$ along with its boundary $\partial R$. After the second step of the algorithm, the $X^2$ and $Z^2$ operators are measured inside $R$ and on $\partial R$. These operators condense the anyons in the even boundary subgroup $\L_{even}$. ]{%
    \includegraphics[width=0.40\textwidth]{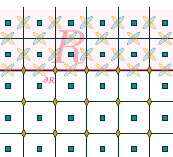}}
  \hfill
  \subfloat[\label{fig:z4e2m22} In the third step, the stabilizers that do not commute with the measured operators are removed. These are any plaquette or star operators that share an edge with a measured $X^2$ or $Z^2$.]{%
    \includegraphics[width=0.40\textwidth]{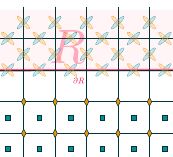}}
  \caption{The first three steps of the algorithm to construct the even boundary of $\Z_4$.}
\end{figure}
First, let's construct the even boundary of the $\Z_4$ surface code \eqref{eq:z4Ls} on the spatial boundary $\partial R$ located at the pink path in Fig.~\ref{fig:z4e2m21}, taking the region $R$ extending to infinity as before. Second,  we measure the ribbons creating the anyons of the Lagrangian subgroup $\L_{\text{Even}}=\{1,e^2,m^2,e^2m^2\}$. In this case, the condensation algebra is generated by the $X^2$ (yellow tilted ellipses) and $Z^2$ (blue tilted ellipses) operators. The configuration after the first two steps is shown in Fig.~\ref{fig:z4e2m21}. Third, we remove any star or plaquette stabilizers that share any edge with the stabilizers $Z^2$ or $X^2$, respectively. After the third step, we reach the configuration in Fig.~\ref{fig:z4e2m22}. Fourth, the products of the old stabilizers that we promote to generators are $(A(v))^2$ (purple four-sided star) and $(B(p))^2$ (big dotted square).
\begin{equation}
  \text{\includegraphics[width=0.48\textwidth]{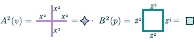}}
  \label{eq:z4evenbdrystab}
\end{equation}

These stabilizers have to be added to the Hamiltonian, as shown in Fig.~\ref{fig:z4e2m23}.
\begin{figure}[htbp]
 \includegraphics[width=0.40\textwidth]{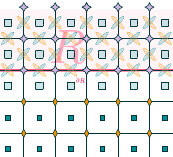}\caption{\label{fig:z4e2m23} In the fourth step, the products of removed stabilizers that commute with the measured operators are added. These are the square of plaquette $B^2(p)$ denoted by big dotted light blue squares, and the square of stars $A^2(v)$ denoted by four-sided mauve stars.}
\end{figure}

Fifth, we now remove the trivialized qudits. Since a pair of degree two constraints is simultaneously applied to a four-dimensional degree of freedom, all edges in the bulk of the region $R$, or on the pink boundary $\partial R$, will condense to the vacuum. In general, the qudits inside $R$ or on $\partial R$ will always condense. We reach Fig.~\ref{fig:z4e2m24}.  

\begin{figure}[htbp]
  \centering
  \subfloat[\label{fig:z4e2m24} In the fifth step, we erase trivialized qudits. These will be all qudits inside $R$ or on $\partial R$.  ]{%
    \includegraphics[width=0.24\textwidth]{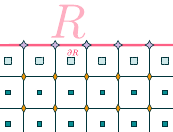}}
  \hfill
  \subfloat[\label{fig:z4e2m25} In the sixth step, boundary stabilizers are restricted to their non-trivialized support. $B^2(p)$ act on three edges, while $A^2(v)$ became an $X^2$ edge operator. ]{%
    \includegraphics[width=0.24\textwidth]{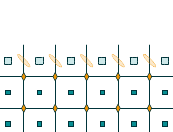}}
  \caption{Steps five and six in the construction of the even boundary for the $\Z_4$ surface code.}
\end{figure}

Observe that the support of the new star $A^2(v)$ operators on the boundary $\partial R$ is reduced to a single edge. Additionally, the new plaquette operators $B^2(p)$ now have support on three edges. After restricting the stabilizers on the boundary to their non-trivialized supports, no more qudits are condensed, and we are done. We arrive at Fig.~\ref{fig:z4e2m25} which describes the even boundary that condenses the $\{1,e^2,m^2,e^2m^2\}$ anyons. This gives the boundary in Eq.~\eqref{eq:hz4bdry}
\begin{equation}
     H_{\Z_4,\text{Even}} = H_{\Z_4} -\sum_{p,e\in \partial \M} (B^2(p)+ X^2(e))+H.C.
\end{equation}

\end{Example}
In these examples, we only cared to check the commutation of operators that are not Hermitian. However, the shift and clock operators in any dimension are unitary, and the following remark shows that one needs only to consider the operator or its Hermitian conjugate.
\begin{Remark}
    Assume we have two commuting unitary operators $[A,B]=0$, $AA^\dagger=A^\dagger A=\mathbb{I}$ and $BB^\dagger=B^\dagger B=\mathbb{I}$. Then it is simple to show that $[A,B^\dagger]=[A^\dagger,B^\dagger]=0$.
\end{Remark}
Finally, to construct different adjacent boundaries, one just applies the algorithm sequentially. For example, after obtaining the even boundary for the $\Z_4$ surface code on one spatial boundary, one can proceed to construct the smooth boundary on any other spatial boundary.

\begin{Example}[Boundary for the Doubled Semion (DS) phase]
 The DS phase \eqref{eq:HDS} has only one boundary with vacuum, where the boson $b$ condenses. 
 
 \begin{figure}[htbp]
  \centering
  \subfloat[\label{fig:dsbdry1} Region $R$ along with its boundary $\partial R$. After the second step of the algorithm, the $X^2$ and $Z^2$ operators are measured inside $R$ and on $\partial R$.  We removed the $X^2Z^2$ for visual clarity (they are also now dependent on the edge stabilizers $X^2$ and $Z^2$). ]{%
    \includegraphics[width=0.40\textwidth]{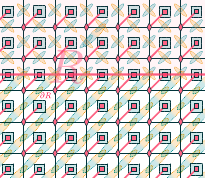}}
  \hfill
  \subfloat[\label{fig:dsbdry2} In the third step, the stabilizers that do not commute with the measured operators are removed. These are the $F$ operators \eqref{eq:HDS} that share an edge with a measured $X^2$ or $Z^2$.]{%
    \includegraphics[width=0.40\textwidth]{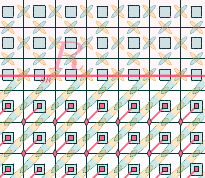}}
  \caption{The first three steps of the algorithm to construct the unique boundary of DS.}
\end{figure}
First, let's choose a spatial boundary $\partial R$ located at the pink path in Fig.~\ref{fig:dsbdry1}, taking the region $R$ extending to infinity. Second,  we measure the ribbons creating the anyons of the Lagrangian subgroup $\L_{\text{DS}}=\{1,b\}$. In this case, the condensation subgroup is generated by the $X^2$ (yellow tilted ellipses) or $Z^2$ (blue tilted ellipses) operators.
\begin{equation}
  \text{\includegraphics[width=0.48\textwidth]{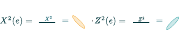}}
  \label{eq:dsbdrystab}
\end{equation}
The configuration after the first two steps is shown in Fig.~\ref{fig:dsbdry1}. Third, we remove any $F$ stabilizers that share any edge with the stabilizers $Z^2$ or $X^2$. The $X^2 $ and plaquettes $B^2(p)$ will continue to commute with the new operators. In fact, after adding the edge operators $X^2$ and $Z^2$, we can remove the old $X^2 Z^2$ or $B^2(p)$ stabilizers in the bulk of $R$. After the third step, we reach the configuration in Fig.~\ref{fig:dsbdry2}. Fourth, the products of the old stabilizers that we promote to generators are $A^2(v) = F^2(v,p)B^2(p)$ (purple four-sided star).
\begin{equation}
  \text{\includegraphics[width=0.35\textwidth]{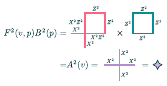}}
  \label{eq:dsbdrystab2}
\end{equation}

These stabilizers have to be added to the Hamiltonian, as shown in Fig.~\ref{fig:dsbdry3}.
\begin{figure}[htbp]
 \includegraphics[width=0.40\textwidth]{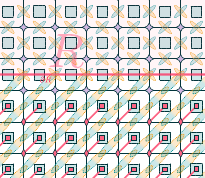}\caption{\label{fig:dsbdry3} In the fourth step, the products of removed stabilizers that commute with the measured operators are added. These are the square of star operators $A^2(v)$ denoted by four-sided mauve stars.}
\end{figure}

Fifth, we now remove the trivialized qudits. All edges in the bulk of the region $R$, or on the pink boundary $\partial R$, will condense to the vacuum. We reach Fig.~\ref{fig:dsbdry4}.  

\begin{figure}[htbp]
  \centering
  \subfloat[\label{fig:dsbdry4} In the fifth step, we erase trivialized qudits. These will be all qudits inside $R$ or on $\partial R$.  ]{%
    \includegraphics[width=0.24\textwidth]{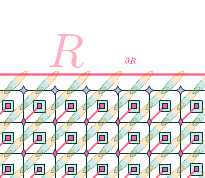}}
  \hfill
  \subfloat[\label{fig:dsbdry5} In the sixth step, boundary stabilizers are restricted to their non-trivialized support. $A^2(v)$ acts on three edges, while the two-edge operator $X^2 Z^2(e)$ became a $Z^2$ single-edge operator. ]{%
    \includegraphics[width=0.24\textwidth]{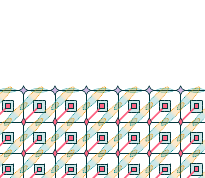}}
  \caption{Steps five and six in the construction of the unique boundary for the DS stabilizer code.}
\end{figure}

Observe that the support of the $X^2Z^2$ operators on the boundary $\partial R$ is reduced to a single edge. Additionally, the new star operators $A^2(v)$ now have support on three edges. After restricting the stabilizers on the boundary to their non-trivialized supports, no more qudits are condensed, and we are done. We arrive at Fig.~\ref{fig:dsbdry5}, which describes the unique DS boundary that condenses the $\{1,b\}$ anyons. This gives the boundary Hamiltonian:
\begin{equation}
     H_{DS-bdry} = H_{\text{DS}} -\sum_{v,e\in \partial \M} (A^2(v)+ X^2(e))+H.C.
\end{equation}
\end{Example}
\subsection{Examples of Domain Walls Construction }\label{sec:DWalg}
The same algorithm to construct boundaries can be used to construct domain walls through the folding trick \cite{Beigi2011}. In this section, we provide representative examples of this procedure. One advantage of constructing domain walls via stabilizer gauging is to avoid defining star operators that explicitly depend on the 2-cocycle of the boundary. Since the second cohomology group of Abelian cyclic groups is trivial, $H^2(Z_m,U(1))=0$, we have not encountered this situation in the prior examples. Naturally, a non-trivial 2-cocycle will appear when considering the more generic defects (domain walls) of and between topological orders.  In particular, $H^2(\Z_n\times \Z_n,U(1))=\Z_n$. If we denote the group action additively, then we have the following 2-cocycles: $\omega_s((a_1,a_2),(b_1,b_2))=e^{\frac{2\pi is}{n} {(a_1b_2-a_2b_1)}}$ for $s \in \{0,1,\dots,n-1\}$.

\begin{figure}[htbp]
 \includegraphics[width=0.48\textwidth]{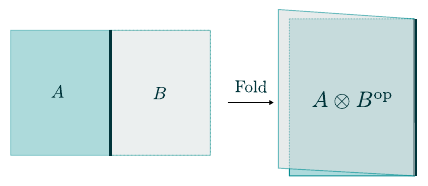}
 \caption{\label{fig:folding} The folding trick shows how the domain wall between phases $A$ and $B$ is the same as a boundary of the phase $A \otimes B^{\text{op}}$. $B^{\text{op}}$ is the opposite phase of $B$, meaning that it has undergone reflection in the direction perpendicular to the wall.}
\end{figure}

The domain walls of a quantum double based on the group $G$ are the boundaries of the quantum double of the group $G\times G$. This is known as the folding trick \cite{Beigi2011} and is shown schematically in Fig.~\ref{fig:folding}.

Thus, the algorithm works for the domain walls as well. We just need to account for folding. This is done by first placing the two topological orders on top of each other with the top layer mirrored (i.e. with a reflection around the domain wall). We treat the double layer as a single layer, realizing the quantum double $D(G \times G)$. If it is a quantum double, it will have the Hamiltonian:
\begin{equation}\label{eq:hz4layers}
H=-\sum_{v}\bigl( A_1(v)+A_2(v) \bigr)-\sum_{p}\bigl( B_1(p)+B_2(p) \bigr) +H.C.
\end{equation}
where the subscript denotes the layer (the top layer is given index 2). Concretely, we have the following stabilizers:

\begin{equation}
  \text{\includegraphics[width=0.48\textwidth]{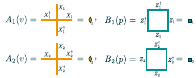}}
  \label{eq:avop}
\end{equation}

\begin{figure}[htbp]
\includegraphics[width=0.48\textwidth]{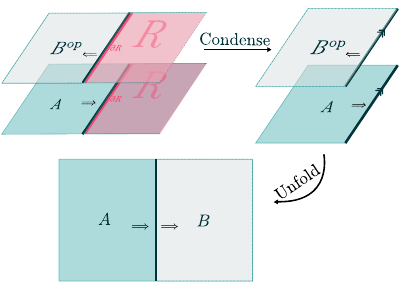}
\caption{\label{fig:condensedefect} Constructing a domain wall using condensation. Before running the algorithm, we fold the top layer ($B$) into its mirrored version ($B^{\text{op}}$). An arrow is drawn to show how the orientation flips from $B^{\text{op}}$ to $B$ when folding (unfolding). After running the algorithm, we unfold the top layer to have a domain wall between the two phases.}
\end{figure}

Note how the stabilizers for the top layer are mirrored versions of the first layer stabilizers. We proceed as before, choosing a region $R$ with a boundary $\partial R$ and following algorithm~\ref{alg:Bdry} with the appropriate Lagrangian subgroup. Finally, to have the boundary as a defect between the two adjacent topological orders, we unfold the two layers. This procedure is depicted schematically in Fig.~\ref{fig:condensedefect}.

As an example, let us construct the $e \leftrightarrow m$ domain wall of the $\Z_4$ surface code. Note that, changing the definition of the generalized Pauli operators to the correct dimension, this procedure naturally generalizes to any $\Z_n$.

\begin{Example}[Domain Wall $(e\leftrightarrow m)$ for $\Z_4$ (or $\Z_n$)] 
\label{Example:emdw}

Before starting the algorithm, we place the two copies of the surface code on top of each other. Denote the anyons of the bottom layer by $\{1,e_1,m_1,\dots\}$ and for the top layer by $\{1,e_2,m_2,\dots\}$. The top layer will be a mirrored version of the bottom layer. This is to ensure we obtain theories with the same orientation after unfolding,  as in Fig.~\ref{fig:condensedefect}.

We then begin the algorithm. First, we choose a region $R$ extending to infinity. The boundary $\partial R$ lives on the direct lattice in layer $1$ but on a dual line in layer $2$. This boundary was just chosen for convenience due to the nature of condensed anyons, but any other choice will work, Fig.~\ref{fig:z4etom0}. Second, we add the shortest ribbon operators creating anyons of the Lagrangian subgroup generated by $\L_{e\leftrightarrow m} = \operatorname{Span}\bigl(\{e_1m_2^{-1},e_2m_1^{-1}\}\bigr)$. These operators are generated by:

\begin{equation}
  \text{\includegraphics[width=.48\textwidth]{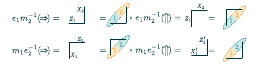}}
  \label{eq:etomribbss}
\end{equation}
Here, the arrows indicate the direction of propagation of the anyons (the direction of the ribbons creating them).
 
\begin{figure}[htbp]
\centering
 
  \includegraphics[width=\linewidth]{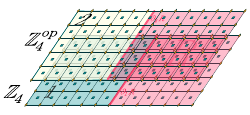}\caption{Region of condensation $R$ for the domain wall $W^{e\to m}$ of the $\Z_4$ surface code. The stabilizers of the top layer are reflected around the domain wall as defined in Eq.~\eqref{eq:hz4layers}. The boundary $\partial R$ lives on the direct lattice in layer $1$ and on the dual lattice in layer $2$. }
  \label{fig:z4etom0}
 
\end{figure}

Note that we needed at least a ribbon with two edges to realize the dyons. And we had to take into account the mirrored nature of layer $2$. For example, the conventions for right and left are reversed in layer $2$ compared to layer $1$. It can be easily checked that they all commute with each other. The configuration after the first two steps is shown in Fig.~\ref{fig:z4etom1}.
\begin{figure}[htbp]
\centering
 
  \includegraphics[width=0.40\textwidth]{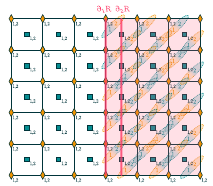}\caption{Top view of the two layers after the first two steps of the algorithm. The short ribbon operators are added to the region of condensation $R$. Note how the boundary $\partial R = \partial_1R \ \cup \partial_2R $ has a component on the direct lattice on layer $1$ and on the dual lattice on layer $2$ denoted by the lines $\partial_1R$ and $\partial_1R$ respectively. The stars and plaquettes here are the summation of the operators of the two layers, Eq.~\eqref{eq:hz4layers}. Only a subset of the measured ribbon operators is shown as they overlap.  }
  \label{fig:z4etom1}
 
\end{figure}

In the third step, we remove the old stabilizers that do not commute with the new ones. These will include all the old star and plaquette stabilizers of both layers in the bulk of the region $R$. In addition, we remove the star operators of both layers on the boundary $\partial_1 R$. We remove the plaquettes $B_1(p)$ of the second layer adjacent to the boundary $\partial_1R$. The configuration after the third step is shown in Fig.~\ref{fig:z4etom2}.
\begin{figure}[htbp]
\centering
 
  \includegraphics[width=0.40\textwidth]{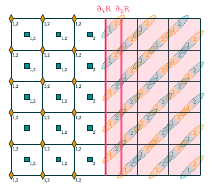}\caption{After the third step, the stabilizers that do not commute with the ribbon operators are removed. Note how the plaquettes adjacent to the boundary $\partial_1R$ now live in the second layer only.  }
  \label{fig:z4etom2}
 
\end{figure}

In the fourth step, we form new stabilizers that are products of the old ones and commute with the new stabilizers. These are given by products of the removed stars with their adjacent removed plaquettes:

\begin{equation}
  \text{\includegraphics[width=.35\textwidth]{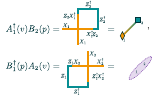}}
  \label{eq:fishy}
\end{equation}

After adding the new stabilizers to the model, we reach the configuration in Fig.~\ref{fig:z4etom3}

\begin{figure}[htbp]
\centering
 
  \includegraphics[width=0.40\textwidth]{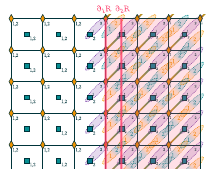}\caption{After the fourth step, products of old stabilizers that commute with the ribbons are added, Eq.~\eqref{eq:fishy}.  }
  \label{fig:z4etom3}
 
\end{figure}
In the fifth step, we erase trivialized qudits. These are all the qudits in the bulk of $R$ or on its two-component boundary $\partial_1R$ and $\partial_2R$. Note that the edges of the second layer that lie above $\partial_1 R$ will not be condensed. After condensing qudits, we also removed all stabilizers that only act on trivialized qudits for clarity, Fig.~\ref{fig:z4etom4}.

\begin{figure}[htbp]
\centering
 
  \includegraphics[width=0.40\textwidth]{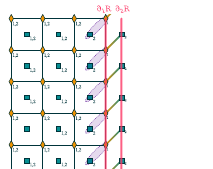}\caption{After the fifth step, the qudits in $R$ or on the boundary $\partial R = \partial_1 R \ \cup \partial_2 R$ are erased. In particular, on $\partial_1R$ only one qudit lives now. }
  \label{fig:z4etom4}
 
\end{figure}
 
The sixth step is to restrict the stabilizers at the boundary to their non-trivialized support. This results in the following stabilizers:
\begin{equation}
  \text{\includegraphics[width=.48\textwidth]{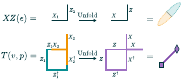}}
  \label{eq:cleanfishy}
\end{equation}

Finally, we unfold layer $2$ to be put on the right. This will result in the stabilizer of the bulk of the two layers being the same again, redoing the horizontal reflection. At each edge, now only one qudit lives, so we can drop the layers' index unambiguously. Importantly, right at the $\partial_1R$ boundary, only the qudits in the second layer survive. We reach the $W^{e\leftrightarrow m}$ domain wall for $\Z_4$. This is also the domain wall for this duality for any $\Z_n$ by changing the definitions of the generalized Pauli matrices accordingly. In particular, for $\Z_2$ we interpret the $X$ and $Z$ as the ordinary 2-dimensional Pauli matrices, which are Hermitian.

\begin{figure}[htbp]
\centering
 
  \includegraphics[width=0.40\textwidth]{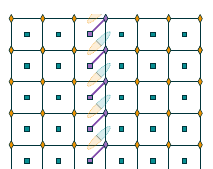}\caption{In the last step of the algorithm, we restrict the stabilizers to their non-trivialized support. After unfolding, we have the twist operators $T(v,p)$ (purple) that live on a vertex and an adjacent plaquette, and also the $XZ(e)$ (ellipse) operators connecting two edges.}
  \label{fig:z4etom4}
 
\end{figure}
\end{Example}

It is clear that one can also construct finite domain walls by only condensing a smaller region than considered in Fig.~\ref{fig:condensedefect}. In particular, we will be interested in constructing certain domain walls between a parent bulk theory (e.g. $D(\Z_4)$) and a finite patch within it that is obtained by condensation (e.g. DS). In this case, one can directly apply the condensation without having to use the folded picture. Let us use this to illustrate one way the domain wall between $D(\Z_4)$ and DS can be constructed.

\begin{Example}[Domain wall between $D(\Z_4)$ and DS]\label{ex:z4dsdwsimple}
\begin{figure}[!htbp]
\includegraphics[width=0.40\textwidth]{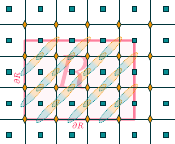}
\caption{\label{fig:dsz4a1} After the two steps of the algorithm, the edge operators $X^2Z^2$ creating $e^2m^2$ are measured in the region $R$ and also on the boundary $\partial R$.}
\end{figure}
Since DS is obtained from $D(\Z_4)$ by condensation, one can condense anyons creating DS in only a finite patch of space. These are the $e^2m^2$ anyons of $\Z_4$. First, we choose a region $R$ with boundary $\partial R$. We want to construct the domain wall that doesn't condense any of the DS phase anyons. Second, we add the short ribbon operators $X^2 Z^2$ that create the $e^2 m^2$ anyons inside $R$ or on $\partial R$. After the first two steps, we reach Fig.~\ref{fig:dsz4a1}.
\begin{figure}[htbp]
 \includegraphics[width=0.40\textwidth]{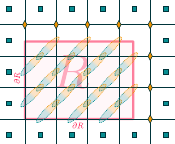}\caption{\label{fig:dsz4a2} After the third step, old stabilizers that do not commute with the measured ribbons are removed. These are the star and plaquette operators of $D(\Z_4)$ that share at least one non-commuting edge with $X^2Z^2$. }
\end{figure}

Third, we remove any $A(v)$ or $B(p)$ stabilizers that do not commute with the measured stabilizers $X^2Z^2$. After the third step, we reach the configuration in Fig.~\ref{fig:dsz4a2}. Fourth, the products of the old stabilizers that we promote to generators are $F(v,p)=A(v)B(p)$ (a connected pink star and a plaquette) and also the $B^2(p)$ operators (big dashed square). These stabilizers have to be added to the Hamiltonian as shown in Fig.~\ref{fig:dsz4a3}.
\begin{figure}[!htbp]
 \includegraphics[width=0.40\textwidth]{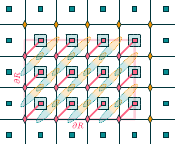}\caption{\label{fig:dsz4a3} In the fourth step, the products of removed stabilizers that commute with the measured operators are added. These are the $F(v,p)$ (a connected pink star and a plaquette) and the $B^2(p)$ operators (big dashed square).}
\end{figure}
 
Fifth, we now remove the trivialized qudits. In this case, no edges are condensed. Subsequently, the sixth step is also trivial. This example shows that in a certain situation, one does not need to do the folding. In App.~\ref{sec:dsz4dwnotsimple}, we corroborate this by deriving the domain walls using the folding method.
\end{Example}

\section{Calculating Logical GSD}\label{sec:GSD}

To confirm our Abelian (twisted) quantum doubles are quantum error-correcting codes, we follow Kitaev's original logic~\citenum{Kitaev2003} and rigorously compute the ground state degeneracy. This topological degeneracy corresponds to the dimension of the logical code space that is acted upon by logical operators as previously described in Sec.~\ref{sec:composite}.

The ground state projector has, by definition, the ground states in its image and all excited states in its kernel. Thus, the most straightforward method to compute the ground state degeneracy is to explicitly construct the ground state projector and take its trace~\footnote{The same logic applies for computing the dimensionality of any excited state sector}. However, due to the tensor product structure, this is not scalable in practice.

In this section, we detail two complementary methods to efficiently calculate, modify, and reason about the $\GSD$ for composite topological codes. The first method utilizes the microscopic stabilizer description, where both local (e.g. stabilizer order) and global (e.g. boundary conditions) information is used to calculate the GSD. This is detailed below in Sec.\ref{sec:GSD}. Alternatively, one can use a coarse-grained macroscopic description to compute the GSD using knowledge of logical ribbon operators and pants decomposition as detailed in Sec.\ref{sec:lego_code}, which follows Ref.~\citenum{Tan2015}.

\subsection{GSD using Microscopic Stabilizers}\label{sec:micro}

In this subsection, we assume we are given a concrete stabilizer layout for a system, and then compute its GSD. Begin with an unconstrained Hilbert space of dimension $D$. In the simple and most common case, all qudits will have the same dimension $d$ and thus, $D=d^{\#E}$ where $\#E$ is the number of edges in the system, each supporting a single qudit~\cite{quditedge}. We then consider the $n$ mutually commuting stabilizer group generators $\left\{O_i \right\}$, each with order $m_i$ (i.e. $O^{m_i}_i=1$). Note that, in general, for higher-dimensional qudits ($d>2$), these generators do not have to be Hermitian.

Next, one must determine and eliminate the $l$ lingering dependencies between the presumed generators. That is, remove $l$ of the presumed generators, which were actually generated from products of the others and are not truly independent. In the examples we consider below, the dependencies are organized with respect to generators of a given order. Since each \textit{independent} generator $O_i$ simultaneously constrains the ground state manifold, by a factor $1/m_i$, the ground state degeneracy is 
\begin{eqnarray}
    \GSD=\frac{D}{\prod_{|m|=2,3,\cdots m_{max}} m^{n_m-l_m}}.
\end{eqnarray} 
Here, $n_m$ and $l_m$ are the numbers of generators and dependencies that have order $m$, respectively.

By definition, the $\GSD$ of topological theories does not depend on the system size nor its microscopic details, e.g., the particular triangulation of the surface. As we will see in the examples below, we can therefore use any minimal example to calculate the $\GSD$ and then extend the result for a larger or more complicated, but topologically equivalent, system.

\textit{Updating GSD}--If the code is re-gauged to add boundaries or domain walls, as described in Alg.~\ref{alg:Bdry}, we may sequentially update the GSD by only accounting for changes. Let us summarize the results of applying Alg.~\ref{alg:Bdry} as they relate to constraint dimensions and dependencies. This summary will culminate in equations to compute the updated $\GSD'$ as a function of the old one. 
Suppose we know the $\GSD$ for a certain configuration realizing stabilizer group $\S_H$ with generators $s_H = \{O_i \}$ with $O_i^{m_i}=1$ and dependencies $l_m$. We can apply the algorithm to obtain a different configuration with Hilbert space dimension $D'$, stabilizer group $\S_{H'}$, generators $  s_{H'} = \{ O'_j\}$, with $O^{m'_j}_j=1$ and dependencies $l'_m$.
After the algorithm is complete, a set of new generators $O_j'\in s_{H'}$ will be added to the new Hamiltonian $H'$:
\begin{equation}s_\text{new} = \{ O'_j \in s_{H'}\backslash s_H \}\end{equation}

In addition, a set of generators $O_i \in s_H$ will be removed from the old Hamiltonian $H$. 
\begin{equation}
s_\text{rem.}=\{O_i \in s_H \backslash s_{H'}\}\end{equation}
Finally, the erasure of qubits may result in a change of the dimension of the Hilbert space. These three components (two sets of generators in addition to the changes in the total dimension of the Hilbert space) are all the data needed from the algorithm to relate the new $\GSD'$ to the old one. 

The introduction of new, or removal of old, degrees of freedom is accounted for by the ratio of total Hilbert space dimensions $r_D = D'/D$. 
 
Order by order, the total change in the constraints is computed by taking the difference of the number of order $m$ constraints $\Delta n_m=n'_m-n_m$ and also tracking the change in dependencies $\Delta l_m = l'_m-l_m$. 
 
After this bookkeeping, we can write the new degeneracy in terms of the old one as:
\begin{equation}
\label{eq:gsdupdate}
    \GSD' = \GSD \frac{r_D}{\prod_{|m|=2,3,\cdots m_{max}} m^{\Delta n_m- \Delta l_m}}.
\end{equation}

\subsubsection{Examples using Minimal Stabilizer Models}

\begin{figure}[htbp]
 \includegraphics[width=0.30\textwidth]{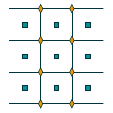}\caption{\label{fig:z2surfacecode} $D(\Z_2)$ surface code with opposite smooth and rough boundary conditions, and which has $\GSD = 2$. }
\end{figure}

To get a feeling for Eq.~\ref{eq:gsdupdate}, let us start with example~\ref{ex:roughcyliner} where we modify the surface code in Fig.~\ref{fig:z2surfacecode} with $\GSD = 2$ to make a cylinder with two rough boundaries.

\begin{Example}[$\Z_2$ on a cylinder with two rough boundaries]\label{ex:roughcyliner}
Starting from the surface code, suppose we want to build the configuration with $\Z_2$ on a cylinder with two rough boundaries as shown in Fig.~\ref{fig:z2cylinderi0}.

   \begin{figure}[!ht]
 \includegraphics[width=0.30\textwidth]{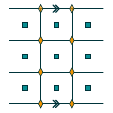}\caption{\label{fig:z2cylinderi0}Minimal stabilizer model for the $\Z_2$ on a cylinder with two rough boundaries. The double arrows indicate periodic boundary conditions. Consequently, the semi-transparent diamonds should not be counted. }
\end{figure}

Let us track the changes from the layout in Fig.~\ref{fig:z2surfacecode}. The edges went from $18$ to $15$ and thus, $$r_D=2^{18-15}=2^{-3}.$$ There are only order-$2$ operators. The number of plaquettes did not change while the number of stars decreased by $2$, so $$\Delta n_2=-2.$$ Finally, while the dependencies were $l_2=0$ in the surface code, here the product of all plaquettes gives us the identity $\prod_p B(p) = \mathbb I$, so we have $l'_2=1$. This gives $$\Delta l_2=1.$$ We then compute the GSD from a direct application of Eq.~\eqref{eq:gsdupdate} as:
\begin{equation}
    GSD' = \frac{r_D}{2^{\Delta n_2 - \Delta l_2}} =  \frac{2^{-3}}{2^{-2-(1)}} \GSD = 2
\end{equation}

 \end{Example}

\begin{Example}  [Toric code through folding a cylinder]
A minimal example for the toric code is depicted in Fig.~\ref{fig:toricgsd}. The toric code can be made starting with a cylinder with $9$ faces, $18$ edges, and $8$ vertices, as in Fig.~\ref{fig:toricgsd}, turning it into a torus equates to enforcing periodic boundary conditions on the remaining two boundaries. This process will involve removing $3$ edges, removing $1$ star operator as it will become dependent on the others, and removing $3$ plaquettes. We then have: 

\begin{equation}
\begin{aligned}
     \GSD_\text{Toric} &= \frac{2^{\Delta(\#E)}}{2^{\Delta({\#A_v})}2^{\Delta({\#B_p})}}\times \GSD_{\text{Cylinder}} \\
     &=\frac{2^{-3}}{2^{-1}2^{-3}} \times 2=4
\end{aligned} 
\end{equation}

   \begin{figure}[htbp]
 \includegraphics[width=0.30\textwidth]{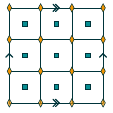}\caption{\label{fig:toricgsd} Minimal stabilizer model for toric code.}
\end{figure}
\end{Example}

This method of constructing a minimal model to count the degeneracy becomes especially useful for more complicated configurations. All one needs to do is to make sure no product of the operators gives identity. This is the same condition that all the operators are independent. An interesting domain wall which was discussed is the $e \leftrightarrow m$ of the toric code~\ref{Example:emdw}. We can compute the GSD for this model using this simple model construction.

\begin{Example}
  [$\Z_2$   on a Torus with a contractible $e\leftrightarrow m$ twist]
A minimal example for the toric code with a contractible twist is depicted in Fig.~\ref{fig:z2torustwist1con}. We perform Alg.~\ref{alg:Bdry} from the toric code and carefully count what changes. We remove $2$ $A$ stabilizers and $2$ $B$ stabilizers and add one $T$, $U$ and $Q$. They are all of order $m=2$. In total, we have $\Delta n_2=3-4=-1$.
\begin{figure}[htbp]
  \centering
     \includegraphics[width=0.30\textwidth]{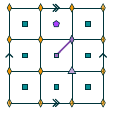}
   \caption{Toric code with a single finite twist}
  \label{fig:z2torustwist1con}
\end{figure}
Now, let us examine how the dependencies are modified. Since the product of the remaining $A$ operators, similarly for all $B$ operators, no longer constitutes the global identity, the two old dependencies are removed. However, note that the product of all $A$s, $B$s, $Q$s, $U$s, and $T$s does constitute the identity, so there is one dependency of order $2$. Altogether, we have $\Delta l_2=1-2=-1$ and since no qubits were added or removed $r_D=1$. The exponent on $m$ in Eq.~\ref{eq:gsdupdate} then goes as $\Delta n_m-\Delta l_m=0$ such that
\begin{equation}
\label{eq:1twist}
    \GSD_\text{1-Twist} = \frac{\GSD_\text{TC}}{2^{-1+1}} = 4. 
\end{equation}
This is correct because the codespace is the same as before, although a global phase factor $-1$ may now be applied due to the presence of the lone twist~\cite{Bombintwist2010}.
\end{Example}

What about an additional twist? We can again use Eq.~\ref{eq:gsdupdate} to obtain a formula for $N$ twists as we now show.
\begin{Example}[$\Z_2$ on a torus with $N$ twists]
Take the configuration with one twist, described by Eq.~\ref{eq:1twist}, as our initial configuration with four-fold GSD. Adding another twist (not connected to the first one) does not change the dimension or number of degrees of freedom ($r_D=1$). We will again add one $1$ T, one $U$, and one $Q$ operators Eq.~\eqref{eq:cleanfishy}, remove $2$ star operators, and remove $2$ plaquette operators. 
\begin{figure}[htbp]
\includegraphics[width=0.48\textwidth]{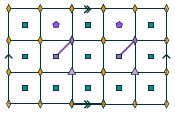}\caption{\label{fig:z2torustwistcons} Stabilizers for the Toric code with two twists.}
\end{figure}

Now, let us count the dependency. Since we already had one twist, which combined the two dependencies of the toric code into a single dependency, the total number of dependencies does not change this time, $\Delta l_2=0$. The new degeneracy is:
\begin{equation}
    \GSD_\text{2-Twists} = \text{Eq.~\ref{eq:1twist}}\times 1\times 2\times 1=8.
 \end{equation} 
We can recursively apply the same calculation when adding more twists, such that for $N$ twists:
\begin{equation}
    \GSD_\text{N-Twists} = \GSD_\text{Toric (N-1)-Twists} \times 2 = 2^{N+1}.
\end{equation}
This result is interpreted as the logical Hilbert space of $N$ twists, coming with $N-1$ qubits, appended to the original toric code's four-fold degeneracy.

\end{Example}

\begin{Example}[$\Z_2$ on a Torus with a non-contractible twist]
\label{ex:z2twistnoncon}
Let us restart with a clean version of the 18-qubit, 9-plaquette, and 9-star toric code. The new configuration with the non-contractible twist is shown in Fig.~\ref{fig:z2torustwistnoncon}.

\begin{figure}[htbp]
 \includegraphics[width=0.30\textwidth]{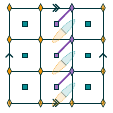}\caption{\label{fig:z2torustwistnoncon} Stabilizer model for the $\Z_2$ toric code with $1$ non-contractible twist (an $e\leftrightarrow m$ domain wall along one handle of the torus.)}
\end{figure}

We summarize the changes from the toric code using the following table:

\begin{table}[!htbp]
\centering
\begin{tabular}{|c|c|c|c|c|c|c|}
\hline
   Order & \multicolumn{4}{|c|}{Operators}  & $\Delta n_m$ & $\Delta l_m$ \\ \hline
m=2      & +3 XZ   & +3 $F$  & -3 $A$  & -3 $B$ &0 &-1 \\ \hline
 
\end{tabular}
\caption{Summary of changes from toric code to toric code with a non-contractible twist }
\end{table}

We then immediately read off the degeneracy as:
\begin{equation}
    \GSD_\text{non-con. Twist} = \frac{1}{2^{0-(-1)}}\times \GSD_\text{Toric Code}=2
\end{equation}
This example follows the result of Ref.~\citenum{pandey2022} where it was shown that adding a non-contractible twist gauges the toric code's $\widetilde{X_1}\widetilde{Z_2}$ degree of freedom, hence adding one \textit{logical} constraint resulting in a \textit{single} logical qubit~\cite{pandey2022}.

\end{Example} 
\begin{Remark}
    In all the examples discussed so far, changing from $\Z_2$ to $\Z_m$ we just change the order of operators from $2$ to $m$ in Eq.\eqref{eq:gsdupdate}.
\end{Remark}
Having developed Alg.~\ref{alg:Bdry} to gauge stabilizer codes, and Eq.~\ref{eq:gsdupdate} to count and update degeneracies, we are now in a position to prove the results presented in Sec.~\ref{sec:patch_code}, which concern twisted quantum doubles coexisting. Since the twisted quantum doubles can be formed by condensation from quantum doubles~\cite{Tyler2022}, the degeneracy of such systems can be proved with minimal illustrative models. Continuing with the arguments above, we begin by considering a $D(\Z_4)$ toric code with $GSD=4^2=16$. We now proceed to condense finite spatial regions and determine the GSDs.

 \begin{Example}[$\Z_4$ with a contractible $DS$ patch]\label{Example:z4dscont}

Start from the $D(\Z_4)$ toric code and condense just one region that contains the DS phase. Any model with a single contractible (meaning it does not wrap around either handle of the torus) patch will have the same degeneracy as this construction. Before condensing the DS patch, we had $\prod_vA_v=1$ and $\prod_p B_p=1$, which meant that one of the plaquettes and one of the star operators is dependent for a total of two order four dependencies ($l_4=2$) and zero order two dependencies ($l_2=0$) such that $l=l_2+l_4=2$. To condense a DS patch, we need to add two $X^2Z^2$ ribbons, as in Fig.~\ref{fig:z4dstorusii}. Afterward, we remove three-star operators and three plaquette operators that do not commute with the $X^2Z^2$ ribbons.
We then add products of the removed operators that commute with the new stabilizers. These will be the $F(v,p)$ and $B^2(p)$ operators Eq.~\eqref{eq:Z4DSops0}. We will add three $F$ operators on the removed vertices and three $B^2(p)$ operators on the removed plaquettes.
\begin{figure}[htbp]
  \centering
 
    \includegraphics[width=0.35\textwidth]{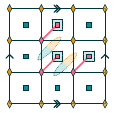}
 
  \caption{Finite contractible DS patch embedded in a $D(\Z_4)$ toric code. }
  \label{fig:z4dstorusii}
\end{figure}
 
Note that we have two new constraint relations again. First, we have $\prod_{v,p}A(v)B(p)F(v,p)=1$ because this is the same as multiplying all star and plaquette operators before adding the DS patch. Second, we have $\left(\prod_{p\in DS} B^2(p)\right) \left(\prod_{p \notin DS}B(p)\right)^2=1$.  Here, the product is over the new plaquettes, which live inside or adjacent to the right or top of the DS patch, with the square of the rest of the old plaquettes that were not removed. Thus, $l'=2$ and $r_l=2/2=1$. 
The crucial observation is that the first constraint will remove an operator with order four, for example, one $F(v,p)$, while the second constraint means we are double-counting an operator of order $2$, taken here as $B^2(p)$.  
 
\footnote{Note also that if we have two operators multiplying to identity $AB=1$ with orders $m,n$ i.e. $A^m=B^n=1$, then $m=n$ and they have the same order. Simply because $(AB)^m=1=A^mB^m=1B^m$ similarly with $n$.}. The counting of stabilizers is then summarized in the following table.

\begin{table}[h]
\centering
\begin{tabular}{|c|c|c|c|c|c|}
\hline
Order & \multicolumn{3}{|c|}{Operators} & $\Delta n_m$ & $\Delta l_m$ \\ \hline
$m=2$ & +2 $X^{2}Z^{2}$ &  \multicolumn{2}{|c|}{+3 $B^{\,2}$}  & +5 & +1 \\ \hline
$m=4$ &  +3 $F$        & -3 $A$        &-3 $B$  &             -3 & -1 \\ \hline
\end{tabular}
\caption{Changes from a $\Z_4$ code to a $\Z_4$ code with a contractible DS patch.}
\end{table}

We then immediately read off the degeneracy as:
\begin{equation}
    \GSD_\text{$\Z_4$-1-DS} = \frac{1}{2^{5-1} 4^{-3-(-1)}}\times \GSD_\text{$\Z_4$}=16
\end{equation}

Intuitively, the new contractible patch does not obstruct any of the old loops (logical operators) and does not add any linearly independent ground states either.

\end{Example} 
We can add more DS patches, and the situation becomes similar to the case of $\Z_2$ with many twists.

\begin{Example}[$\Z_4$ on a Torus with $N$ $DS$ patches]\label{Example:z4noncontpatches}
  Starting with example~\ref{Example:z4dscont} with one $DS$ patch, we can add one more patch as shown in Fig.~\ref{fig:z4dstorusiin}. 
  \begin{figure}[htbp]
 \includegraphics[width=0.48\textwidth]{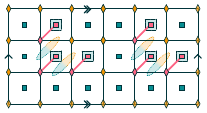}\caption{\label{fig:z4dstorusiin} 2 contractible  DS patches }\end{figure}
 
 In the old configuration we had the relations: $$\prod_{v,p}A(v)B(p)F(v,p)=1, \ \bigl(\prod_{p\in DS} B^2(p)\bigr) \bigl(\prod_{p \notin DS}B(p)\bigr)^2=1$$ As before, in the second constraint, the product is over the new plaquettes which live inside or to the right or top of the DS patch and the square of the rest of the old plaquettes that were not removed. This means $l_2=1$ and $l_4=1$. To form the second patch, we first remove three star operators $A(v)$ and three plaquette operators $B(p)$. We proceed by adding two $X^2Z^2$ operators, three new $F(v,p)$ operators, and three new plaquette operators $B^2(p)$. The constraints will remain the same. Consequently, $l'_2=1$ $l'_4=1$ and $\Delta  l_2= \Delta l_4=0$. The counting of the stabilizers and dependencies is then summarized in the following table:
\begin{table}[h]
\centering
\begin{tabular}{|c|c|c|c|c|c|}
\hline
Order & \multicolumn{3}{|c|}{Operators} & $\Delta n_m$ & $\Delta l_m$ \\ \hline
$m=2$ & +2 $X^{2}Z^{2}$ &  \multicolumn{2}{|c|}{+3 $B^{\,2}$}  & +5 & 0 \\ \hline
$m=4$ &  +3 $F$        & -3 $A$        &-3 $B$  &             -3 & 0\\ \hline
\end{tabular}
\caption{Table summarizing changes from $\Z_4$ code to $\Z_4$ and DS half and half contractible code.}
\end{table}

The ground state degeneracy for one contractible patch of DS is then:
 \begin{equation}
    \GSD_\text{$\Z_4$-2-DS} = \frac{1}{2^{5-0} 4^{-3-(0)}}\times \GSD_\text{$\Z_4$-1-DS}=32
\end{equation}

Adding more patches will not change the constraints, as we saw. For $N$ patches, we then have the following degeneracy, which is valid for $N\geq1$:
 \begin{equation}
    \GSD_\text{$\Z_4$-N-DS} = 4^{\frac{N+3}{2}}
\end{equation}

\end{Example}

\begin{Example}[Torus with half $\Z_4$ and half $DS$ phase]\label{Example:z4dsnoncont}

Let us start with a $\Z_4$ on a torus. We want to calculate the degeneracy of a phase where half of the torus is a $\Z_4$ QD, and the other half is in the Doubled Semion (DS) phase. Since here the DS lives on a cylinder, the boundary conditions will be different from just a DS patch.
\begin{figure}[!htbp]
\centering
\begin{subfigure}[b]{0.19\textwidth}
\centering
\includegraphics[width=\textwidth]{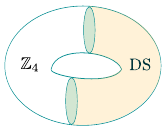}
\caption{Sketch for the configuration with $\Z_4$ toric code with the DS phase occupying half the torus.}
\label{fig:z4dstorus}
\end{subfigure}
\hfill
\begin{subfigure}[b]{0.28\textwidth}
\centering
\includegraphics[width=\textwidth]{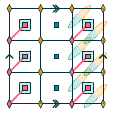}
\caption{Minimal lattice model for the configuration on the right. Compare with Example~\ref{Example:z4noncontpants}, which uses a different method.}
\label{fig:z4dstorus}
\end{subfigure}
\caption{$\Z_4$ toric code with half the torus as DS phase.}
\end{figure}

In this case, we also need to specify the order of the operators since they are different.

\begin{table}[h]
\centering
\begin{tabular}{|c|c|c|c|c|c|}
\hline
Order & \multicolumn{3}{|c|}{Operators} & $\Delta n_m$ & $\Delta l_m$ \\ \hline
$m=2$ & +6 $X^{2}Z^{2}$ &  \multicolumn{2}{|c|}{+6 $B^{\,2}$}  & +12 & 1 \\ \hline
$m=4$ &  +6 $F$        & -6 $A$        &-6 $B$  &             -6 & -1\\ \hline
\end{tabular}
\caption{Summary of changes from $\Z_4$ toric code to $\Z_4$ sharing the torus with DS phase. }
\end{table}

We then immediately read off the degeneracy as:
\begin{equation}
    \GSD_\text{$\Z_4$-DS} = \frac{1}{2^{12}4^{-6-(-1)}}\times \GSD_\text{$\Z_4$}=8
\end{equation}

\end{Example} 
This can be understood intuitively since the full $\Z_4$ loops can no longer traverse the torus as some of them get identified or confined when they pass through the DS phase. Another way to calculate the $\GSD$ will make this clearer in~\ref{sec:lego_code}. We saw in example~\ref{Example:z4dscont} that when the DS phase occupied a contractible region, the degeneracy was the same as without the DS phase. The stabilizer counting using minimal models is then a powerful tool for exploring modifications to topological orders in a sequential way. Paired with Alg.~\ref{alg:Bdry}, one can imagine using a computer program to automate the construction of topological quantum error-correction codes.

\subsection{GSD using Macroscopic Pants Decomposition}\label{sec:lego_code}

Dimensional counting is useful for deriving the logical properties, such as ground state degeneracy, from microscopic configurations, as we saw in the last subsection. This subsection offers a complementary, top-down rather than bottom-up, and intuitive methodology, involving only high-level macroscopic information, for calculating the $\GSD$ by counting the number of non-contractible logical loops of anyons in our model. This is based on Ref.~\citenum{Tan2015}. We use these methods to explore and validate results about the introduced composite codes.

\begin{Example}[$D(\Z_2)$ on a Cylinder with two smooth boundaries] 

Let's take the cylinder with both smooth boundary conditions, as shown in Fig.~\ref{fig:z2cylinderi}. While there are two non-contractible loops of anyons around the circumference of the cylinder, only the $Z$ loop is non-trivial. Further, an $X$ string between the two smooth boundaries of the cylinder is non-contractible and non-trivial. The $Z$ loop and the $X$ string anticommute, and we have $\GSD=2$. Evidently, all we needed was the information about the topological surface the theory lives on, along with the types of boundaries it has. The specific microscopic details were not used.
    \begin{figure}[htbp]
 \includegraphics[width=0.48\textwidth]{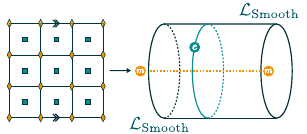}\caption{\label{fig:z2cylinderi} Left: microscopic stabilizer model for the $\Z_2$ code on a cylinder with two smooth boundaries. Right: the macroscopic topological data needed to compute the $\GSD$ of the model.}
\end{figure}
\end{Example}

The heuristic method of counting loops can be readily generalized to any orientable surface with or without boundaries. The exact stabilizer formulation is not needed to compute the ground state degeneracy once the topological data are fixed. The subtlety is that one should avoid overcounting or counting dependent loops. This issue can be tackled for a broad class of topological codes, as we now discuss.

In topology, it is important to systematically classify how arbitrary surfaces are constructed in terms of simpler primitive structures. For example, how standard spheres and holes can be glued to form any connected and simply-connected 2-complex is a well-studied problem~\cite{Bakalov2000}. In this spirit, this section uses the results of Ref.~\citenum{Tan2015} to validate results in Sec.~\ref{sec:patch_code}.

\begin{figure}[htbp]
\includegraphics[width=0.48\textwidth]{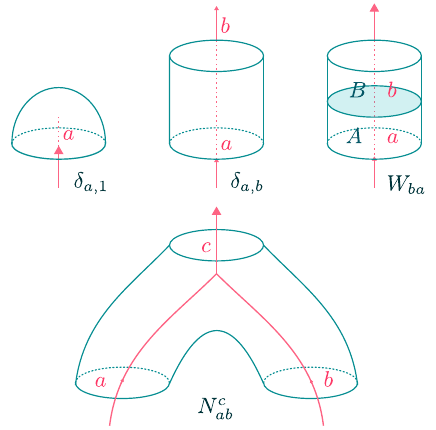}
\caption{\label{fig:pants} Any orientable 2D manifold can be decomposed into caps, cylinders, and pants. Each component will be associated with a tensor with indices in the anyons of the bulk theory. }
\end{figure}

The construction relies on two facts. First, as shown in Fig.~\ref{fig:pants}, any orientable 2D manifold can be decomposed into caps, cylinders, and pants \cite{hatcherpants1999,Tan2015}. Second, there is a correspondence between domain walls (boundaries) and a tunneling or transfer matrix that describes the domain wall's action on anyons \cite{Tan2015}. 

\begin{figure}[htbp]
\includegraphics[width=0.48\textwidth]{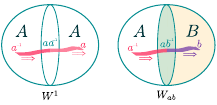}
\caption{\label{fig:wabmatrix} Left: the trivial domain wall denoted by $W^1$ transports $a$ to $a$. Alternatively, this transparent domain wall condenses $aa^{-1}$.  Right: a domain wall $W$ can transport anyon $a$ to $b$, thus condensing $ab^{-1}$. The $W_{ab}$ matrix entry is the dimension of the fusion space of the anyon $ab^{-1}$ on the domain wall $W$ on a sphere. Intuitively, it measures in how many independent ways the anyon $a$ in phase $A$ tunnels into an anyon $b$ in phase $B$. }
\end{figure}

To define the tunneling matrix $W$ \cite{Tan2015}, consider the case where two phases $A$ and $B$ share a domain wall $W$ between them as shown in Fig.~\ref{fig:wabmatrix}. Note that we will use $W$ for the domain wall and its associated matrix.

We define $W_{ab}$ to be the dimension of the fusion space ($\mathcal{V}$) of this configuration on a sphere ($S^2$): $W^1_{ab }\coloneq \operatorname{dim}[\mathcal{V}(S^2,b,W,a^{-1})] \in \N$~\cite{Tan2015}. This is also the number of independent tunneling channels that tunnel anyon $a$ in $A$ to anyon $b$ in $B$. Since gapped domain walls respect the braiding statistics of the two phases, we have the following relations \cite{Tan2015}:
\begin{equation}
{W}{S}^B = {S}^A{W}, \quad {W}{T}^B ={T}^A{W} 
\end{equation}
Here, $S^A$ matrix encodes braiding statistics of anyons in phase $A$. The $T^A$ matrix is a diagonal matrix that encodes their topological twist or spin. We will also only consider stable domain walls whose GSD cannot change due to local perturbations. This gives us the constraint \cite{Tan2015}:
\begin{equation}
{W}_{i a} {W}_{j b} \leq \sum_{k c}\left({N}^D\right)_{i j}^k {W}_{k c}\left({N}^A\right)_{a b}^c
\end{equation}
where the $(N^A)^c_{ab} \in \N$ is the dimension of the fusion channels of anyons $a$ and $b$ into $c$.
Using this construction, and given a certain spatial topology, any surface can be decomposed into caps, cylinders, and $W$ matrices corresponding to different domain walls between phases. This is shown in Fig.~\ref{fig:pants}. It actually suffices to consider caps, domain walls, and pants since a cylinder can be formed with one pants and capping one of its boundaries. The two types of cylinders are included for convenience. The $\GSD$ can be computed by enumerating how many non-trivial independent anyonic loops exist. This will correspond to concatenating the topological components of each configuration, known as the pants' decomposition. 

Systematically, the method can be described as follows. First, we begin by taking as an input a 2D surface $\M$ that can have multiple spatial boundaries $\partial \M_i$. The surface may also be divided into different topological phases with appropriate domain walls $W^{j}$ between them. Boundaries are described as domain walls with the vacuum $W^{i}$ for uniformity. Second, we decompose the 2D surface using pants decomposition into the 2D components described in Fig.~\ref{fig:pants}. This will isolate different bulk phases that have domain walls between them into different elementary 2D components. Multiple pants decompositions exist for the same configuration, and it is not crucial to use the minimal one. The different decompositions are related by a set of topological moves that will naturally not change the topology of the surface nor affect the $\GSD$ calculated here \cite{hatcherpants1999,Tan2015}. We also do not discuss how this decomposition is formally carried out, since in all the examples, finding a decomposition is relatively easy. Third, we attach to every 2D component its appropriate tensor from Fig.~\ref{fig:pants}. The tensors take values in the anyon labels of the bulk (boundary) of their respective topological phases. In this step, we also need to choose labels for the indices. Two indices that are connected should have the same label. In the end, no free indices should remain. Fourth, the $\GSD$ will be the scalar resulting from contracting the indices with the same label. This process is better illustrated with examples.

\subsubsection{Examples using Pants Decomposition}
\label{sec:pants}

\begin{Example}[$\Z_2$ on a Torus with a non-contractible twist II]
As a first example, let's recalculate the case of Toric code with one non-contractible twist, Fig.~\ref{fig:etompants}. The twist is just the domain wall $e\leftrightarrow m$. The matrix of this domain wall is given by:
\begin{equation}
W^{e\leftrightarrow m} \;=\;
\bordermatrix{%
      & 1 & e & m & em \cr
  1   & 1 & 0 & 0 & 0  \cr
  e   & 0 & 0 & 1 & 0  \cr
  m   & 0 & 1 & 0 & 0  \cr
  em  & 0 & 0 & 0 & 1  \cr
}
\end{equation}

We show the order of the basis for clarity. We can decompose the torus in this case as two cylinders. They intersect at two domain walls, one of them is the trivial domain wall $W^1_{ab}=\delta_{a,b}$ and the other one is $W^{e\leftrightarrow m}$. We then have: \begin{equation}
\begin{aligned}
        \GSD &= \sum_{a,b}W^{e\leftrightarrow m}_{a,b}W^1_{ab} =\sum_{a,b} W^{e\leftrightarrow m}_{ab}\delta_{a,b}\\
        &= \operatorname{Tr}(W^{e\leftrightarrow m})=2,
\end{aligned}
\end{equation} agreeing with Ref.~\citenum{pandey2022} and our result using stabilizers model in Ex.~\ref{ex:z2twistnoncon}.
\begin{figure}[htbp]
 \includegraphics[width=0.35\textwidth]{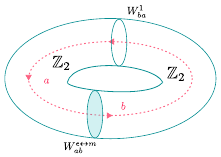}\caption{\label{fig:etompants} Pants decomposition for the $D(\Z_2)$ twisted toric code, i.e., containing one non-contractible twist. It consists of two $D(\Z_2)$ cylinders glued together on one side by the transparent domain wall and by the $W^{e \leftrightarrow m}$ on the other side.}
\end{figure}
\end{Example}

\begin{Example}[$\Z_4$ with a non-contractible DS patch]
\label{Example:z4noncontpants}
    Another example is the case of $\Z_4$ and a non-contractible $DS$ patch on a torus. We have a $4\times16$ domain wall $W$ because $DS$ has 4 anyons $\{1,s,\overline{s},b\}$ while $\Z_4$ has 16 anyons labeled by $e^im^j$ where $0\leq i,j\leq3$. The condensation procedure to obtain $DS$ from $\Z_4$ informs us that the $e^2m^2$ anyons in $\Z_4$ are identified with the vacuum in $DS$, while the semion $s$ condenses with $\{em,e^3m^3\}$ etc. The condensation and tunneling are exactly the same for $DS$ anyons, as each one is its own antiparticle. We then have the $W$ matrix:
\begin{equation}\label{eq:wforz4ds}
W^{\dagger} \;=\;
\bordermatrix{%
           & 1 & em & e^2 & m^2 & e^2m^2 & em^3 & e^3m & e^3m^3  \cr
  1        & 1 & 0  & 0   & 0   & 1      & 0    & 0    & 0       \cr
  s        & 0 & 1  & 0   & 0   & 0      & 0    & 0    & 1       \cr
  \bar s   & 0 & 0  & 0   & 0   & 0      & 1    & 1    & 0       \cr
  b        & 0 & 0  & 1   & 1   & 0      & 0    & 0    & 0    
}
\end{equation}
We only showed the non-zero columns of the matrix. As shown in Fig.~\ref{fig:dsz4pants}, the torus can be decomposed as a $D(\Z_4)$ and a DS cylinder, which meet at the two copies of the same domain wall. The GSD is then:
\begin{equation}
\GSD = \sum_{a,b} W_{ab}{W}^{\dagger}_{ba} = \Tr( WW^{\dagger})=8
\end{equation}
which confirms the results of Ex.~\ref{Example:z4dsnoncont}.
\begin{figure}[htbp]
\includegraphics[width=0.35\textwidth]{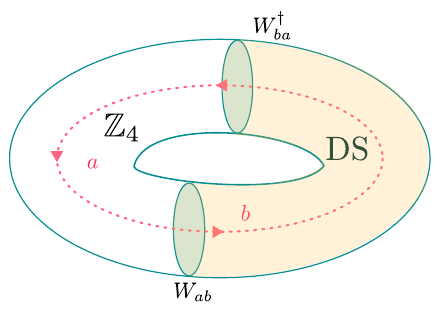}
\caption{\label{fig:dsz4pants} Pants decomposition of the torus with half $\Z_4$ and half DS phases. It consists of two cylinders glued at both ends with the unique domain wall $W$ between the two phases.}
\end{figure}
\end{Example}

One can also treat the example with a contractible $DS$ patch. While this was a relatively complicated example using stabilizer constraints, it has an easier solution using pants decomposition.
\begin{Example}[$D(\Z_4)$ with a one non-contractible DS patch]
   For a one-patch code, as shown in Fig.~\ref{fig:z4dspatchespants}, we decompose the Torus into a $\Z_4$ cylinder, $\Z_4$ pants, and a $DS$ cap. 
   \begin{figure}[htbp]
\includegraphics[width=0.35\textwidth]{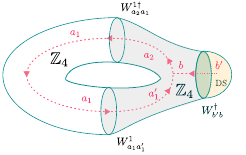}
\caption{\label{fig:z4dspatchespants} Pants decomposition of the $\Z_4$ toric code with one contractible DS patch. Because of the trivial wall $W^1$, we actually have $a_1=a_1'$ and $a_2=a_1$. }
\end{figure}

   All the domain walls between two $\Z_4$ phases are trivial and denoted here by $W^{1}$. The domain wall from $\Z_4$ to $DS$ is denoted by $W$, which is given in Ex.~\ref{Example:z4noncontpants}. Finally, we attach to the pants the $D(\Z_4)$ fusion coefficients $N_{i,j}^k = \del_{i+j,k}$ and to the cap the delta operator. The $\GSD$ is then:
\begin{equation}
\begin{aligned}
\GSD &= \sum_{a,j,i,u,k} W^1_{aj}N_{i,j}^{k}W^{\dagger}_{ui}\del_{u,1}{W^1}_{ka} = \sum_{a,i}  N_{i,a}^{a}W^{\dagger}_{i,1} \\
&=\sum_a N_{1,a}^aW^{\dagger}_{1,1}=\sum_{a}=16 
\end{aligned}
\end{equation}
where in the second step we used $N_{i,a}^{a}=\delta_{i,1}$. In the last step, we just summed over the anyon types of $D(\Z_4)$. This agrees with the results of Ex.~\ref{Example:z4dscont}.

\begin{Example}[$D(\Z_4)$ on a torus with a $N$ DS patches]\label{ex:Ndsz4}
Let us start with the case of two patches. We use the decomposition shown in Fig.~\ref{fig:z4dspatchespants2}.

We can then compute the $\GSD$ of the two patches as:
 \begin{equation}
  \begin{aligned}
  \GSD &= \sum_{a,b} W^1_{a_1a_1'}N_{a_1,b_1}^{a_2}W^{\dagger}_{b_1,b_1'}\delta_{b_1',1}W^{\dagger}_{a_2,a_2'}N_{a_2',b_2}^{a_1} W^{\dagger}_{b_2'b_2}\delta_{b_2',1} \\
        &= \sum_{a,b} W^{\dagger}_{b_1,1}N_{a_1,b_1}^{a_2}W^{\dagger}_{b_2,1}N_{a_2,b_2}^{a_1} \\
        &= \sum_{a,b} W^{\dagger}_{b_1,1}\delta_{a_1+b_1,a_2}W^{\dagger}_{b_2,1} \delta_{b_2,-b_1} \\
         &= \sum_{a_1,b_1} W^{\dagger}_{b_1,1}W^{\dagger}_{-b_1,1}\\
&= 16 \times 2
  \end{aligned}
\end{equation}
In the second line, we simplified using the deltas from the transparent domain walls and from the caps. Since the theory is Abelian, in the third line we replaced $N_{a,b}^c$ with $\delta_{a+b,c}$. In the last line, we observe that $a_1$ has $16$ choices corresponding to the anyons of $\Z_4$. While $b_1$ have only two choices corresponding to $\{1,e^2m^2\}$ because of the domain wall $W$.
\begin{figure}[htbp]
 \includegraphics[width=0.35\textwidth]{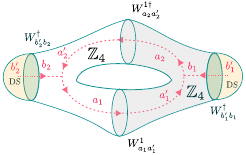}\caption{\label{fig:z4dspatchespants2} Pants decomposition of the $\Z_4$ toric code with two contractible DS patches. Because of the trivial domain walls, we have $a_i=a_i'$. }\end{figure}
\end{Example}

The pattern will continue when adding new patches. If we added $N$-patches, then the ground state degeneracy can be computed as follows (for $N\geq1$):
 
\begin{alignat}{2}          
\GSD  &=\sum_{a,b}\,&&W^{\dagger}_{b_{1},1}  N_{a_{1},b_{1}}^{a_{2}}W^{\dagger}_{b_{2},1}N_{a_{2},b_{2}}^{a_{3}}\dots \\
        &&&\times W^{\dagger}_{b_{N-1},1} N_{a_{N-1},b_{N-1}}^{a_{N}}W^{\dagger}_{b_{N},1}   N_{a_{N},b_{N}}^{a_{1}} \nonumber \\
         &= \sum_{a,b}\,&&W^{\dagger}_{b_{1},1}\,\delta_{a_{2},a_{1}+b_{1}}\,
              W^{\dagger}_{b_{2},1}\,\delta_{a_{3},a_{2}+b_{2}}
              \dots \nonumber \\
              &&& \times W^{\dagger}_{b_{N-1},1}\,\delta_{a_{N},a_{N-1}+b_{N-1}}\,
              W^{\dagger}_{b_{N},1}\,\delta_{b_{N},a_{1}-a_{N}} \nonumber \\[2pt]
      &= 16 \times &&2^{N-1}.\nonumber
\end{alignat}
 
 We note that $W^{\dagger}_{x,1}$ is only non-zero when $x=1$ or $x=e^2m^2$ as can be seen from Eq.~\eqref{eq:wforz4ds}. We then have only two choices for the indices $b_i$, and there are $N-1$ of them (since $b_N$ is dependent). This gives $2^{N-1}$ choices. After picking the free $N-1$ b's, picking any arbitrary choice for $a_1$ from the $16$ anyons of $\Z_4$ will determine the rest of $a's$ in addition to $b_N$. This gives $16\times2^{N-1}$ choices. This agrees with Ex.~\ref{Example:z4noncontpatches}.
 
\end{Example}

Cases with open boundaries can also be treated similarly since we can cap them with the vacuum. This is just the fact that boundaries of a phase are domain walls between the phase and the vacuum. However, this restricts us to the case when the boundary is of one type. Equivalently, it is not clear how to include the case where domain walls intersect.

\begin{Example}[DS Patches in a $\Z_4$ bulk with smooth boundaries]

The Pants decomposition can also be used with a single open boundary condition. In this case, one needs to add a cap that has the vacuum phase. Let us treat the configuration of two DS patches inside a $D(\Z_4)$, which itself has a smooth boundary. This is the same degeneracy as Fig.~\ref{fig:ds_patches} with two patches rather than four. The pants decomposition is shown in Fig.~\ref{fig:surfacedspants}. One can imagine stretching (topologically deforming) the vacuum cap until it becomes an outer boundary.

 \begin{equation}
  \begin{aligned}
  \GSD &= \sum_{a,b,c,a',b',c'} \delta_{a',1}W^\dagger_{a'a}N_{a,b}^c {W_{b'b}^{\text{smooth}}}^\dagger \delta_{b',1} W_{cc'} \delta_{c',1} \\
  &= \sum_{a,b,c} W^\dagger_{1a}\delta_{b,c-a} {W_{1b}^{\text{smooth}}}^\dagger  W_{c1} \\
   &= \sum_{a,b,c} W^\dagger_{1a}\delta_{b,1}\delta_{c-a,b} {W_{1b}^{\text{smooth}}}^\dagger  W_{c1} \\
    &= \sum_{a} W^\dagger_{1a} W_{-a1} = 2 \\
  \end{aligned}
\end{equation}
In the second line, we enforced the delta functions. In the third line, since $a,c \in \{1,e^2m^2\}$, the only pure flux $b$ that can appear as $(a-c)$ is the vacuum $1$. Finally, summing over the two choices of $a$ gives us $\GSD =2$. The case with $N$ patches can be treated similar to Example~\ref{ex:Ndsz4}
\begin{figure}[htbp]
\includegraphics[width=0.35\textwidth]{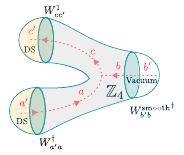}
\caption{\label{fig:surfacedspants} Pants decomposition of the $\Z_4$ toric code with two contractible DS patches and a smooth boundary. }
\end{figure}
\end{Example}

\section{Conclusion}
\label{sec:fin}

In this work, we have defined a new family of Abelian quantum error correcting codes. In addition to conventional boundaries of quantum doubles \cite{Bravyi1998,Beigi2011}, our codes contain finite-sized condensates of a twisted quantum double of the parent quantum double. We have explicitly illustrated this using the group $G=\Z_4$ and generalizing the DS global condensation\cite{Tyler2022} to a spatially local one. We derived the logical operators and showed that, in addition to whatever background logical operators of the parent $D(\Z_4)$, $N-1$ logical qubits are added via $N$ patches. This result explicitly shows how topological qubits of distinct dimensions (two and four) can co-exist in a single code. We have described why, from the perspective of simplifying and improving error correction in the near term, this approach is desirable.

To prove our results, we derived technical machinery in two complementary directions. 
First, we provided a concrete algorithm to gauge one stabilizer code into another. This was based on the concept of anyon condensation and, in an experimental setting, can be interpreted as measuring a particular set of stabilizers. Building on this algorithm, we have derived the ground state degeneracies by modding out the constraints from the total Hilbert space dimension. In addition to this microscopic lattice perspective, we have described how the number of logical operators, as loops of ribbon operators, can be counted and constructed in terms of topological primitives such as cylinders, pants, and caps.

Our work paves the path for multiple new research directions. Our first motivation is that these codes be realized in experiments. In this light, it will be interesting to compare the error rates and explore the tradeoffs exhibited between $\Z_2$ codes, DS codes, and composite $\Z_4-$DS codes. In addition to being experimentally explored, these questions can be explored by QECC simulations, which more holistically include the realistic costs of real-time decoding and the burden this places on code performance in experiments. 

Lastly, our work paves the path for more theoretical analysis of codes and a grand classification of quantum topologies. For example, our results here have, for the sake of brevity and providing concrete examples, been limited to low-dimensional Abelian cyclic groups. In addition to exploring non-Abelian groups, one may also perform a similar analysis with quantum doubles generated by higher-dimensional cyclic groups. In such a case, there will be room for more types of anyon condensation (both in its variety and in the number of times this procedure can be performed. It will be interesting to see how such concepts can be judiciously applied to lower the experimentalist's resource burden in QECC and also to engineer more exotic quasi-particle types for computation. 

During the final stages of writing this manuscript, the authors became aware of Ref.~\cite{Z4DS2025}, which analyzes phase transitions between DS and $\Z_4$ phases. While complementary, our work focuses on the quantum error-correcting codes derived from the judicious use of these phases. It will be interesting to investigate the intersection of these two concepts in future studies.

\begin{acknowledgments}
We thank Shawn Cui, Birgit Kaufmann, Ralph Kaufmann, Phil Lotshaw, Paul Kairys, Sohan Ghosh, and Tyler Ellison for useful discussions and comments. We thank Sabre Kais for access to the Bell computing cluster at Purdue University. M.M., A.G., and E.D. are supported by the U.S. Department of Energy, Office of Science, Advanced Scientific Research Program, Early Career Award under contract number ERKJ420. This research was performed during M.M.'s participation in the Graduate Research at ORNL (GRO) Educational Program. This research used resources of the Compute and Data Environment for Science (CADES) at the Oak Ridge National Laboratory, which is supported by the Office of Science of the U.S. Department of Energy under Contract No. DE-AC05-00OR22725.
\end{acknowledgments}

\appendix 

\section{Construction of Domain Wall between DS-$D(\Z_4)$}\label{sec:dsz4dwnotsimple}
In this appendix, we give the derivation of the domain wall between the DS phase and the $\Z_4$ phase using Alg.~\ref{alg:Bdry}. This proves the more heuristic method given in Ex.~\ref{ex:z4dsdwsimple}. Before starting the algorithm, we place the $D(\Z_4)$ on top while keeping the DS at the bottom. Denote the anyons of the bottom layer by $\{1,s_1,\overline{s}_1,b_1\}$ and for the top layer by $\{1,e_2,m_2,\dots\}$. The top layer will be a mirrored version of the bottom layer as in Fig.~\ref{fig:condensedefect}. The Hamiltonian is a mirrored version of the $\Z_4$ Hamiltonian Eq.~\eqref{eq:avop}. Here, the mirrored nature of the top layer will be fully exploited. Importantly, the anyon $e_2m_2$ with twist $\T(e_2m_2)=i$ in the unfolded picture is mapped to the anyon $e_2m_2^{-1}$ with $\T^{\text{op}}(e_2m_2)=i$ in the folded version because the handedness changes with mirror reflection.

\begin{figure}[htbp]
    \centering
        \begin{subfigure}[b]{0.23\textwidth}
        \includegraphics[width=\textwidth]{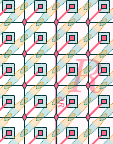}
     
    \end{subfigure}
    \hfill
    \begin{subfigure}[b]{0.23\textwidth}
      \includegraphics[width=\textwidth]{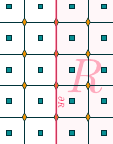}
    \end{subfigure}\caption{\label{fig:dsz4b1} Left: region of condensation $R$ with its boundary $\partial R$ in layer $1$ (the bottom un-mirrored DS layer). Right: region of condensation $R$ with its boundary $\partial R$ in layer $2$ (the top mirrored  $\Z_4$ layer). }
\end{figure}

We then begin the algorithm. First, we choose a region $R$ extending to infinity. The boundary $\partial R$ lives on the direct lattice in both layers, Fig.~\ref{fig:dsz4b1}. Second, we add the shortest ribbon operators creating anyons of the Lagrangian subgroup generated by $\L_{\Z_4-DS} =\{s_1 e_2m_2^{-1},\overline{s}_1 e_2m_2,b_1 e_2^2,b_1m_2^2,1_1e^2m_2, \dots \}\bigr)$. We show a subset of the generators for illustration:

\begin{equation}
  \text{\includegraphics[width=.48\textwidth]{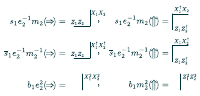}}
  \label{eq:dsz4ribbs}
\end{equation}
Here, the arrows indicate the direction of propagation of the anyons (the direction of the ribbons creating them).
 
After the second step, the ribbons in Eq.~\eqref{eq:dsz4ribbs} are added in $R$ and on $\partial R$. In the third step, we remove the old stabilizers that do not commute with the new ones. These will include all the stabilizers of both layers in the bulk of the region $R$. In addition, we remove the star operators $A_2(v)$ of layer $2$ on the boundary $\partial_1 R$. We remove the plaquettes $B_2(p)$ of the second layer adjacent to the boundary $\partial_1R$. We remove the $X^2Z^2$ operators of the DS touching $\partial R$. The configuration after the third step is shown in Fig.~\ref{fig:z4etom2}.
  
  \begin{figure}[htbp]
    \centering
        \begin{subfigure}[b]{0.23\textwidth}
        \includegraphics[width=\textwidth]{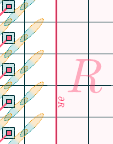}
        
    \end{subfigure}
    \hfill
    \begin{subfigure}[b]{0.23\textwidth}
      \includegraphics[width=\textwidth]{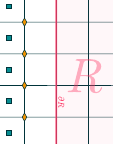} 
    \end{subfigure}\caption{After the third step, the stabilizers that do not commute with the measured ribbon operators are removed. These include all stabilizers of both layers in $R$. In layer $1$ (left), $B^2(p)$ and $X^2Z^2$ touching the boundary $\partial R$ were removed. In layer $2$ (right), $B_2(p)$ and $A_2(v)$ touching $\partial R$ were removed.  }
  \label{fig:dsz4b3}
\end{figure}

In the fourth step, we form new stabilizers that are products of the old ones and commute with the new stabilizers. These are given by products of the removed stars with their adjacent removed plaquettes, along with the $F$ operators of layer $1$. In addition, the product of the plaquette squared of the two layers.

\begin{equation}
  \text{\includegraphics[width=.49\textwidth]{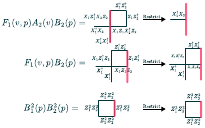}}
  \label{eq:realfish}
\end{equation}

We denoted the placement of the stabilizer relative to the boundary. After adding the new stabilizers to the model, we reach the configuration in Fig.~\ref{fig:z4etom3}

  \begin{figure}[htbp]
    \centering
        \begin{subfigure}[b]{0.23\textwidth}
        \includegraphics[width=\textwidth]{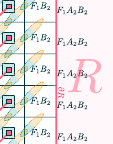}
        
    \end{subfigure}
    \hfill
    \begin{subfigure}[b]{0.23\textwidth}
      \includegraphics[width=\textwidth]{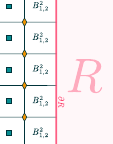} 
    \end{subfigure}\caption{After the fifth step, the products of old stabilizers that commute with the ribbons are added Eq.~\eqref{eq:realfish}. We only show here the stabilizers near the boundary. In addition, qudits in $R$ or on $\partial R$ were erased.   }
  \label{fig:dsz4b5}
\end{figure}

In the fifth step, we erase trivialized qudits. These are all the qudits in the bulk of $R$ or on its boundary $\partial R$. After condensing qudits, we also removed all stabilizers that only act on trivialized qudits for clarity, Fig.~\ref{fig:dsz4b5}.

The sixth step is to restrict the stabilizers at the boundary to their non-trivialized support. This results in the stabilizers shown in Eq.~\eqref{eq:realfish}. Finally, we unfold layer $2$ to be put on the right. This will result in the stabilizer of the bulk of layer $2$ ($\Z_4$) to be the same again as the un-mirrored $D(\Z_4)$ redoing the horizontal reflection. At each edge, now only one qudit lives, so we can drop the layers' index unambiguously. This results in the following stabilizers:
\begin{equation}
  \text{\includegraphics[width=.49\textwidth]{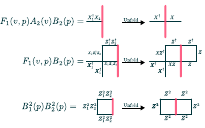}}
  \label{eq:realfish2}
\end{equation}

 Note how the vertical edges corresponding to $\partial R$ are now condensed. These are perfectly valid stabilizers for the domain wall between the DS and $D(\Z_4)$, however, it is desirable to have lower-weight stabilizers. This is done by taking the first stabilizer $X_1^\dagger X_2=1$ on the Hilbert space level. This will transform the two qudits into one qudit. A mapping of two-qudit operators to this new degree of freedom stitches the lattice back, giving the familiar stabilizers.  
\begin{equation}
  \text{\includegraphics[width=.49\textwidth]{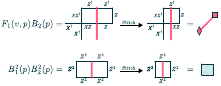}}
  \label{eq:realfish3}
\end{equation}

\begin{figure}[htbp]
\centering
 
  \includegraphics[width=0.35\textwidth]{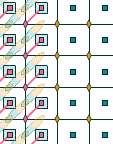}\caption{After stitching the stabilizers at the boundary together, we retrieve the domain wall between the DS and $D(\Z_4)$ phases.}
  \label{fig:dsz4b7}
 
\end{figure}

After stitching the lattice back, we find the domain wall between the DS and $\Z_4$ phases shown in Fig.~\ref{fig:dsz4b7}. This proves the results using the more heuristic method used in Ex.~\ref{ex:z4dsdwsimple} with direct condensation.

\section{Example of a $0D$ Defect Construction}\label{sec:appinvdw}
In this appendix, we use Alg.\ref{alg:Bdry} along with the folding method to construct the inverse domain wall of $D(\Z_4)$ code (${(e,m) \to (e^{-1},m^{-1})}$). More importantly, this exemplifies how finite domain walls ending with $0$D defects can be constructed systematically. 

\begin{figure}[htbp]
\centering
 
  \includegraphics[width=0.35\textwidth]{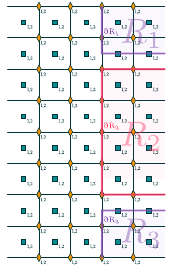}\caption{Regions used to create the finite $(e,m) \to (e^-{1},m^{-1})$ Defect of $D(\Z_4)$. The three boundaries do not overlap and completely cover the left half-plane of the doubled layer. Stabilizers have two subscripts corresponding to both layers.}
  \label{fig:defectee1}
 
\end{figure}

Before beginning the algorithm, we place two copies of the $D(\Z_4)$ and denote the bottom layer with index $1$ and the top layer with index $2$. First, we select three regions this time. For regions $R_1$ and $R_3$ we will create the trivial domain wall, while for region $R_2$ we will create the inverse domain wall. The regions are shown in Fig.~\ref{fig:defectee1}

\begin{figure}[htbp]
\centering
   \includegraphics[width=0.35\textwidth]{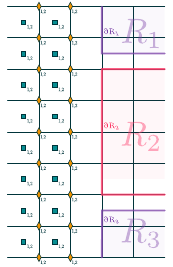}\caption{After the third step, the stabilizers that did not commute with measured operators are removed. These are all stabilizers from both layers that share at least one edge with the bulks or the boundaries of the three regions.}
  \label{fig:defectee2}
 \end{figure}

Second, we need to measure the ribbon operators corresponding to the Lagrangian subgroup of the respective region and domain wall. In regions $R_1$ and $R_3$, we will measure the ribbons of the Lagrangian subgroup creating the trivial domain wall $\L_{\text{trivial}}=\{1,e_1e_2^{-1},m_1m_2^{-1}\}$ while for region $R_2$ we measure the Lagrangian subgroup creating the inverse domain wall. The ribbons are all one-edge operators \eqref{eq:defecteeribbs1}.
\begin{equation}
  \text{\includegraphics[width=.40\textwidth]{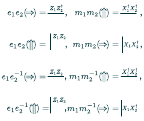}}
  \label{eq:defecteeribbs1}
\end{equation}
Third, we remove old stabilizers that do not commute with the measured operators. These will be all stabilizers in the bulk of the three regions. In addition, we remove star operators on the boundaries $\partial R_1$, $\partial R_2$, and $\partial R_3$ along with plaquettes adjacent to the three boundaries. After this step, we reach the configuration shown in Fig.~\ref{fig:defectee2} 

In the fourth step, we need to add products of removed stabilizers that commute with the measured operators. These will be products of star operators of both layers $A_1(v)A_2(v)$ or $A_1(v)A^\dagger_2(v)$ for the trivial and the inverse domain wall, respectively. Similarly products of plaquettes of each layer  $B_1(p)B_2(p)$ or $B_1(p)B^\dagger_2(p)$ will be added. Note that the operators of the second layer are mirrored since it is folded as shown in Eq.~\eqref{eq:defecteeribbs2}.
\begin{equation}
  \text{\includegraphics[width=.40\textwidth]{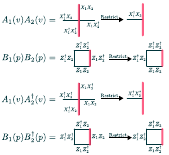}}
  \label{eq:defecteeribbs2}
\end{equation}

\begin{figure}[htbp]
\centering
   \includegraphics[width=0.35\textwidth]{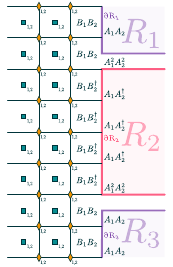}\caption{The configuration after the fourth and fifth step. In the fourth step, products of old stabilizers that commute with the new measured operators are added. We only added them near the boundary, as the rest of them will be condensed. After the fifth step, trivialized qudits are erased.}
  \label{fig:defectee3}
 \end{figure}

These operators have to be added in place of the old, removed stabilizers everywhere. However, they are most important near the boundary, as they will create the domain wall. Thus, we do not need to figure out those terms except at the boundaries. Importantly, right at the $0$D defect, the stars will need to satisfy both boundaries. This is the first time the order of the group will matter. So far, everything was applicable to all $D(\Z_n)$. In $D(\Z_4)$, the square of stars $A_1^2(v)A_2^2(v)$ will commute with both boundaries. Fifth, we erase trivialized qudits in the bulk of the three regions along with their boundaries. In Fig.~\ref{fig:defectee3}, the new plaquette and star operators are shown at the three boundaries, while trivialized qudits are erased.

 Sixth, we need to restrict the operators to their non-trivialized support. This results in the operators of the last column of Eq.~\eqref{eq:defecteeribbs2}. Since there are no more qudits to condense, we are done. We can unfold the two layers, dropping the layer index. The operators at the boundary will transform to:
\begin{equation}
  \text{\includegraphics[width=.40\textwidth]{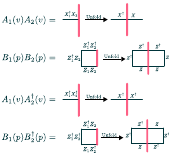}}
  \label{eq:defecteeribbs3}
\end{equation}

This is a valid finite-domain wall mapping anyons to their anyons with a $0$D defect at each end. We see that qudits on the vertical line have been condensed. If it is desirable to have lower-weight stabilizers, we can stitch the two sides together. This is formally done by imposing the $XX^\dagger$ or $XX$ two-edge operators at the Hilbert space level. This stitching will transform the two qudits into one. The action of any operator that acted on the qudits can now be mapped into this new degree of freedom. For the stabilizers right at the boundary, we have:
\begin{equation}
  \text{\includegraphics[width=.40\textwidth]{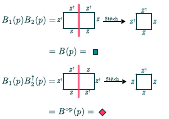}}
  \label{eq:defecteeribbs4}
\end{equation}

\begin{figure}[htbp]
\centering
   \includegraphics[width=0.35\textwidth]{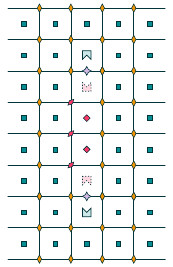}\caption{The final stabilizers for the defect after unfolding and stitching. Two weight-five operators now live at each end of the finite domain wall. The stabilizers defined in Eq.~\eqref{eq:defecteeribbs5}.}
  \label{fig:defectee4}
 \end{figure}

The star operators that share an edge with the trivial wall will remain unchanged, while those sharing an edge with the inverse domain will result in new star operators $A^{\text{op}}(v)$. Right at the $0$D defect, we have the order $2$ operator $X^2X^2$, which cannot stitch the two sides back. We will be left with four weight-five operators denoted by $D^{(1)}$, $D^{(2)}$, $E^{(1)}$, and $E^{(2)}$.
\begin{equation}
  \text{\includegraphics[width=.48\textwidth]{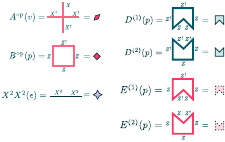}}
  \label{eq:defecteeribbs5}
\end{equation}
Heuristically, to make the inverse wall, we choose a vertical line in the direct lattice $l$ and an adjacent vertical line in the dual lattice $\widetilde{l}$. For the plaquettes intersecting $\widetilde{l}$, we take the adjoint of their Pauli operators acting on edges intersecting $l$. For the star operators intersecting $l$, we apply the adjoint of their Pauli operators acting on edges intersecting $\widetilde{l}$. With these new operators, we reach a stabilizer code for the defect that still requires weight-five operators at the two ends. The new stabilizer placement is shown in Fig.~\ref{fig:defectee4} after the unfolding and stitching.

\section{Belief Propagation}\label{sec:belief_propagation_app}

We will describe here the belief-propagation implementation in more detail. It follows the case for qubits closely \cite{roffeDecodingQuantumLowdensity2020,panteleevDegenerateQuantumLDPC2021} generalized to the four-dimensional qudits with extra modifications due to the fact that $\Z_4$ is not a field. We operate in the usual setting where the problem is stated as a Tanner graph \cite{tanner1981recursive} where qudit and check nodes are denoted by the variables $v$ and $u$, respectively. The check node $u_i$ is connected to the data node $v_j$ with an edge if $H_{ij} \neq 0$. Furthermore, generalizing the qubit case, the edges are now weighted with exactly $H_{ij}$.

Given as input a syndrome vector $s$ of the same length as the number of code checks, the goal is to estimate the error vector $x$, which has the same length as the number of qudit nodes. For each check node $u_i$ that has $V(u_i)$ neighboring data qudits, it naturally satisfies the equation $\sum_{j=1}^{V(u_i)}H_{ij}x_{v_j} = s_{u_i}$. Here, the goal is to find the estimated error $x$ that satisfies all the constraints. This is just again the usual $H.x = s$, but recast in terms of the graph variables. The check-to-data messages used some sort of an XOR operation in the qubit case (that can be captured with a simple product operation since it was over $\Z_2$) \cite{roffeDecodingQuantumLowdensity2020}. This naturally generalizes to convolution for the $\Z_4$ case. The convolution of two vectors $f$ and $g$ given the syndrome $s$ is defined as: \begin{equation}
(f \oplus g)(s)=\sum_{t \in \mathbb{Z}_4} f(t) \cdot g(s-t \bmod 4)
\end{equation}

Here, we use the function notation for the entries of the vectors. For example, for a vector $g$, $g(0)$ is its first element. It will also be useful to define iterated convolution (usually over the data qudit neighbors of a check node ($V(u_i)$): $$\bigotimes_{k \neq j} \phi_k=\phi_1 \oplus \cdots \oplus \phi_{j-1} \oplus \phi_{j+1} \oplus \cdots \oplus \phi_{V\left(u_i\right)}.$$ This naively scales like $O(4^{V(u_i)})$. However, it can be computed using the forward-backward check-node update $O(16 \times V(u_i))$ or fast Fourier transform $O(4\log(4)\times V(u_i))$ \cite{cooley1965algorithm,wymeerschLogdomainDecodingLDPC2004,declercqDecodingAlgorithmsNonbinary2007}. We will describe the convolution using the simpler naive approach, while numerically, we have used the forward-backward check-node update.

Let us now begin the algorithm. First, we take as an input the probability of physical errors $p$, assuming for simplicity that they are uniform and over a pure Pauli channel (e.g., pure X noise). The messages will be 4-dimensional vectors carrying the information for each power of $X$. The probability of failure for any physical nose is then $P(v) = (1-p, \frac{p}{3}, \frac{p}{3}, \frac{p}{3})$ assuming the noise model described in Sec.~\ref{sec:noise_model}. The initial messages from the data qudits to the check nodes are then $m_{v \to u}(a) = P(v,a)$ and from check to data qudits is just $m_{u \to v}(a) = \frac{1}{4}$ since we do not yet have any preference for the estimated error.

In the next iteration, the check node $u_i$ projects the messages it has into contributions to the syndrome it sees from its neighbor data qudit $u_i$. This is captured in successive steps. We first calculate the probability that the qudit contributes $y$ value to the syndrome: $\phi_j(a)=\sum_{\substack{y \in Z_4 \\ H_{i j} \cdot y\equiv a\pmod 4}} m_{v_j \rightarrow u_i}(y)$. The check nodes send their estimated errors to their neighboring data qudits by doing a convolution over all the neighbors except the receiver:
\begin{equation}
m_{u_i \rightarrow v_j}(a) \propto\left(\bigotimes_{\substack{k=1 \\ k \neq j}}^{V\left(u_i\right)} \phi_k\right)\left(s_{u_i}-H_{i j} \cdot a \bmod 4\right)
\end{equation}

The message is normalized such that $\sum_{b \in \mathbb{Z}_4} m_{u_i \rightarrow v_j}(b)=1$. After the data node receives its messages, it updates its belief to

\begin{equation}
q_{v_j}(a) \propto P\left(v_j, a\right) \prod_{u_i \in U\left(v_j\right)} m_{u_i \rightarrow v_j}(a).
\end{equation}

Here, $U(v_j)$ denotes the check nodes neighboring the data node $v_j$. After that, the data node sends a ratio of its cumulative belief to the received belief from each check node:
\begin{equation}
m_{v_j \rightarrow u_i}(a)=\frac{q_{v_j}(a)}{m_{u_i \rightarrow v_j}(a)}
\end{equation}

Optionally, we apply a damping scheme to improve numerical stability. We have found that this parameter is not very crucial near the thresholds. However, a non-zero value close to 1 was empirically found to help convergence.

\begin{equation}
m_{v_j \rightarrow u_i}^{(t)}(a)=\alpha \cdot m_{v_j \rightarrow u_i}^{\text {new }}(a)+(1-\alpha) \cdot m_{v_j \rightarrow u_i}^{(t-1)}(a)
\end{equation}

Finally, we use the qudit's belief to compute a hard decision estimate of the error: $\hat{x}_{v_j}= \arg\!\max (q_{v_j}) $. This new vector takes the index with the highest probability for all data qudits. Finally, we check convergence $H.\hat{x} = s$ every few iterations to reduce computational cost. If the estimated error satisfies the syndrome, the algorithm halts; otherwise, it continues with new iterations. When the algorithm fails to converge after a preset maximum number of iterations $T_{\text{max}}$ is reached, it passes the soft decisions vectors $q$ to the OSD step described in Sec.~\ref{sec:bposd_decoder}. The algorithm is summarized in Alg.~\ref{alg:BP}. 

\begin{algorithm}[H]
\caption{Qudit Belief Propagation Decoder}\label{alg:BP}
\begin{algorithmic}[1]
  \Procedure{Qudit-BP}{$H$, $s$, $p$, $\alpha$, $T_{\text{max}}$}
    \State Initialize priors $P(v,x) \gets (1-p, p/3, p/3, p/3)$ for all data qudits $v$.
    \State Set initial messages: $m_{v \to u}(a) \gets P(v,a)$ and $m_{u \to v}(a) \gets 1/4$.
    \For{$t = 1$ \textbf{to} $T_{\text{max}}$}
        \State \textbf{Check-to-Data:} Project incoming messages:\\ $\phi_k(y) = \sum_{H_{ik}z \equiv y \pmod 4} m_{v_k \to u_i}(z)$.
        \State Convolve to find check messages:\\ $m_{u_i \to v_j}(a) \propto \left( \bigotimes_{k \neq j} \phi_k \right) (s_{u_i} - H_{ij}a \bmod 4)$.
        \State Normalize $m_{u_i \to v_j}(a)$.
        \State \textbf{Data-to-Check:} Compute qudit belief:\\ $q_{v_j}(x) \propto P(v_j, x) \prod_{u_c} m_{u_c \to v_j}(x)$.
        \State Divide out check message: \\ $m_{v_j \to u_i}^{\text{new}}(x) \gets \frac{q_{v_j}(x)}{m_{u_i \to v_j}(x)}$.
        \State Update with damping: \\   $m_{v_j \to u_i}(x) \gets \alpha \cdot m_{v_j \to u_i}^{\text{new}}(x) + (1-\alpha) \cdot m_{v_j \to u_i}^{\text{old}}(x)$.
        \State \textbf{Hard Decision:} Estimate error:\\ $\hat{x}_{v_j} \gets \arg\!\max_x q_{v_j}(x)$.
        \If{$H\hat{x} \equiv s \pmod 4$}
            \State \Return $\hat{x}$ \Comment{BP converged}
        \EndIf
    \EndFor
    \State \Return Soft decisions $q$ \Comment{Pass to OSD-0}
  \EndProcedure
\end{algorithmic}
\end{algorithm}

We note that for the depolarizing channel, one has vectors of length $16$ instead of $4$. We could also use $\log$ probabilities to avoid underflow. However, near the threshold, this is not an issue, and using raw probabilities is slightly faster. We finally note that during the preparation of this work, a recent paper appeared that shows promise in modifying the Tanner graph in order to avoid the OSD step altogether~\cite{hack2026achievingthresholdsstandalonebelief}. It will be interesting to see if an adaptation of this method to the qudit code could provide further improvements.

\bibliography{Pauli,zotero_references}

@article{Drinfeld1988,
  title = {Quantum groups},
  volume = {41},
  ISSN = {1573-8795},
  url = {http://dx.doi.org/10.1007/BF01247086},
  DOI = {10.1007/bf01247086},
  number = {2},
  journal = {Journal of Soviet Mathematics},
  publisher = {Springer Science and Business Media LLC},
  author = {Drinfel’d,  V. G.},
  year = {1988},
  month = apr,
  pages = {898–915}
}

@article{DijkgraafWitten1990,
  author  = {Dijkgraaf, Robbert and Witten, Edward},
  title   = {Topological gauge theories and group cohomology},
  journal = {Communications in Mathematical Physics},
  volume  = {129},
  number  = {2},
  pages   = {393--429},
  year    = {1990},
  doi     = {10.1007/BF02096898},
}

@article{Gould_1993,
title={Quantum double finite group algebras and their representations}, volume={48}, DOI={10.1017/S0004972700015707}, number={2}, journal={Bulletin of the Australian Mathematical Society}, author={Gould, M.D.}, year={1993}, pages={275–301}}

@article{Kitaev2003,
  title = {Fault-tolerant quantum computation by anyons},
  volume = {303},
  ISSN = {0003-4916},
  url = {http://dx.doi.org/10.1016/S0003-4916(02)00018-0},
  DOI = {10.1016/s0003-4916(02)00018-0},
  number = {1},
  journal = {Annals of Physics},
  publisher = {Elsevier BV},
  author = {Kitaev,  A.Yu.},
  year = {2003},
  month = jan,
  pages = {2–30}
}

@article{Bombin2008family,
  title = {Family of non-Abelian Kitaev models on a lattice: Topological condensation and confinement},
  author = {Bombin, H. and Martin-Delgado, M. A.},
  journal = {Phys. Rev. B},
  volume = {78},
  issue = {11},
  pages = {115421},
  numpages = {28},
  year = {2008},
  month = {Sep},
  publisher = {American Physical Society},
  doi = {10.1103/PhysRevB.78.115421},
  url = {https://link.aps.org/doi/10.1103/PhysRevB.78.115421}
}

@phdthesis{deWild1995,
  author        = {Mark de Wild Propitius},
  title         = {{Topological interactions in broken gauge theories}},
  school        = {University of Amsterdam},
  year          = {1995},
  type          = {PhD thesis},
  address       = {Amsterdam, The Netherlands},
  eprint        = {hep-th/9511195},
  archivePrefix = {arXiv},
  primaryClass  = {hep-th},
  note          = {168 pages. Available at \url{https://arxiv.org/abs/hep-th/9511195}}
}

@article{Levin2013,
  title = {Protected Edge Modes without Symmetry},
  author = {Levin, Michael},
  journal = {Phys. Rev. X},
  volume = {3},
  issue = {2},
  pages = {021009},
  numpages = {18},
  year = {2013},
  month = {May},
  publisher = {American Physical Society},
  doi = {10.1103/PhysRevX.3.021009},
  url = {https://link.aps.org/doi/10.1103/PhysRevX.3.021009}
}

@misc{Bravyi1998,
  doi = {10.48550/ARXIV.QUANT-PH/9811052},
  url = {https://arxiv.org/abs/quant-ph/9811052},
  author = {Bravyi,  S. B. and Kitaev,  A. Yu.},
  title = {Quantum codes on a lattice with boundary},
  publisher = {arXiv},
  year = {1998},
  copyright = {Assumed arXiv.org perpetual,  non-exclusive license to distribute this article for submissions made before January 2004}
}

@article{Kong2017,
  title = {Boundary-bulk relation in topological orders},
  volume = {922},
  ISSN = {0550-3213},
  url = {http://dx.doi.org/10.1016/j.nuclphysb.2017.06.023},
  DOI = {10.1016/j.nuclphysb.2017.06.023},
  journal = {Nuclear Physics B},
  publisher = {Elsevier BV},
  author = {Kong,  Liang and Wen,  Xiao-Gang and Zheng,  Hao},
  year = {2017},
  month = sep,
  pages = {62–76}
}

@article{Kong2014,
  title = {Anyon condensation and tensor categories},
  volume = {886},
  ISSN = {0550-3213},
  url = {http://dx.doi.org/10.1016/j.nuclphysb.2014.07.003},
  DOI = {10.1016/j.nuclphysb.2014.07.003},
  journal = {Nuclear Physics B},
  publisher = {Elsevier BV},
  author = {Kong,  Liang},
  year = {2014},
  month = sep,
  pages = {436–482}
}

@article{Kapustin2011,
  title = {Topological boundary conditions in abelian Chern–Simons theory},
  volume = {845},
  ISSN = {0550-3213},
  url = {http://dx.doi.org/10.1016/j.nuclphysb.2010.12.017},
  DOI = {10.1016/j.nuclphysb.2010.12.017},
  number = {3},
  journal = {Nuclear Physics B},
  publisher = {Elsevier BV},
  author = {Kapustin,  Anton and Saulina,  Natalia},
  year = {2011},
  month = apr,
  pages = {393–435}
}

@article{Beigi2011,
  title = {The Quantum Double Model with Boundary: Condensations and Symmetries},
  volume = {306},
  ISSN = {1432-0916},
  url = {http://dx.doi.org/10.1007/s00220-011-1294-x},
  DOI = {10.1007/s00220-011-1294-x},
  number = {3},
  journal = {Communications in Mathematical Physics},
  publisher = {Springer Science and Business Media LLC},
  author = {Beigi,  Salman and Shor,  Peter W. and Whalen,  Daniel},
  year = {2011},
  month = jun,
  pages = {663–694}
}

@article{Kitaev2012,
  title = {Models for Gapped Boundaries and Domain Walls},
  volume = {313},
  ISSN = {1432-0916},
  url = {http://dx.doi.org/10.1007/s00220-012-1500-5},
  DOI = {10.1007/s00220-012-1500-5},
  number = {2},
  journal = {Communications in Mathematical Physics},
  publisher = {Springer Science and Business Media LLC},
  author = {Kitaev,  Alexei and Kong,  Liang},
  year = {2012},
  month = jun,
  pages = {351–373}
}

@misc{Kong2022,
  doi = {10.48550/ARXIV.2205.05565},
  url = {https://arxiv.org/abs/2205.05565},
  author = {Kong,  Liang and Zhang,  Zhi-Hao},
  title = {An invitation to topological orders and category theory},
  publisher = {arXiv},
  year = {2022},
  copyright = {arXiv.org perpetual,  non-exclusive license}
}

@article{Haah2013,
  title = {Commuting Pauli Hamiltonians as Maps between Free Modules},
  volume = {324},
  ISSN = {1432-0916},
  url = {http://dx.doi.org/10.1007/s00220-013-1810-2},
  DOI = {10.1007/s00220-013-1810-2},
  number = {2},
  journal = {Communications in Mathematical Physics},
  publisher = {Springer Science and Business Media LLC},
  author = {Haah,  Jeongwan},
  year = {2013},
  month = oct,
  pages = {351–399}
}

@misc{Cong2016,
  doi = {10.48550/ARXIV.1609.02037},
  url = {https://arxiv.org/abs/1609.02037},
  author = {Cong,  Iris and Cheng,  Meng and Wang,  Zhenghan},
  title = {Topological Quantum Computation with Gapped Boundaries},
  publisher = {arXiv},
  year = {2016},
  copyright = {arXiv.org perpetual,  non-exclusive license}
}

@article{Cong2017,
  title = {Hamiltonian and Algebraic Theories of Gapped Boundaries in Topological Phases of Matter},
  volume = {355},
  ISSN = {1432-0916},
  url = {http://dx.doi.org/10.1007/s00220-017-2960-4},
  DOI = {10.1007/s00220-017-2960-4},
  number = {2},
  journal = {Communications in Mathematical Physics},
  publisher = {Springer Science and Business Media LLC},
  author = {Cong,  Iris and Cheng,  Meng and Wang,  Zhenghan},
  year = {2017},
  month = jul,
  pages = {645–689}
}

@article{Tyler2022,
  title = {Pauli Stabilizer Models of Twisted Quantum Doubles},
  author = {Ellison, Tyler D. and Chen, Yu-An and Dua, Arpit and Shirley, Wilbur and Tantivasadakarn, Nathanan and Williamson, Dominic J.},
  journal = {PRX Quantum},
  volume = {3},
  issue = {1},
  pages = {010353},
  numpages = {37},
  year = {2022},
  month = {Mar},
  publisher = {American Physical Society},
  doi = {10.1103/PRXQuantum.3.010353},
  url = {https://link.aps.org/doi/10.1103/PRXQuantum.3.010353}
}

@misc{operatoralg2024,
  doi = {10.48550/ARXIV.2410.11942},
  url = {https://arxiv.org/abs/2410.11942},
  author = {Liang,  Zijian and Yang,  Bowen and Iosue,  Joseph T. and Chen,  Yu-An},
  title = {Operator algebra and algorithmic construction of boundaries and defects in (2+1)D topological Pauli stabilizer codes},
  publisher = {arXiv},
  year = {2024},
  copyright = {arXiv.org perpetual,  non-exclusive license}
}

@misc{bdry2025,
  doi = {10.48550/ARXIV.2504.19512},
  url = {https://arxiv.org/abs/2504.19512},
  author = {Li,  Mu and Yang,  Xiaohan and Dong,  Xiao-Yu},
  title = {Gapped Boundaries of Kitaev's Quantum Double Models: A Lattice Realization of Anyon Condensation from Lagrangian Algebras},
  publisher = {arXiv},
  year = {2025},
  copyright = {Creative Commons Attribution Share Alike 4.0 International}
}

@misc{sptsewing2024,
  doi = {10.48550/ARXIV.2411.11967},
  url = {https://arxiv.org/abs/2411.11967},
  author = {Li,  Yabo and Song,  Zijian and Kubica,  Aleksander and Kim,  Isaac H.},
  title = {Domain walls from SPT-sewing},
  publisher = {arXiv},
  year = {2024},
  copyright = {arXiv.org perpetual,  non-exclusive license}
}

@article{Tan2015,
  title = {Gapped Domain Walls, Gapped Boundaries, and Topological Degeneracy},
  author = {Lan, Tian and Wang, Juven C. and Wen, Xiao-Gang},
  journal = {Phys. Rev. Lett.},
  volume = {114},
  issue = {7},
  pages = {076402},
  numpages = {5},
  year = {2015},
  month = {Feb},
  publisher = {American Physical Society},
  doi = {10.1103/PhysRevLett.114.076402},
  url = {https://link.aps.org/doi/10.1103/PhysRevLett.114.076402}
}

@article{Yuting2013,
  title = {Twisted quantum double model of topological phases in two dimensions},
  author = {Hu, Yuting and Wan, Yidun and Wu, Yong-Shi},
  journal = {Phys. Rev. B},
  volume = {87},
  issue = {12},
  pages = {125114},
  numpages = {33},
  year = {2013},
  month = {Mar},
  publisher = {American Physical Society},
  doi = {10.1103/PhysRevB.87.125114},
  url = {https://link.aps.org/doi/10.1103/PhysRevB.87.125114}
}

@article{Wilczek82,
  title = {Quantum Mechanics of Fractional-Spin Particles},
  author = {Wilczek, Frank},
  journal = {Phys. Rev. Lett.},
  volume = {49},
  issue = {14},
  pages = {957--959},
  numpages = {0},
  year = {1982},
  month = {Oct},
  publisher = {American Physical Society},
  doi = {10.1103/PhysRevLett.49.957},
  url = {https://link.aps.org/doi/10.1103/PhysRevLett.49.957}
}

@article{Bakalov2000,
  title = {On the Lego-Teichm\"{u}ller game},
  volume = {5},
  ISSN = {1531-586X},
  url = {http://dx.doi.org/10.1007/BF01679714},
  DOI = {10.1007/bf01679714},
  number = {3},
  journal = {Transformation Groups},
  publisher = {Springer Science and Business Media LLC},
  author = {Bakalov,  B. and Kirillov,  A.},
  year = {2000},
  month = sep,
  pages = {207–244}
}

@article{Canright1989,
  title = {Superconductive pairing of fermions and semions in two dimensions},
  author = {Canright, G. S. and Girvin, S. M. and Brass, A.},
  journal = {Phys. Rev. Lett.},
  volume = {63},
  issue = {20},
  pages = {2295--2298},
  numpages = {0},
  year = {1989},
  month = {Nov},
  publisher = {American Physical Society},
  doi = {10.1103/PhysRevLett.63.2295},
  url = {https://link.aps.org/doi/10.1103/PhysRevLett.63.2295}
}

@article{Ani2020,
author = {Krishna, Anirudh and Poulin, David},
doi = {10.1103/PhysRevResearch.2.023116},
issn = {26431564},
journal = {Physical Review Research},
number = {2},
title = {{Topological wormholes: Nonlocal defects on the toric code}},
volume = {2},
year = {2020}
}

@Article{Kobayashi2023,
	title={{Fermionic defects of topological phases and logical gates}},
	author={Ryohei Kobayashi},
	journal={SciPost Phys.},
	volume={15},
	pages={028},
	year={2023},
	publisher={SciPost},
	doi={10.21468/SciPostPhys.15.1.028},
	url={https://scipost.org/10.21468/SciPostPhys.15.1.028},
}

@article{Dennis2002,
  title = {Topological quantum memory},
  volume = {43},
  ISSN = {1089-7658},
  url = {http://dx.doi.org/10.1063/1.1499754},
  DOI = {10.1063/1.1499754},
  number = {9},
  journal = {Journal of Mathematical Physics},
  publisher = {AIP Publishing},
  author = {Dennis,  Eric and Kitaev,  Alexei and Landahl,  Andrew and Preskill,  John},
  year = {2002},
  month = sep,
  pages = {4452–4505}
}

@article{Benhemou2022,
  title = {Non-Abelian statistics with mixed-boundary punctures on the toric code},
  author = {Benhemou, Asmae and Pachos, Jiannis K. and Browne, Dan E.},
  journal = {Phys. Rev. A},
  volume = {105},
  issue = {4},
  pages = {042417},
  numpages = {7},
  year = {2022},
  month = {Apr},
  publisher = {American Physical Society},
  doi = {10.1103/PhysRevA.105.042417},
  url = {https://link.aps.org/doi/10.1103/PhysRevA.105.042417}
}

@article{Shor1995,
  title   = {Scheme for reducing decoherence in quantum computer memory},
  author  = {Shor, Peter W.},
  journal = {Physical Review A},
  volume  = {52},
  number  = {R4},
  pages   = {R2493--R2496},
  year    = {1995},
  doi     = {10.1103/PhysRevA.52.R2493}
}

@phdthesis{Gottesman1997,
  author       = {Gottesman, Daniel},
  title        = {Stabilizer Codes and Quantum Error Correction},
  school       = {California Institute of Technology},
  address      = {Pasadena, CA},
  year         = {1997},
  url          = {https://arxiv.org/abs/quant-ph/9705052},
  archivePrefix= {arXiv},
  eprint       = {quant-ph/9705052}
}

@article{Bombintwist2010,
  title = {Topological Order with a Twist: Ising Anyons from an Abelian Model},
  author = {Bombin, H.},
  journal = {Phys. Rev. Lett.},
  volume = {105},
  issue = {3},
  pages = {030403},
  numpages = {4},
  year = {2010},
  month = {Jul},
  publisher = {American Physical Society},
  doi = {10.1103/PhysRevLett.105.030403},
  url = {https://link.aps.org/doi/10.1103/PhysRevLett.105.030403}
}

@article{tqdbdry,
  title = {Twisted quantum double model of topological order with boundaries},
  author = {Bullivant, Alex and Hu, Yuting and Wan, Yidun},
  journal = {Phys. Rev. B},
  volume = {96},
  issue = {16},
  pages = {165138},
  numpages = {11},
  year = {2017},
  month = {Oct},
  publisher = {American Physical Society},
  doi = {10.1103/PhysRevB.96.165138},
  url = {https://link.aps.org/doi/10.1103/PhysRevB.96.165138}
}

@article{levinwen2005,
  title = {String-net condensation: A physical mechanism for topological phases},
  author = {Levin, Michael A. and Wen, Xiao-Gang},
  journal = {Phys. Rev. B},
  volume = {71},
  issue = {4},
  pages = {045110},
  numpages = {21},
  year = {2005},
  month = {Jan},
  publisher = {American Physical Society},
  doi = {10.1103/PhysRevB.71.045110},
  url = {https://link.aps.org/doi/10.1103/PhysRevB.71.045110}
}

@misc{pandey2022,
      title={Topological Characterization with a Twist, Condensation, and Reflection}, 
      author={Tushar Pandey and Eugene Dumitrescu},
      year={2022},
      eprint={2209.11126},
      archivePrefix={arXiv},
      primaryClass={quant-ph},
      url={https://arxiv.org/abs/2209.11126}, 
}

@misc{hatcherpants1999,
  doi = {10.48550/ARXIV.MATH/9906084},
  url = {https://arxiv.org/abs/math/9906084},
  author = {Hatcher,  Allen},
  title = {Pants Decompositions of Surfaces},
  publisher = {arXiv},
  year = {1999},
  copyright = {Assumed arXiv.org perpetual,  non-exclusive license to distribute this article for submissions made before January 2004}
}

@article{Lindon2023,
  title = {Complete Unitary Qutrit Control in Ultracold Atoms},
  author = {Lindon, Joseph and Tashchilina, Arina and Cooke, Logan W. and LeBlanc, Lindsay J.},
  journal = {Phys. Rev. Appl.},
  volume = {19},
  issue = {3},
  pages = {034089},
  numpages = {11},
  year = {2023},
  month = {Mar},
  publisher = {American Physical Society},
  doi = {10.1103/PhysRevApplied.19.034089},
  url = {https://link.aps.org/doi/10.1103/PhysRevApplied.19.034089}
}

@article{Kiktenko2020,
  title = {Scalable quantum computing with qudits on a graph},
  author = {Kiktenko, E. O. and Nikolaeva, A. S. and Xu, Peng and Shlyapnikov, G. V. and Fedorov, A. K.},
  journal = {Phys. Rev. A},
  volume = {101},
  issue = {2},
  pages = {022304},
  numpages = {7},
  year = {2020},
  month = {Feb},
  publisher = {American Physical Society},
  doi = {10.1103/PhysRevA.101.022304},
  url = {https://link.aps.org/doi/10.1103/PhysRevA.101.022304}
}

@article{Campbell2012,
  doi       = {10.1103/physrevx.2.041021},
  url       = {https://doi.org/10.1103/physrevx.2.041021},
  year      = {2012},
  month     = {dec},
  publisher = {American Physical Society},
  volume    = {2},
  number    = {4},
  pages     = {041021},
  author    = {Earl T. Campbell and Hussain Anwar and Dan E. Browne},
  title     = {Magic-State Distillation in All Prime Dimensions Using Quantum Reed–Muller Codes},
  journal   = {Physical Review X}
}

@article{Campbell2014,
  doi       = {10.1103/physrevlett.113.230501},
  url       = {https://doi.org/10.1103/physrevlett.113.230501},
  year      = {2014},
  month     = {dec},
  publisher = {American Physical Society},
  volume    = {113},
  number    = {23},
  pages     = {230501},
  author    = {Earl T. Campbell},
  title     = {Enhanced Fault-Tolerant Quantum Computing in {\it d}-Level Systems},
  journal   = {Physical Review Letters}
}

@article{Bocharov2017,
  doi       = {10.1103/PhysRevA.96.012306},
  url       = {https://doi.org/10.1103/PhysRevA.96.012306},
  year      = {2017},
  month     = {jul},
  publisher = {American Physical Society},
  volume    = {96},
  number    = {1},
  pages     = {012306},
  author    = {Alex Bocharov and Martin Roetteler and Krysta M. Svore},
  title     = {Factoring with qutrits: Shor’s algorithm on ternary and metaplectic quantum architectures},
  journal   = {Physical Review A}
}

@inproceedings{Gokhale2019,
  doi       = {10.1145/3307650.3322253},
  url       = {https://doi.org/10.1145/3307650.3322253},
  year      = {2019},
  publisher = {ACM},
  pages     = {554--566},
  author    = {Pranav Gokhale and Yiqing Ding and Travis Tomesh and Martin Suchara and Margaret Martonosi and Frederic T. Chong},
  title     = {Asymptotic Improvements to Quantum Circuits via Qutrits},
  booktitle = {Proceedings of the 46th International Symposium on Computer Architecture (ISCA)}
}

@article{Chu2023,
  doi       = {10.1038/s41567-022-01813-7},
  url       = {https://doi.org/10.1038/s41567-022-01813-7},
  year      = {2023},
  month     = {jan},
  publisher = {Springer Nature},
  volume    = {19},
  pages     = {126--131},
  author    = {Ji Chu and Xiaoyu He and Yuxuan Zhou and others},
  title     = {Scalable algorithm simplification using quantum AND logic},
  journal   = {Nature Physics}
}

@article{Meth2025,
  doi       = {10.1038/s41567-025-02797-w},
  url       = {https://doi.org/10.1038/s41567-025-02797-w},
  year      = {2025},
  publisher = {Springer Nature},
  volume    = {21},
  pages     = {570--576},
  author    = {Michael Meth and Jinglei Zhang and Jan F. Haase and Claire Edmunds and Lukas Postler and Andrew J. Jena and Alex Steiner and Luca Dellantonio and Rainer Blatt and Peter Zoller and Thomas Monz and Philipp Schindler and Christine Muschik and Martin Ringbauer},
  title     = {Simulating two-dimensional lattice gauge theories on a qudit quantum computer},
  journal   = {Nature Physics}
}

@article{Sawaya2020,
  doi       = {10.1038/s41534-020-0278-0},
  url       = {https://doi.org/10.1038/s41534-020-0278-0},
  year      = {2020},
  publisher = {Springer Nature},
  volume    = {6},
  pages     = {49},
  author    = {Nicolas P. D. Sawaya and Tim Menke and Thi Ha Kyaw and Sonika Johri and Al{\'a}n Aspuru‑Guzik and Gian G. Guerreschi},
  title     = {Resource‑efficient digital quantum simulation of $d$‑level systems for photonic, vibrational, and spin‑$s$ Hamiltonians},
  journal   = {npj Quantum Information}
}

@article{FernandezDeFuentes2024,
  doi       = {10.1038/s41467-024-45368-y},
  url       = {https://doi.org/10.1038/s41467-024-45368-y},
  year      = {2024},
  publisher = {Springer Nature},
  volume    = {15},
  pages     = {1380},
  author    = {Irene Fern{\'a}ndez De Fuentes and Tim Botzem and Mark A. I. Johnson and others},
  title     = {Navigating the 16‑dimensional Hilbert space of a high‑spin donor qudit with electric and magnetic fields},
  journal   = {Nature Communications}
}

@article{Vilas2024,
  doi       = {10.1038/s41586-024-07199-1},
  url       = {https://doi.org/10.1038/s41586-024-07199-1},
  year      = {2024},
  publisher = {Springer Nature},
  volume    = {628},
  pages     = {282--286},
  author    = {Nathaniel B. Vilas and Paige Robichaud and Christian Hallas and Grace K. Li and Lo{\"i}c Anderegg and John M. Doyle},
  title     = {An optical tweezer array of ultracold polyatomic molecules},
  journal   = {Nature}
}

@article{Chaudhury2007,
  doi       = {10.1103/PhysRevLett.99.163002},
  url       = {https://doi.org/10.1103/PhysRevLett.99.163002},
  year      = {2007},
  publisher = {American Physical Society},
  volume    = {99},
  pages     = {163002},
  author    = {S. Chaudhury and S. T. Merkel and A. Silberfarb and I. H. Deutsch and P. S. Jessen},
  title     = {Quantum control of the hyperfine spin of a Cs atom ensemble},
  journal   = {Physical Review Letters}
}

@article{Kues2017,
  doi       = {10.1038/nature22986},
  url       = {https://doi.org/10.1038/nature22986},
  year      = {2017},
  publisher = {Springer Nature},
  volume    = {546},
  pages     = {622--626},
  author    = {Michael Kues and Christian Reimer and Pawel Roztocki and Luis Romero Cort{\'e}s and Stefano Sciara and Bastian Wetzel and Yingnan Zhang and Alfredo Cino and S.‑T. Chu and B. E. Little and others},
  title     = {On‑chip generation of high‑dimensional entangled quantum states and their coherent control},
  journal   = {Nature}
}

@article{Chi2022,
  doi       = {10.1038/s41467-022-28860-w},
  url       = {https://doi.org/10.1038/s41467-022-28860-w},
  year      = {2022},
  publisher = {Springer Nature},
  volume    = {13},
  pages     = {1166},
  author    = {Yuntian Chi and Jing Huang and Zhenda Zhang and others},
  title     = {A programmable qudit‑based quantum processor},
  journal   = {Nature Communications}
}

@article{Nguyen2024,
  doi       = {10.1038/s41467-024-46373-0},
  url       = {https://doi.org/10.1038/s41467-024-46373-0},
  year      = {2024},
  publisher = {Springer Nature},
  volume    = {15},
  pages     = {7117},
  author    = {L. B. Nguyen and Y. Chi and Z. Zhang and Y. Sun and X. Chen and Z. Zhai and B. Tang and Y. Yang},
  title     = {Empowering a qudit‑based quantum processor by traversing the dual bosonic ladder},
  journal   = {Nature Communications}
}

@article{Roy2025,
  doi       = {10.1103/PhysRevX.15.021009},
  url       = {https://doi.org/10.1103/PhysRevX.15.021009},
  year      = {2025},
  publisher = {American Physical Society},
  volume    = {15},
  pages     = {021009},
  author    = {S. Roy and X. Wu and H. J. Kim and J. Lee and Y. Zhao and others},
  title     = {Synthetic high angular momentum spin dynamics in a microwave oscillator},
  journal   = {Physical Review X}
}

@article{Wang2025,
  doi       = {10.1103/PhysRevApplied.23.034046},
  url       = {https://doi.org/10.1103/PhysRevApplied.23.034046},
  year      = {2025},
  publisher = {American Physical Society},
  volume    = {23},
  pages     = {034046},
  author    = {Z. Wang and R. W. Parker and E. Champion and M. S. Blok},
  title     = {High‑\(E_J/E_C\) transmon qudits with up to 12 levels},
  journal   = {Physical Review Applied}
}

@article{Leupold2018,
  doi       = {10.1103/PhysRevLett.120.180401},
  url       = {https://doi.org/10.1103/PhysRevLett.120.180401},
  year      = {2018},
  publisher = {American Physical Society},
  volume    = {120},
  pages     = {180401},
  author    = {F. M. Leupold and P. Jurcevic and C. Hempel and G. A. Kazakov and M. Giustina and R. Blatt and C. F. Roos},
  title     = {Sustained state‑independent quantum contextual correlations from a single ion},
  journal   = {Physical Review Letters}
}

@article{Ringbauer2022,
  doi       = {10.1038/s41567-022-01652-5},
  url       = {https://doi.org/10.1038/s41567-022-01652-5},
  year      = {2022},
  publisher = {Springer Nature},
  volume    = {18},
  pages     = {1053--1057},
  author    = {Martin Ringbauer and Jan Hapla and Christian T. Schmiegelow and Kai Kieling and Rainer Blatt and Christian F. Roos},
  title     = {A universal qudit quantum processor with trapped ions},
  journal   = {Nature Physics}
}

@article{Adambukulam2024,
  doi       = {10.1103/PhysRevLett.132.060603},
  url       = {https://doi.org/10.1103/PhysRevLett.132.060603},
  year      = {2024},
  publisher = {American Physical Society},
  volume    = {132},
  pages     = {060603},
  author    = {C. Adambukulam and B. Johnson and A. Morello and A. Laucht},
  title     = {Hyperfine spectroscopy and fast, all‑optical arbitrary state initialization and readout of a single, ten‑level \{73\}Ge vacancy nuclear spin qudit in diamond},
  journal   = {Physical Review Letters}
}

@article{Soltamov2019,
  doi       = {10.1038/s41467-019-09685-9},
  url       = {https://doi.org/10.1038/s41467-019-09685-9},
  year      = {2019},
  publisher = {Springer Nature},
  volume    = {10},
  pages     = {1678},
  author    = {V. A. Soltamov and P. V. Klimov and N. A. Zakharenko and E. B. Monakhov and P. G. Baranov},
  title     = {Excitation and coherent control of spin qudit modes in silicon carbide at room temperature},
  journal   = {Nature Communications}
}

@article{Brock2025,
  doi       = {10.1038/s41586-025-08899-y},
  url       = {https://doi.org/10.1038/s41586-025-08899-y},
  year      = {2025},
  month     = {may},
  publisher = {Springer Nature},
  volume    = {641},
  pages     = {612--618},
  author    = {Benjamin L. Brock and Shraddha Singh and Alec Eickbusch and Volodymyr V. Sivak and Andy Z. Ding and Luigi Frunzio and Steven M. Girvin and Michel H. Devoret},
  title     = {Quantum error correction of qudits beyond break-even},
  journal   = {Nature}
}

@misc{landahl2021,
  doi = {10.48550/ARXIV.2110.10280},
  url = {https://arxiv.org/abs/2110.10280},
  author = {Landahl,  Andrew J. and Morrison,  Benjamin C. A.},
  title = {Logical fermions for fault-tolerant quantum simulation},
  publisher = {arXiv},
  year = {2021},
  copyright = {arXiv.org perpetual,  non-exclusive license}
}

@article{Bluvstein2023,
  title = {Logical quantum processor based on reconfigurable atom arrays},
  volume = {626},
  ISSN = {1476-4687},
  url = {http://dx.doi.org/10.1038/s41586-023-06927-3},
  DOI = {10.1038/s41586-023-06927-3},
  number = {7997},
  journal = {Nature},
  publisher = {Springer Science and Business Media LLC},
  author = {Bluvstein,  Dolev and Evered,  Simon J. and Geim,  Alexandra A. and Li,  Sophie H. and Zhou,  Hengyun and Manovitz,  Tom and Ebadi,  Sepehr and Cain,  Madelyn and Kalinowski,  Marcin and Hangleiter,  Dominik and Bonilla Ataides,  J. Pablo and Maskara,  Nishad and Cong,  Iris and Gao,  Xun and Sales Rodriguez,  Pedro and Karolyshyn,  Thomas and Semeghini,  Giulia and Gullans,  Michael J. and Greiner,  Markus and Vuletić,  Vladan and Lukin,  Mikhail D.},
  year = {2023},
  month = dec,
  pages = {58–65}
}

@article{Satzinger2021,
  title = {Realizing topologically ordered states on a quantum processor},
  volume = {374},
  ISSN = {1095-9203},
  url = {http://dx.doi.org/10.1126/science.abi8378},
  DOI = {10.1126/science.abi8378},
  number = {6572},
  journal = {Science},
  publisher = {American Association for the Advancement of Science (AAAS)},
  author = {Satzinger,  K. J. and Liu,  Y.-J and Smith,  A. and Knapp,  C. and Newman,  M. and Jones,  C. and Chen,  Z. and Quintana,  C. and Mi,  X. and Dunsworth,  A. and Gidney,  C. and others},
  year = {2021},
  month = dec,
  pages = {1237–1241}
}

@article{google2024,
  title = {Quantum error correction below the surface code threshold},
  volume = {638},
  ISSN = {1476-4687},
  url = {http://dx.doi.org/10.1038/s41586-024-08449-y},
  DOI = {10.1038/s41586-024-08449-y},
  number = {8052},
  journal = {Nature},
  publisher = {Springer Science and Business Media LLC},
author = {Acharya, Rajeev
          and Abanin, Dmitry A.
          and Aghababaie-Beni, Laleh
          and Aleiner, Igor
          and Andersen, Trond I.
          and Ansmann, Markus
          and Arute, Frank
          and Arya, Kunal
          and Asfaw, Abraham
          and Astrakhantsev, Nikita
          and others},
  year = {2024},
  month = dec,
  pages = {920–926}
}

@article{SalesRodriguez2025,
  title = {Experimental demonstration of logical magic state distillation},
  ISSN = {1476-4687},
  url = {http://dx.doi.org/10.1038/s41586-025-09367-3},
  DOI = {10.1038/s41586-025-09367-3},
  journal = {Nature},
  publisher = {Springer Science and Business Media LLC},
author = {Sales Rodriguez, Pedro and
          Robinson, John M. and
          Jepsen, Paul Niklas and
          He, Zhiyang and
          Duckering, Casey and
          Zhao, Chen and
          Wu, Kai-Hsin and
          Campo, Joseph and
          Bagnall, Kevin and
          Kwon, Minho and
          others},

  year = {2025},
  month = jul 
}

@misc{exp1,
  doi = {10.48550/ARXIV.2404.02280},
  url = {https://arxiv.org/abs/2404.02280},
author = {Paetznick, A. and
          da Silva, M. P. and
          Ryan-Anderson, C. and
          Bello-Rivas, J. M. and
          Campora, J. P. and
          Chernoguzov, A. and
          Dreiling, J. M. and
          Foltz, C. and
          Frachon, F. and
          Gaebler, J. P. and
          others},
  title = {Demonstration of logical qubits and repeated error correction with better-than-physical error rates},
  publisher = {arXiv},
  year = {2024},
  copyright = {arXiv.org perpetual,  non-exclusive license}
}

@article{exp2,
  title = {Encoding a magic state with beyond break-even fidelity},
  volume = {625},
  ISSN = {1476-4687},
  url = {http://dx.doi.org/10.1038/s41586-023-06846-3},
  DOI = {10.1038/s41586-023-06846-3},
  number = {7994},
  journal = {Nature},
  publisher = {Springer Science and Business Media LLC},
  author = {Gupta,  Riddhi S. and Sundaresan,  Neereja and Alexander,  Thomas and Wood,  Christopher J. and Merkel,  Seth T. and Healy,  Michael B. and Hillenbrand,  Marius and Jochym-O’Connor,  Tomas and Wootton,  James R. and Yoder,  Theodore J. and Cross,  Andrew W. and Takita,  Maika and Brown,  Benjamin J.},
  year = {2024},
  month = jan,
  pages = {259–263}
}

@article{exp3,
  title = {Suppressing quantum errors by scaling a surface code logical qubit},
  volume = {614},
  ISSN = {1476-4687},
  url = {http://dx.doi.org/10.1038/s41586-022-05434-1},
  DOI = {10.1038/s41586-022-05434-1},
  number = {7949},
  journal = {Nature},
  publisher = {Springer Science and Business Media LLC},
  author = {Acharya,  Rajeev and Aleiner,  Igor and Allen,  Richard and Andersen,  Trond I. and Ansmann,  Markus and Arute,  Frank and Arya,  Kunal and Asfaw,  Abraham and Atalaya,  Juan and Babbush,  Ryan and  others},
  year = {2023},
  month = feb,
  pages = {676–681}
}

@article{exp4,
  title = {A quantum processor based on coherent transport of entangled atom arrays},
  volume = {604},
  ISSN = {1476-4687},
  url = {http://dx.doi.org/10.1038/s41586-022-04592-6},
  DOI = {10.1038/s41586-022-04592-6},
  number = {7906},
  journal = {Nature},
  publisher = {Springer Science and Business Media LLC},
  author = {Bluvstein,  Dolev and Levine,  Harry and Semeghini,  Giulia and Wang,  Tout T. and Ebadi,  Sepehr and Kalinowski,  Marcin and Keesling,  Alexander and Maskara,  Nishad and Pichler,  Hannes and Greiner,  Markus and Vuletić,  Vladan and Lukin,  Mikhail D.},
  year = {2022},
  month = apr,
  pages = {451–456}
}

@article{exp5,
  title = {Real-time quantum error correction beyond break-even},
  volume = {616},
  ISSN = {1476-4687},
  url = {http://dx.doi.org/10.1038/s41586-023-05782-6},
  DOI = {10.1038/s41586-023-05782-6},
  number = {7955},
  journal = {Nature},
  publisher = {Springer Science and Business Media LLC},
  author = {Sivak,  V. V. and Eickbusch,  A. and Royer,  B. and Singh,  S. and Tsioutsios,  I. and Ganjam,  S. and Miano,  A. and Brock,  B. L. and Ding,  A. Z. and Frunzio,  L. and Girvin,  S. M. and Schoelkopf,  R. J. and Devoret,  M. H.},
  year = {2023},
  month = mar,
  pages = {50–55}
}

@article{exp6,
  title = {Realizing repeated quantum error correction in a distance-three surface code},
  volume = {605},
  ISSN = {1476-4687},
  url = {http://dx.doi.org/10.1038/s41586-022-04566-8},
  DOI = {10.1038/s41586-022-04566-8},
  number = {7911},
  journal = {Nature},
  publisher = {Springer Science and Business Media LLC},
  author = {Krinner,  Sebastian and Lacroix,  Nathan and Remm,  Ants and Di Paolo,  Agustin and Genois,  Elie and Leroux,  Catherine and Hellings,  Christoph and Lazar,  Stefania and Swiadek,  Francois and Herrmann,  Johannes and Norris,  Graham J. and Andersen,  Christian Kraglund and M\"{u}ller,  Markus and Blais,  Alexandre and Eichler,  Christopher and Wallraff,  Andreas},
  year = {2022},
  month = may,
  pages = {669–674}
}

@article{exp7,
  title = {Realization of Real-Time Fault-Tolerant Quantum Error Correction},
  author = {Ryan-Anderson, C. and Bohnet, J. G. and Lee, K. and Gresh, D. and Hankin, A. and Gaebler, J. P. and Francois, D. and Chernoguzov, A. and Lucchetti, D. and Brown, N. C. and Gatterman, T. M. and Halit, S. K. and Gilmore, K. and Gerber, J. A. and Neyenhuis, B. and Hayes, D. and Stutz, R. P.},
  journal = {Phys. Rev. X},
  volume = {11},
  issue = {4},
  pages = {041058},
  numpages = {29},
  year = {2021},
  month = {Dec},
  publisher = {American Physical Society},
  doi = {10.1103/PhysRevX.11.041058},
  url = {https://link.aps.org/doi/10.1103/PhysRevX.11.041058}
}

@article{colorcond2024,
  title = {Anyon Condensation and the Color Code},
  author = {Kesselring, Markus S. and Magdalena de la Fuente, Julio C. and Thomsen, Felix and Eisert, Jens and Bartlett, Stephen D. and Brown, Benjamin J.},
  journal = {PRX Quantum},
  volume = {5},
  issue = {1},
  pages = {010342},
  numpages = {56},
  year = {2024},
  month = {Mar},
  publisher = {American Physical Society},
  doi = {10.1103/PRXQuantum.5.010342},
  url = {https://link.aps.org/doi/10.1103/PRXQuantum.5.010342}
}

@article{iqbal2018,
  title = {Study of anyon condensation and topological phase transitions from a ${\mathbb{Z}}_{4}$ topological phase using the projected entangled pair states approach},
  author = {Iqbal, Mohsin and Duivenvoorden, Kasper and Schuch, Norbert},
  journal = {Phys. Rev. B},
  volume = {97},
  issue = {19},
  pages = {195124},
  numpages = {25},
  year = {2018},
  month = {May},
  publisher = {American Physical Society},
  doi = {10.1103/PhysRevB.97.195124},
  url = {https://link.aps.org/doi/10.1103/PhysRevB.97.195124}
}

@article{brown2017,
  title = {Poking Holes and Cutting Corners to Achieve Clifford Gates with the Surface Code},
  author = {Brown, Benjamin J. and Laubscher, Katharina and Kesselring, Markus S. and Wootton, James R.},
  journal = {Phys. Rev. X},
  volume = {7},
  issue = {2},
  pages = {021029},
  numpages = {20},
  year = {2017},
  month = {May},
  publisher = {American Physical Society},
  doi = {10.1103/PhysRevX.7.021029},
  url = {https://link.aps.org/doi/10.1103/PhysRevX.7.021029}
}

@article{Nayak2008,
  title = {Non-Abelian anyons and topological quantum computation},
  author = {Nayak, Chetan and Simon, Steven H. and Stern, Ady and Freedman, Michael and Das Sarma, Sankar},
  journal = {Rev. Mod. Phys.},
  volume = {80},
  issue = {3},
  pages = {1083--1159},
  numpages = {0},
  year = {2008},
  month = {Sep},
  publisher = {American Physical Society},
  doi = {10.1103/RevModPhys.80.1083},
  url = {https://link.aps.org/doi/10.1103/RevModPhys.80.1083}
}

@misc{Z4DS2025,
  doi = {10.48550/ARXIV.2508.08376},
  url = {https://arxiv.org/abs/2508.08376},
  author = {Zhang,  Qi and Xu,  Wen-Tao},
  title = {Continuous topological phase transition between $\mathbb{Z}_2$ topologically ordered phases},
  publisher = {arXiv},
  year = {2025},
  copyright = {Creative Commons Attribution 4.0 International}
}

@article{Burnell2018,
  title = {Anyon Condensation and Its Applications},
  volume = {9},
  ISSN = {1947-5462},
  url = {http://dx.doi.org/10.1146/annurev-conmatphys-033117-054154},
  DOI = {10.1146/annurev-conmatphys-033117-054154},
  number = {1},
  journal = {Annual Review of Condensed Matter Physics},
  publisher = {Annual Reviews},
  author = {Burnell,  F.J.},
  year = {2018},
  month = mar,
  pages = {307–327}
}

@article{Gheorghiu2014,
  title = {Standard form of qudit stabilizer groups},
  volume = {378},
  ISSN = {0375-9601},
  url = {http://dx.doi.org/10.1016/j.physleta.2013.12.009},
  DOI = {10.1016/j.physleta.2013.12.009},
  number = {5–6},
  journal = {Physics Letters A},
  publisher = {Elsevier BV},
  author = {Gheorghiu,  Vlad},
  year = {2014},
  month = jan,
  pages = {505–509}
}

@article{Bombin2012Universal,
  author  = {H{\'e}ctor Bomb{\'\i}n and Guillaume Duclos-Cianci and David Poulin},
  title   = {Universal topological phase of two-dimensional stabilizer codes},
  journal = {New Journal of Physics},
  volume  = {14},
  pages   = {073048},
  year    = {2012},
  doi     = {10.1088/1367-2630/14/7/073048},
  eprint  = {1103.4606},
  archivePrefix = {arXiv}
}

@article{Bombin2014Structure,
  author  = {H{\'e}ctor Bomb{\'\i}n},
  title   = {Structure of 2D Topological Stabilizer Codes},
  journal = {Communications in Mathematical Physics},
  volume  = {327},
  pages   = {387--432},
  year    = {2014},
  doi     = {10.1007/s00220-014-1893-4},
  eprint  = {1107.2707},
  archivePrefix = {arXiv}
}

@article{bombin2013introduction,
  title={An Introduction to Topological Quantum Codes},
  author={Bombin, Hector},
  journal={arXiv preprint arXiv:1311.0277},
  year={2013},
  url={https://arxiv.org/abs/1311.0277}
}

@article{Haah2021Classification,
  author  = {Jeongwan Haah},
  title   = {Classification of translation invariant topological Pauli stabilizer codes for prime dimensional qudits on two-dimensional lattices},
  journal = {Journal of Mathematical Physics},
  volume  = {62},
  number  = {1},
  pages   = {012201},
  year    = {2021},
  doi     = {10.1063/5.0021068},
  eprint  = {1812.11193},
  archivePrefix = {arXiv}
}

@article{kesselring2018boundaries,
  author        = {Kesselring, M. S. and Pastawski, F. and Eisert, J. and Brown, B. J.},
  title         = {The boundaries and twist defects of the color code and their applications to topological quantum computation},
  journal       = {Quantum},
  volume        = {2},
  pages         = {101},
  year          = {2018},
  doi           = {10.22331/q-2018-11-05-101}
}

@article{scruby2020hierarchy,
  author        = {Scruby, T. R. and Browne, D. E.},
  title         = {A Hierarchy of Anyon Models Realised by Twists in Stacked Surface Codes},
  journal       = {Quantum},
  volume        = {4},
  pages         = {251},
  year          = {2020},
  doi           = {10.22331/q-2020-04-14-251}
}

@article{bombin2011clifford,
  author        = {H. Bombin},
  title         = {Clifford gates by code deformation},
  journal       = {New Journal of Physics},
  volume        = {13},
  number        = {4},
  pages         = {043005},
  year          = {2011},
  doi           = {10.1088/1367-2630/13/4/043005}
}

@article{yoder2017surface,
  author        = {Yoder, T. J. and Kim, I. H.},
  title         = {The surface code with a twist},
  journal       = {Quantum},
  volume        = {1},
  pages         = {2},
  year          = {2017},
  doi           = {10.22331/q-2017-04-25-2}
}

@article{Petiziol2024,
  title = {Non-Abelian Anyons in Periodically Driven Abelian Spin Liquids},
  author = {Petiziol, Francesco},
  journal = {Phys. Rev. Lett.},
  volume = {133},
  issue = {3},
  pages = {036601},
  numpages = {7},
  year = {2024},
  month = {Jul},
  publisher = {American Physical Society},
  doi = {10.1103/PhysRevLett.133.036601},
  url = {https://link.aps.org/doi/10.1103/PhysRevLett.133.036601}
}

@article{Bravyi2010,
  title = {Topological quantum order: Stability under local perturbations},
  volume = {51},
  ISSN = {1089-7658},
  url = {http://dx.doi.org/10.1063/1.3490195},
  DOI = {10.1063/1.3490195},
  number = {9},
  journal = {Journal of Mathematical Physics},
  publisher = {AIP Publishing},
  author = {Bravyi,  Sergey and Hastings,  Matthew B. and Michalakis,  Spyridon},
  year = {2010},
  month = sep 
}

@article{Tuckett2019,
  title = {Tailoring Surface Codes for Highly Biased Noise},
  author = {Tuckett, David K. and Darmawan, Andrew S. and Chubb, Christopher T. and Bravyi, Sergey and Bartlett, Stephen D. and Flammia, Steven T.},
  journal = {Phys. Rev. X},
  volume = {9},
  issue = {4},
  pages = {041031},
  numpages = {22},
  year = {2019},
  month = {Nov},
  publisher = {American Physical Society},
  doi = {10.1103/PhysRevX.9.041031},
  url = {https://link.aps.org/doi/10.1103/PhysRevX.9.041031}
}

@article{Wang2020,
  title = {In and around abelian anyon models
                    *},
  volume = {53},
  ISSN = {1751-8121},
  url = {http://dx.doi.org/10.1088/1751-8121/abc6c0},
  DOI = {10.1088/1751-8121/abc6c0},
  number = {50},
  journal = {Journal of Physics A: Mathematical and Theoretical},
  publisher = {IOP Publishing},
  author = {Wang,  Liang and Wang,  Zhenghan},
  year = {2020},
  month = nov,
  pages = {505203}
}

@misc{Etingof2009,
  doi = {10.48550/ARXIV.0909.3140},
  url = {https://arxiv.org/abs/0909.3140},
  author = {Etingof,  Pavel and Nikshych,  Dmitri and Ostrik,  Victor and Meir,  with an appendix by Ehud},
  title = {Fusion categories and homotopy theory},
  publisher = {arXiv},
  year = {2009},
  copyright = {arXiv.org perpetual,  non-exclusive license}
}

@article{Barkeshli2024,
	title={{Higher-group symmetry in finite gauge theory and stabilizer codes}},
	author={Maissam Barkeshli and Yu-An Chen and Po-Shen Hsin and Ryohei Kobayashi},
	journal={SciPost Phys.},
	volume={16},
	pages={089},
	year={2024},
	publisher={SciPost},
	doi={10.21468/SciPostPhys.16.4.089},
	url={https://scipost.org/10.21468/SciPostPhys.16.4.089},
}

@article{Barkeshli2013,
  title = {Theory of defects in Abelian topological states},
  author = {Barkeshli, Maissam and Jian, Chao-Ming and Qi, Xiao-Liang},
  journal = {Phys. Rev. B},
  volume = {88},
  issue = {23},
  pages = {235103},
  numpages = {21},
  year = {2013},
  month = {Dec},
  publisher = {American Physical Society},
  doi = {10.1103/PhysRevB.88.235103},
  url = {https://link.aps.org/doi/10.1103/PhysRevB.88.235103}
}

@article{Barkeshli2023,
	title={{Codimension-2 defects and higher symmetries in (3+1)D topological phases}},
	author={Maissam Barkeshli and Yu-An Chen and Sheng-Jie Huang and Ryohei Kobayashi and Nathanan Tantivasadakarn and Guanyu Zhu},
	journal={SciPost Phys.},
	volume={14},
	pages={065},
	year={2023},
	publisher={SciPost},
	doi={10.21468/SciPostPhys.14.4.065},
	url={https://scipost.org/10.21468/SciPostPhys.14.4.065},
}

@article{Kobayashi2024,
  title = {Cross-Cap Defects and Fault-Tolerant Logical Gates in the Surface Code and the Honeycomb Floquet Code},
  author = {Kobayashi, Ryohei and Zhu, Guanyu},
  journal = {PRX Quantum},
  volume = {5},
  issue = {2},
  pages = {020360},
  numpages = {26},
  year = {2024},
  month = {Jun},
  publisher = {American Physical Society},
  doi = {10.1103/PRXQuantum.5.020360},
  url = {https://link.aps.org/doi/10.1103/PRXQuantum.5.020360}
}

@misc{quditedge,
  note = {the terms edge and qudit are essentially synonymous in this work since each edge always supports a qudit}
}

@article{Laubscher2019,
  title = {Universal quantum computation in the surface code using non-Abelian islands},
  author = {Laubscher, Katharina and Loss, Daniel and Wootton, James R.},
  journal = {Phys. Rev. A},
  volume = {100},
  issue = {1},
  pages = {012338},
  numpages = {13},
  year = {2019},
  month = {Jul},
  publisher = {American Physical Society},
  doi = {10.1103/PhysRevA.100.012338},
  url = {https://link.aps.org/doi/10.1103/PhysRevA.100.012338}
}

@article{Hu2017,
  title = {Boundary Hamiltonian Theory for Gapped Topological Orders},
  volume = {34},
  ISSN = {1741-3540},
  url = {http://dx.doi.org/10.1088/0256-307X/34/7/077103},
  DOI = {10.1088/0256-307x/34/7/077103},
  number = {7},
  journal = {Chinese Physics Letters},
  publisher = {IOP Publishing},
  author = {Hu,  Yuting and Wan,  Yidun and Wu,  Yong-Shi},
  year = {2017},
  month = jun,
  pages = {077103}
}

@article{Hu2018,
  title = {Boundary Hamiltonian theory for gapped topological phases on an open surface},
  volume = {2018},
  ISSN = {1029-8479},
  url = {http://dx.doi.org/10.1007/JHEP01(2018)134},
  DOI = {10.1007/jhep01(2018)134},
  number = {1},
  journal = {Journal of High Energy Physics},
  publisher = {Springer Science and Business Media LLC},
  author = {Hu,  Yuting and Luo,  Zhu-Xi and Pankovich,  Ren and Wan,  Yidun and Wu,  Yong-Shi},
  year = {2018},
  month = jan 
}

@article{Calderbank1996Goodcodes,
  title = {Good quantum error-correcting codes exist},
  author = {Calderbank, A. R. and Shor, Peter W.},
  journal = {Phys. Rev. A},
  volume = {54},
  issue = {2},
  pages = {1098--1105},
  numpages = {0},
  year = {1996},
  month = {Aug},
  publisher = {American Physical Society},
  doi = {10.1103/PhysRevA.54.1098},
  url = {https://link.aps.org/doi/10.1103/PhysRevA.54.1098}
}

@book{nielsen_chuang_2010,
  author    = {Nielsen, Michael A. and Chuang, Isaac L.},
  title     = {Quantum Computation and Quantum Information: 10th Anniversary Edition},
  publisher = {Cambridge University Press},
  year      = {2010},
  address   = {Cambridge},
  isbn      = {978-1-107-00217-3},
  doi       = {10.1017/CBO9780511976667}
}

@article{edmondsMaximumMatchingPolyhedron1965,
    title = {Maximum matching and a polyhedron with 0,1-vertices},
    volume = {69B},
    issn = {0022-4340},
    url = {https://nvlpubs.nist.gov/nistpubs/jres/69B/jresv69Bn1-2p125_A1b.pdf},
    doi = {10.6028/jres.069B.013},
    language = {en},
    number = {1 and 2},
    urldate = {2026-02-26},
    journal = {Journal of Research of the National Bureau of Standards Section B Mathematics and Mathematical Physics},
    author = {Edmonds, Jack},
    month = jan,
    year = {1965},
    pages = {125},
}

@article{gidney2021stim,
  doi = {10.22331/q-2021-07-06-497},
  url = {https://doi.org/10.22331/q-2021-07-06-497},
  title = {Stim: a fast stabilizer circuit simulator},
  author = {Gidney, Craig},
  journal = {{Quantum}},
  issn = {2521-327X},
  publisher = {{Verein zur F{\"{o}}rderung des Open Access Publizierens
                in den Quantenwissenschaften}},
  volume = 5,
  pages = 497,
  month = jul,
  year = 2021
}

@article{higgott2022pymatching,
  title={PyMatching: A Python package for decoding quantum codes with minimum-weight perfect matching},
  author={Higgott, Oscar},
  journal={ACM Transactions on Quantum Computing},
  volume={3},
  number={3},
  pages={1--16},
  year={2022},
  publisher={ACM New York, NY}
}

@misc{wu2023fusion,
  title={Fusion Blossom: Fast MWPM Decoders for QEC}, 
  author={Yue Wu and Lin Zhong},
  year={2023},
  eprint={2305.08307},
  archivePrefix={arXiv},
  primaryClass={quant-ph}
}

@phdthesis{tuckett2020tailoring,
  title={Tailoring surface codes: Improvements in quantum error correction with biased noise},
  author={Tuckett, David Kingsley},
  year={2020},
  school={University of Sydney},
  note={(qecsim: \url{https://github.com/qecsim/qecsim})}
}

@misc{UnionFindCPP,
  author       = {Park, Chae-Yeun and Meinerz, Kai},
  title        = {Open-source C++ implementation of the Union-Find decoder},
  year         = {2020},
  publisher    = {GitHub},
  journal      = {GitHub repository},
  doi          = {10.5281/zenodo.10044828},
  howpublished = {\url{https://github.com/chaeyeunpark/UnionFind}}
}

@software{roffe_2022_ldpc,
  author       = {Roffe, Joschka},
  title        = {{LDPC: Python tools for low density parity check codes}},
  year         = {2022},
  publisher    = {PyPI},
  url          = {https://pypi.org/project/ldpc/}
}

@misc{qiskit_qec_2022,
  author       = {{Qiskit Community}},
  title        = {{Qiskit QEC: Framework for Quantum Error Correction}},
  year         = {2022},
  publisher    = {GitHub},
  journal      = {GitHub repository},
  howpublished = {\url{https://github.com/qiskit-community/qiskit-qec}}
}

@misc{roffe_2021_bposd,
  author       = {Roffe, Joschka},
  title        = {{BPOSD: A Python package for Belief Propagation with Ordered Statistic Decoding}},
  year         = {2021},
  publisher    = {GitHub},
  journal      = {GitHub repository},
  howpublished = {\url{https://github.com/quantumcircuits/bposd}}
}

@manual{cplex,
  author = {{IBM ILOG}},
  title = {{IBM ILOG CPLEX Optimization Studio User's Manual}},
  organization = {International Business Machines Corporation},
  year = {2023},
  url = {https://www.ibm.com/products/ilog-cplex-optimization-studio}
}

@misc{gurobi,
  author = {{Gurobi Optimization, LLC}},
  title = {{Gurobi Optimizer Reference Manual}},
  year = {2024},
  url = {https://www.gurobi.com}
}

@article{norton2000structure,
  title={On the structure of linear and cyclic codes over a finite chain ring},
  author={Norton, Graham H and S{\u{a}}l{\u{a}}gean, Ana},
  journal={Applicable Algebra in Engineering, Communication and Computing},
  volume={10},
  number={6},
  pages={489--506},
  year={2000},
  publisher={Springer},
  doi={10.1007/s002000050138}
}

@book{dummit2004abstract,
  title={Abstract Algebra},
  author={Dummit, David S and Foote, Richard M},
  edition={3rd},
  year={2004},
  publisher={John Wiley and Sons},
  address={Hoboken, NJ}
}

@article{mulders2003mathematical,
  title={On mathematical properties of the {H}owell normal form},
  author={Mulders, Thom and Storjohann, Arne},
  journal={Journal of Symbolic Computation},
  volume={35},
  number={2},
  pages={137--190},
  year={2003},
  publisher={Elsevier},
  doi={10.1016/S0747-7171(02)00118-7}
}

@article{sarkar2024qudit,
  title = {The qudit {P}auli group: non-commuting pairs, non-commuting sets, and structure theorems},
  author = {Sarkar, Rahul and Yoder, Theodore J.},
  journal = {Quantum},
  volume = {8},
  pages = {1307},
  year = {2024},
  month = {Apr},
  publisher = {Verein zur F{\"{o}}rderung des Open Access Publizierens in den Quantenwissenschaften},
  doi = {10.22331/q-2024-04-04-1307}
}

@article{cooley1965algorithm,
  title={An algorithm for the machine calculation of complex Fourier series},
  author={Cooley, James W and Tukey, John W},
  journal={Mathematics of Computation},
  volume={19},
  number={90},
  pages={297--301},
  year={1965},
  publisher={American Mathematical Society}
}

@article{tanner1981recursive,
  title={A recursive approach to low complexity codes},
  author={Tanner, R. Michael},
  journal={IEEE Transactions on Information Theory},
  volume={27},
  number={5},
  pages={533--547},
  year={1981},
  publisher={IEEE},
  doi={10.1109/TIT.1981.1056404}
}

@book{pearl1988probabilistic,
  title={Probabilistic Reasoning in Intelligent Systems: Networks of Plausible Inference},
  author={Pearl, Judea},
  year={1988},
  publisher={Morgan Kaufmann},
  address={San Mateo, CA}
}

@article{richardson2001capacity,
  title={The capacity of low-density parity-check codes under message-passing decoding},
  author={Richardson, Thomas J and Urbanke, Ruediger L},
  journal={IEEE Transactions on Information Theory},
  volume={47},
  number={2},
  pages={599--618},
  year={2001},
  publisher={IEEE},
  doi={10.1109/18.910577}
}

@incollection{yedidia2003understanding,
  title={Understanding belief propagation and its generalizations},
  author={Yedidia, Jonathan S and Freeman, William T and Weiss, Yair},
  booktitle={Exploring artificial intelligence in the new millennium},
  pages={239--269},
  year={2003},
  publisher={Morgan Kaufmann}
}

@article{weiss2000correctness,
  title={Correctness of local probability propagation in graphical models with loops},
  author={Weiss, Yair},
  journal={Neural computation},
  volume={12},
  number={1},
  pages={1--41},
  year={2000},
  publisher={MIT Press}
}

@misc{hack2026achievingthresholdsstandalonebelief,
      title={Achieving Thresholds via Standalone Belief Propagation on Surface Codes}, 
      author={Pedro Hack and Luca Menti and Francisco Lazaro and Alexandru Paler},
      year={2026},
      eprint={2603.05381},
      archivePrefix={arXiv},
      primaryClass={quant-ph},
      url={https://arxiv.org/abs/2603.05381}
}

@misc{huang2025hybrid,
      title={Hybrid Lattice Surgery: Non-Clifford Gates via Non-Abelian Surface Codes}, 
      author={Sheng-Jie Huang and Alison Warman and Sakura Schafer-Nameki and Yanzhu Chen},
      year={2025},
      eprint={2510.20890},
      archivePrefix={arXiv},
      primaryClass={quant-ph},
      url={https://arxiv.org/abs/2510.20890}
}

@misc{manjunath2026universal,
      title={Universal quantum computation with group surface codes}, 
      author={Naren Manjunath and Vieri Mattei and Apoorv Tiwari and Tyler D. Ellison},
      year={2026},
      eprint={2603.05502},
      archivePrefix={arXiv},
      primaryClass={quant-ph},
      url={https://arxiv.org/abs/2603.05502}
}

@misc{chubbGeneralTensorNetwork2021,
	title = {General tensor network decoding of {2D} {Pauli} codes},
	url = {http://arxiv.org/abs/2101.04125},
	doi = {10.48550/arXiv.2101.04125},
	urldate = {2026-03-25},
	publisher = {arXiv},
	author = {Chubb, Christopher T.},
	month = oct,
	year = {2021},
	note = {arXiv:2101.04125 [quant-ph]},
}

@article{ferrisTensorNetworksQuantum2014,
	title = {Tensor {Networks} and {Quantum} {Error} {Correction}},
	volume = {113},
	url = {https://link.aps.org/doi/10.1103/PhysRevLett.113.030501},
	doi = {10.1103/PhysRevLett.113.030501},
	number = {3},
	urldate = {2026-03-25},
	journal = {Physical Review Letters},
	publisher = {American Physical Society},
	author = {Ferris, Andrew J. and Poulin, David},
	month = jul,
	year = {2014},
	pages = {030501},
}

@inproceedings{wymeerschLogdomainDecodingLDPC2004,
	title = {Log-domain decoding of {LDPC} codes over {GF}(q)},
	volume = {2},
	url = {https://ieeexplore.ieee.org/document/1312606/},
	doi = {10.1109/ICC.2004.1312606},
	urldate = {2026-03-25},
	booktitle = {2004 {IEEE} {International} {Conference} on {Communications}},
	author = {Wymeersch, H. and Steendam, H. and Moeneclaey, M.},
	month = jun,
	year = {2004},
	pages = {772--776 Vol.2},
}

@article{declercqDecodingAlgorithmsNonbinary2007,
	title = {Decoding {Algorithms} for {Nonbinary} {LDPC} {Codes} {Over} {GF}(q)},
	volume = {55},
	issn = {1558-0857},
	url = {https://ieeexplore.ieee.org/document/4155118/},
	doi = {10.1109/TCOMM.2007.894088},
	number = {4},
	urldate = {2026-03-25},
	journal = {IEEE Transactions on Communications},
	author = {Declercq, David and Fossorier, Marc},
	month = apr,
	year = {2007},
	pages = {633--643},
}

@misc{MultipathSummationDecoding,
	title = {Multi-path {Summation} for {Decoding} {2D} {Topological} {Codes} – {Quantum}},
	url = {https://quantum-journal.org/papers/q-2018-10-19-102/},
	urldate = {2026-03-12},
}

@misc{beniTesseractSearchBasedDecoder2025,
	title = {Tesseract: {A} {Search}-{Based} {Decoder} for {Quantum} {Error} {Correction}},
	shorttitle = {Tesseract},
	url = {http://arxiv.org/abs/2503.10988},
	doi = {10.48550/arXiv.2503.10988},
	urldate = {2026-03-11},
	publisher = {arXiv},
	author = {Beni, Laleh Aghababaie and Higgott, Oscar and Shutty, Noah},
	month = aug,
	year = {2025},
	note = {arXiv:2503.10988 [quant-ph]},
}

@article{takadaIsingModelFormulation2024,
	title = {Ising model formulation for highly accurate topological color codes decoding},
	volume = {6},
	issn = {2643-1564},
	url = {https://link.aps.org/doi/10.1103/PhysRevResearch.6.013092},
	doi = {10.1103/PhysRevResearch.6.013092},
	language = {en},
	number = {1},
	urldate = {2026-03-11},
	journal = {Physical Review Research},
	author = {Takada, Yugo and Takeuchi, Yusaku and Fujii, Keisuke},
	month = jan,
	year = {2024},
	pages = {013092},
}

@misc{cainCorrelatedDecodingLogical2025,
	title = {Correlated decoding of logical algorithms with transversal gates},
	url = {http://arxiv.org/abs/2403.03272},
	doi = {10.48550/arXiv.2403.03272},
	urldate = {2026-03-11},
	publisher = {arXiv},
	author = {Cain, Madelyn and Zhao, Chen and Zhou, Hengyun and Meister, Nadine and Ataides, J. Pablo Bonilla and Jaffe, Arthur and Bluvstein, Dolev and Lukin, Mikhail D.},
	month = apr,
	year = {2025},
	note = {arXiv:2403.03272 [quant-ph]},
}

@misc{landahlFaulttolerantQuantumComputing2011,
	title = {Fault-tolerant quantum computing with color codes},
	url = {http://arxiv.org/abs/1108.5738},
	doi = {10.48550/arXiv.1108.5738},
	urldate = {2026-02-25},
	publisher = {arXiv},
	author = {Landahl, Andrew J. and Anderson, Jonas T. and Rice, Patrick R.},
	month = aug,
	year = {2011},
	note = {arXiv:1108.5738 [quant-ph]},
}

@misc{sakashitaFastAccurateDecoder2026,
	title = {Fast and {Accurate} {Decoder} for the {XZZX} {Code} {Using} {Simulated} {Annealing}},
	url = {http://arxiv.org/abs/2509.17837},
	doi = {10.48550/arXiv.2509.17837},
	urldate = {2026-03-11},
	publisher = {arXiv},
	author = {Sakashita, Tatsuya},
	month = mar,
	year = {2026},
	note = {arXiv:2509.17837 [quant-ph]},
}

@misc{javedLowComplexityLinearProgramming2024,
	title = {Low-{Complexity} {Linear} {Programming} {Based} {Decoding} of {Quantum} {LDPC} codes},
	url = {http://arxiv.org/abs/2311.18488},
	doi = {10.48550/arXiv.2311.18488},
	urldate = {2026-02-25},
	publisher = {arXiv},
	author = {Javed, Sana and Garcia-Herrero, Francisco and Vasic, Bane and Flanagan, Mark F.},
	month = jan,
	year = {2024},
	note = {arXiv:2311.18488 [cs]},
}

@article{feldmanUsingLinearProgramming2005,
	title = {Using linear programming to {Decode} {Binary} linear codes},
	volume = {51},
	issn = {1557-9654},
	url = {https://ieeexplore.ieee.org/document/1397933/},
	doi = {10.1109/TIT.2004.842696},
	number = {3},
	urldate = {2026-02-27},
	journal = {IEEE Transactions on Information Theory},
	author = {Feldman, J. and Wainwright, M.J. and Karger, D.R.},
	month = mar,
	year = {2005},
	pages = {954--972},
}

@article{andristErrorThresholdsAbelian2015,
	title = {Error {Thresholds} for {Abelian} {Quantum} {Double} {Models}: {Increasing} the bit-flip {Stability} of {Topological} {Quantum} {Memory}},
	volume = {91},
	issn = {1050-2947, 1094-1622},
	shorttitle = {Error {Thresholds} for {Abelian} {Quantum} {Double} {Models}},
	url = {http://arxiv.org/abs/1406.5974},
	doi = {10.1103/PhysRevA.91.042331},
	number = {4},
	urldate = {2026-02-25},
	journal = {Physical Review A},
	author = {Andrist, Ruben S. and Wootton, James R. and Katzgraber, Helmut G.},
	month = apr,
	year = {2015},
	note = {arXiv:1406.5974 [quant-ph]},
	pages = {042331},
}

@article{fossorierSoftdecisionDecodingLinear1995,
	title = {Soft-decision decoding of linear block codes based on ordered statistics},
	volume = {41},
	issn = {1557-9654},
	url = {https://ieeexplore.ieee.org/document/412683},
	doi = {10.1109/18.412683},
	number = {5},
	urldate = {2026-03-10},
	journal = {IEEE Transactions on Information Theory},
	author = {Fossorier, M.P.C. and Lin, Shu},
	month = sep,
	year = {1995},
	pages = {1379--1396},
}

@article{duclos-cianciKitaevsZ_dCodesThreshold2013,
	title = {Kitaev's {Z}\_d-{Codes} {Threshold} {Estimates}},
	volume = {87},
	issn = {1050-2947, 1094-1622},
	url = {http://arxiv.org/abs/1302.3638},
	doi = {10.1103/PhysRevA.87.062338},
	number = {6},
	urldate = {2026-02-25},
	journal = {Physical Review A},
	author = {Duclos-Cianci, Guillaume and Poulin, David},
	month = jun,
	year = {2013},
	note = {arXiv:1302.3638 [quant-ph]},
	pages = {062338},
}

@article{kschischangFactorGraphsSumproduct2001,
	title = {Factor graphs and the sum-product algorithm},
	volume = {47},
	issn = {1557-9654},
	url = {https://ieeexplore.ieee.org/document/910572},
	doi = {10.1109/18.910572},
	number = {2},
	urldate = {2026-03-10},
	journal = {IEEE Transactions on Information Theory},
	author = {Kschischang, F.R. and Frey, B.J. and Loeliger, H.-A.},
	month = feb,
	year = {2001},
	pages = {498--519},
}

@article{mackayGoodErrorcorrectingCodes1999,
	title = {Good error-correcting codes based on very sparse matrices},
	volume = {45},
	issn = {1557-9654},
	url = {https://ieeexplore.ieee.org/document/748992},
	doi = {10.1109/18.748992},
	number = {2},
	urldate = {2026-03-10},
	journal = {IEEE Transactions on Information Theory},
	author = {MacKay, D.J.C.},
	month = mar,
	year = {1999},
	pages = {399--431},
}

@article{fossorierIterativeReliabilitybasedDecoding2001,
	title = {Iterative reliability-based decoding of low-density parity check codes},
	volume = {19},
	issn = {1558-0008},
	url = {https://ieeexplore.ieee.org/document/924874},
	doi = {10.1109/49.924874},
	number = {5},
	urldate = {2026-03-10},
	journal = {IEEE Journal on Selected Areas in Communications},
	author = {Fossorier, M.P.C.},
	month = may,
	year = {2001},
	pages = {908--917},
}

@article{panteleevDegenerateQuantumLDPC2021,
	title = {Degenerate {Quantum} {LDPC} {Codes} {With} {Good} {Finite} {Length} {Performance}},
	volume = {5},
	issn = {2521-327X},
	url = {http://arxiv.org/abs/1904.02703},
	doi = {10.22331/q-2021-11-22-585},
	urldate = {2026-03-10},
	journal = {Quantum},
	author = {Panteleev, Pavel and Kalachev, Gleb},
	month = nov,
	year = {2021},
	note = {arXiv:1904.02703 [quant-ph]},
	pages = {585},
}

@article{delfosseAlmostlinearTimeDecoding2021,
	title = {Almost-linear time decoding algorithm for topological codes},
	volume = {5},
	url = {https://quantum-journal.org/papers/q-2021-12-02-595/},
	doi = {10.22331/q-2021-12-02-595},
	urldate = {2026-02-26},
	journal = {Quantum},
	publisher = {Verein zur Förderung des Open Access Publizierens in den Quantenwissenschaften},
	author = {Delfosse, Nicolas and Nickerson, Naomi H.},
	month = dec,
	year = {2021},
	pages = {595},
}

@article{eastin_restrictions_2009,
	title = {Restrictions on {Transversal} {Encoded} {Quantum} {Gate} {Sets}},
	volume = {102},
	copyright = {http://link.aps.org/licenses/aps-default-license},
	issn = {0031-9007, 1079-7114},
	url = {https://link.aps.org/doi/10.1103/PhysRevLett.102.110502},
	doi = {10.1103/PhysRevLett.102.110502},
	number = {11},
	urldate = {2026-02-26},
	journal = {Physical Review Letters},
	author = {Eastin, Bryan and Knill, Emanuel},
	month = mar,
	year = {2009},
	pages = {110502},
}

@article{bravyiQuantumSelfCorrection3D2013,
	title = {Quantum {Self}-{Correction} in the {3D} {Cubic} {Code} {Model}},
	volume = {111},
	copyright = {http://link.aps.org/licenses/aps-default-license},
	issn = {0031-9007, 1079-7114},
	url = {https://link.aps.org/doi/10.1103/PhysRevLett.111.200501},
	doi = {10.1103/PhysRevLett.111.200501},
	number = {20},
	urldate = {2026-02-25},
	journal = {Physical Review Letters},
	author = {Bravyi, Sergey and Haah, Jeongwan},
	month = nov,
	year = {2013},
	pages = {200501},
}

@article{edmondsPathsTreesFlowers1965,
	title = {Paths, {Trees}, and {Flowers}},
	volume = {17},
	issn = {0008-414X, 1496-4279},
	url = {https://www.cambridge.org/core/journals/canadian-journal-of-mathematics/article/paths-trees-and-flowers/08B492B72322C4130AE800C0610E0E21},
	doi = {10.4153/CJM-1965-045-4},
	urldate = {2026-02-26},
	journal = {Canadian Journal of Mathematics},
	author = {Edmonds, Jack},
	month = jan,
	year = {1965},
	pages = {449--467},
}

@article{higgottSparseBlossomCorrecting2025,
	title = {Sparse {Blossom}: correcting a million errors per core second with minimum-weight matching},
	volume = {9},
	shorttitle = {Sparse {Blossom}},
	url = {https://quantum-journal.org/papers/q-2025-01-20-1600/},
	doi = {10.22331/q-2025-01-20-1600},
	urldate = {2026-02-26},
	journal = {Quantum},
	publisher = {Verein zur Förderung des Open Access Publizierens in den Quantenwissenschaften},
	author = {Higgott, Oscar and Gidney, Craig},
	month = jan,
	year = {2025},
	pages = {1600},
}

@article{barberRealtimeScalableFast2025,
	title = {A real-time, scalable, fast and highly resource efficient decoder for a quantum computer},
	volume = {8},
	issn = {2520-1131},
	url = {http://arxiv.org/abs/2309.05558},
	doi = {10.1038/s41928-024-01319-5},
	number = {1},
	urldate = {2026-02-26},
	journal = {Nature Electronics},
	author = {Barber, Ben and Barnes, Kenton M. and Bialas, Tomasz and Buğdaycı, Okan and Campbell, Earl T. and Gillespie, Neil I. and Johar, Kauser and Rajan, Ram and Richardson, Adam W. and Skoric, Luka and Topal, Canberk and Turner, Mark L. and Ziad, Abbas B.},
	month = jan,
	year = {2025},
	note = {arXiv:2309.05558 [quant-ph]},
	pages = {84--91},
}

@article{krastanovDeepNeuralNetwork2017,
	title = {Deep {Neural} {Network} {Probabilistic} {Decoder} for {Stabilizer} {Codes}},
	volume = {7},
	copyright = {2017 The Author(s)},
	issn = {2045-2322},
	url = {https://www.nature.com/articles/s41598-017-11266-1},
	doi = {10.1038/s41598-017-11266-1},
	number = {1},
	urldate = {2026-02-26},
	journal = {Scientific Reports},
	publisher = {Nature Publishing Group},
	author = {Krastanov, Stefan and Jiang, Liang},
	month = sep,
	year = {2017},
	pages = {11003},
}

@misc{cauneDemonstratingRealtimeLowlatency2024,
	title = {Demonstrating real-time and low-latency quantum error correction with superconducting qubits},
	url = {http://arxiv.org/abs/2410.05202},
	doi = {10.48550/arXiv.2410.05202},
	urldate = {2026-02-26},
	publisher = {arXiv},
	author = {Caune, Laura and Skoric, Luka and Blunt, Nick S. and Ruban, Archibald and McDaniel, Jimmy and Valery, Joseph A. and Patterson, Andrew D. and Gramolin, Alexander V. and Majaniemi, Joonas and Barnes, Kenton M. and Bialas, Tomasz and Buğdaycı, Okan and Crawford, Ophelia and Gehér, György P. and Krovi, Hari and Matekole, Elisha and Topal, Canberk and Poletto, Stefano and Bryant, Michael and Snyder, Kalan and Gillespie, Neil I. and Jones, Glenn and Johar, Kauser and Campbell, Earl T. and Hill, Alexander D.},
	month = oct,
	year = {2024},
	note = {arXiv:2410.05202 [quant-ph]},
}

@article{steaneErrorCorrectingCodes1996,
	title = {Error {Correcting} {Codes} in {Quantum} {Theory}},
	volume = {77},
	copyright = {http://link.aps.org/licenses/aps-default-license},
	issn = {0031-9007, 1079-7114},
	url = {https://link.aps.org/doi/10.1103/PhysRevLett.77.793},
	doi = {10.1103/PhysRevLett.77.793},
	number = {5},
	urldate = {2026-02-25},
	journal = {Physical Review Letters},
	author = {Steane, A. M.},
	month = jul,
	year = {1996},
	pages = {793--797},
}

@article{liyanageFPGAbasedDistributedUnionFind2024,
	title = {{FPGA}-based {Distributed} {Union}-{Find} {Decoder} for {Surface} {Codes}},
	volume = {5},
	issn = {2689-1808},
	url = {http://arxiv.org/abs/2406.08491},
	doi = {10.1109/TQE.2024.3467271},
	urldate = {2026-02-26},
	journal = {IEEE Transactions on Quantum Engineering},
	author = {Liyanage, Namitha and Wu, Yue and Tagare, Siona and Zhong, Lin},
	year = {2024},
	note = {arXiv:2406.08491 [quant-ph]},
	pages = {1--18},
}

@article{chamberlandDeepNeuralDecoders2018,
	title = {Deep neural decoders for near term fault-tolerant experiments},
	volume = {3},
	issn = {2058-9565},
	url = {https://doi.org/10.1088/2058-9565/aad1f7},
	doi = {10.1088/2058-9565/aad1f7},
	number = {4},
	urldate = {2026-02-26},
	journal = {Quantum Science and Technology},
	publisher = {IOP Publishing},
	author = {Chamberland, Christopher and Ronagh, Pooya},
	month = jul,
	year = {2018},
	pages = {044002},
}

@article{breuckmannHyperbolicSemiHyperbolicSurface2017,
	title = {Hyperbolic and {Semi}-{Hyperbolic} {Surface} {Codes} for {Quantum} {Storage}},
	volume = {2},
	issn = {2058-9565},
	url = {http://arxiv.org/abs/1703.00590},
	doi = {10.1088/2058-9565/aa7d3b},
	number = {3},
	urldate = {2026-02-25},
	journal = {Quantum Science and Technology},
	author = {Breuckmann, Nikolas P. and Vuillot, Christophe and Campbell, Earl and Krishna, Anirudh and Terhal, Barbara M.},
	month = sep,
	year = {2017},
	note = {arXiv:1703.00590 [quant-ph]},
	pages = {035007},
}

@article{watsonFastFaulttolerantDecoder2015,
	title = {A fast fault-tolerant decoder for qubit and qudit surface codes},
	volume = {92},
	issn = {1050-2947, 1094-1622},
	url = {http://arxiv.org/abs/1411.3028},
	doi = {10.1103/PhysRevA.92.032309},
	number = {3},
	urldate = {2026-02-25},
	journal = {Physical Review A},
	author = {Watson, Fern H. E. and Anwar, Hussain and Browne, Dan E.},
	month = sep,
	year = {2015},
	note = {arXiv:1411.3028 [quant-ph]},
	pages = {032309},
}

@article{bao_magic_2022,
	title = {Magic state distillation from entangled states},
	volume = {105},
	issn = {2469-9926, 2469-9934},
	url = {https://link.aps.org/doi/10.1103/PhysRevA.105.022602},
	doi = {10.1103/PhysRevA.105.022602},
	number = {2},
	urldate = {2026-02-26},
	journal = {Physical Review A},
	author = {Bao, Ning and Cao, ChunJun and Su, Vincent Paul},
	month = feb,
	year = {2022},
	pages = {022602},
}

@article{bennettMixedstateEntanglementQuantum1996,
	title = {Mixed-state entanglement and quantum error correction},
	volume = {54},
	copyright = {http://link.aps.org/licenses/aps-default-license},
	issn = {1050-2947, 1094-1622},
	url = {https://link.aps.org/doi/10.1103/PhysRevA.54.3824},
	doi = {10.1103/PhysRevA.54.3824},
	number = {5},
	urldate = {2026-02-25},
	journal = {Physical Review A},
	author = {Bennett, Charles H. and DiVincenzo, David P. and Smolin, John A. and Wootters, William K.},
	month = nov,
	year = {1996},
	pages = {3824--3851},
}

@misc{guPowerLimitationsLinear2025,
	title = {Power and {Limitations} of {Linear} {Programming} {Decoder} for {Quantum} {LDPC} {Codes}},
	url = {http://arxiv.org/abs/2508.04769},
	doi = {10.48550/arXiv.2508.04769},
	urldate = {2026-02-25},
	publisher = {arXiv},
	author = {Gu, Shouzhen and Soleimanifar, Mehdi},
	month = aug,
	year = {2025},
	note = {arXiv:2508.04769 [quant-ph]},
}

@inproceedings{fawziLinearProgrammingDecoder2021,
	address = {Riva del Garda, Italy},
	title = {Linear programming decoder for hypergraph product quantum codes},
	copyright = {https://ieeexplore.ieee.org/Xplorehelp/downloads/license-information/IEEE.html},
	isbn = {978-1-7281-5962-1},
	url = {https://ieeexplore.ieee.org/document/9457611/},
	doi = {10.1109/ITW46852.2021.9457611},
	urldate = {2026-02-25},
	booktitle = {2020 {IEEE} {Information} {Theory} {Workshop} ({ITW})},
	publisher = {IEEE},
	author = {Fawzi, Omar and Groues, Lucien and Leverrier, Anthony},
	month = apr,
	year = {2021},
	pages = {1--5},
}

@article{bravyi_universal_2005,
	title = {Universal quantum computation with ideal {Clifford} gates and noisy ancillas},
	volume = {71},
	copyright = {http://link.aps.org/licenses/aps-default-license},
	issn = {1050-2947, 1094-1622},
	url = {https://link.aps.org/doi/10.1103/PhysRevA.71.022316},
	doi = {10.1103/PhysRevA.71.022316},
	number = {2},
	urldate = {2026-02-26},
	journal = {Physical Review A},
	author = {Bravyi, Sergey and Kitaev, Alexei},
	month = feb,
	year = {2005},
	pages = {022316},
}

@article{bravyi_classification_2013,
	title = {Classification of {Topologically} {Protected} {Gates} for {Local} {Stabilizer} {Codes}},
	volume = {110},
	copyright = {http://link.aps.org/licenses/aps-default-license},
	issn = {0031-9007, 1079-7114},
	url = {https://link.aps.org/doi/10.1103/PhysRevLett.110.170503},
	doi = {10.1103/PhysRevLett.110.170503},
	number = {17},
	urldate = {2026-02-26},
	journal = {Physical Review Letters},
	author = {Bravyi, Sergey and König, Robert},
	month = apr,
	year = {2013},
	pages = {170503},
}

@article{bravyiTradeoffsReliableQuantum2010,
	title = {Tradeoffs for {Reliable} {Quantum} {Information} {Storage} in {2D} {Systems}},
	volume = {104},
	copyright = {http://link.aps.org/licenses/aps-default-license},
	issn = {0031-9007, 1079-7114},
	url = {https://link.aps.org/doi/10.1103/PhysRevLett.104.050503},
	doi = {10.1103/PhysRevLett.104.050503},
	number = {5},
	urldate = {2026-02-25},
	journal = {Physical Review Letters},
	author = {Bravyi, Sergey and Poulin, David and Terhal, Barbara},
	month = feb,
	year = {2010},
	pages = {050503},
}

@article{roffeDecodingQuantumLowdensity2020,
	title = {Decoding across the quantum low-density parity-check code landscape},
	volume = {2},
	issn = {2643-1564},
	url = {https://link.aps.org/doi/10.1103/PhysRevResearch.2.043423},
	doi = {10.1103/PhysRevResearch.2.043423},
	number = {4},
	urldate = {2026-02-26},
	journal = {Physical Review Research},
	author = {Roffe, Joschka and White, David R. and Burton, Simon and Campbell, Earl},
	month = dec,
	year = {2020},
	pages = {043423},
}

@misc{wolanskiAmbiguityClusteringAccurate2025,
	title = {Ambiguity {Clustering}: an accurate and efficient decoder for {qLDPC} codes},
	shorttitle = {Ambiguity {Clustering}},
	url = {http://arxiv.org/abs/2406.14527},
	doi = {10.48550/arXiv.2406.14527},
	urldate = {2026-02-26},
	publisher = {arXiv},
	author = {Wolanski, Stasiu and Barber, Ben},
	month = jan,
	year = {2025},
	note = {arXiv:2406.14527 [quant-ph]},
}

@article{ziadLocalClusteringDecoder2025,
	title = {Local clustering decoder as a fast and adaptive hardware decoder for the surface code},
	volume = {16},
	issn = {2041-1723},
	url = {https://www.nature.com/articles/s41467-025-66773-x},
	doi = {10.1038/s41467-025-66773-x},
	number = {1},
	urldate = {2026-02-26},
	journal = {Nature Communications},
	author = {Ziad, Abbas B. and Zalawadiya, Ankit and Topal, Canberk and Camps, Joan and Gehér, György P. and Stafford, Matthew P. and Turner, Mark L.},
	month = dec,
	year = {2025},
	pages = {11048},
}

@article{bravyiEfficientAlgorithmsMaximum2014,
	title = {Efficient algorithms for maximum likelihood decoding in the surface code},
	volume = {90},
	copyright = {http://link.aps.org/licenses/aps-default-license},
	issn = {1050-2947, 1094-1622},
	url = {https://link.aps.org/doi/10.1103/PhysRevA.90.032326},
	doi = {10.1103/PhysRevA.90.032326},
	number = {3},
	urldate = {2026-02-26},
	journal = {Physical Review A},
	author = {Bravyi, Sergey and Suchara, Martin and Vargo, Alexander},
	month = sep,
	year = {2014},
	pages = {032326},
}

@article{chen_efficient_2026,
	title = {Efficient {Magic} {State} {Cultivation} on {R} {P} 2},
	volume = {7},
	issn = {2691-3399},
	url = {https://link.aps.org/doi/10.1103/9kys-3whh},
	doi = {10.1103/9kys-3whh},
	number = {1},
	urldate = {2026-02-26},
	journal = {PRX Quantum},
	author = {Chen, Zi-Han and Chen, Ming-Cheng and Lu, Chao-Yang and Pan, Jian-Wei},
	month = jan,
	year = {2026},
	pages = {010315},
}

@article{bravyiHighthresholdLowoverheadFaulttolerant2024,
	title = {High-threshold and low-overhead fault-tolerant quantum memory},
	volume = {627},
	copyright = {2024 The Author(s)},
	issn = {1476-4687},
	url = {https://www.nature.com/articles/s41586-024-07107-7},
	doi = {10.1038/s41586-024-07107-7},
	number = {8005},
	urldate = {2026-02-25},
	journal = {Nature},
	publisher = {Nature Publishing Group},
	author = {Bravyi, Sergey and Cross, Andrew W. and Gambetta, Jay M. and Maslov, Dmitri and Rall, Patrick and Yoder, Theodore J.},
	month = mar,
	year = {2024},
	pages = {778--782},
}

@article{varonaDeterminationSemionCode2020,
	title = {Determination of the {Semion} {Code} {Threshold} using {Neural} {Decoders}},
	volume = {102},
	issn = {2469-9926, 2469-9934},
	url = {http://arxiv.org/abs/2002.08666},
	doi = {10.1103/PhysRevA.102.032411},
	number = {3},
	urldate = {2026-02-25},
	journal = {Physical Review A},
	author = {Varona, Santiago and Martin-Delgado, Miguel Angel},
	month = sep,
	year = {2020},
	note = {arXiv:2002.08666 [quant-ph]},
	pages = {032411},
}

@article{knillTheoryQuantumErrorcorrecting1997,
	title = {Theory of quantum error-correcting codes},
	volume = {55},
	copyright = {http://link.aps.org/licenses/aps-default-license},
	issn = {1050-2947, 1094-1622},
	url = {https://link.aps.org/doi/10.1103/PhysRevA.55.900},
	doi = {10.1103/PhysRevA.55.900},
	number = {2},
	urldate = {2026-02-25},
	journal = {Physical Review A},
	author = {Knill, Emanuel and Laflamme, Raymond},
	month = feb,
	year = {1997},
	pages = {900--911},
}

@inproceedings{liLPDecodingQuantum2018,
	title = {{LP} {Decoding} of {Quantum} {Stabilizer} {Codes}},
	issn = {2157-8117},
	url = {https://ieeexplore.ieee.org/document/8437905/},
	doi = {10.1109/ISIT.2018.8437905},
	urldate = {2026-02-26},
	booktitle = {2018 {IEEE} {International} {Symposium} on {Information} {Theory} ({ISIT})},
	author = {Li, July X. and Vontobel, Pascal O.},
	month = jun,
	year = {2018},
	note = {ISSN: 2157-8117},
	pages = {1306--1310},
}

@misc{kabirSdimQuditStabilizer2025,
	title = {Sdim: {A} {Qudit} {Stabilizer} {Simulator}},
	shorttitle = {Sdim},
	url = {http://arxiv.org/abs/2511.12777},
	doi = {10.48550/arXiv.2511.12777},
	urldate = {2026-02-26},
	publisher = {arXiv},
	author = {Kabir, Adeeb and Nguyen, Steven and Ghosh, Sohan and Kiran, Tijil and Kim, Isaac H. and Huang, Yipeng},
	month = nov,
	year = {2025},
	note = {arXiv:2511.12777 [quant-ph]},
}

@misc{seniorScalableRealtimeNeural2025,
	title = {A scalable and real-time neural decoder for topological quantum codes},
	url = {http://arxiv.org/abs/2512.07737},
	doi = {10.48550/arXiv.2512.07737},
	urldate = {2026-02-26},
	publisher = {arXiv},
	author = {Senior, Andrew W. and Edlich, Thomas and Heras, Francisco J. H. and Zhang, Lei M. and Higgott, Oscar and Spencer, James S. and Applebaum, Taylor and Blackwell, Sam and Ledford, Justin and Žemgulytė, Akvilė and Žídek, Augustin and Shutty, Noah and Cowie, Andrew and Li, Yin and Holland, George and Brooks, Peter and Beattie, Charlie and Newman, Michael and Davies, Alex and Jones, Cody and Boixo, Sergio and Neven, Hartmut and Kohli, Pushmeet and Bausch, Johannes},
	month = dec,
	year = {2025},
	note = {arXiv:2512.07737 [quant-ph]},
}

@misc{zhangLearningNeuralDecoding2025,
	title = {Learning {Neural} {Decoding} with {Parallelism} and {Self}-{Coordination} for {Quantum} {Error} {Correction}},
	url = {http://arxiv.org/abs/2509.03815},
	doi = {10.48550/arXiv.2509.03815},
	urldate = {2026-02-26},
	publisher = {arXiv},
	author = {Zhang, Kai and Wang, Situ and Kong, Linghang and Zhang, Fang and Ji, Zhengfeng and Chen, Jianxin},
	month = sep,
	year = {2025},
	note = {arXiv:2509.03815 [quant-ph]},
}

@misc{huEfficientUniversalNeuralNetwork2025,
	title = {Efficient and {Universal} {Neural}-{Network} {Decoder} for {Stabilizer}-{Based} {Quantum} {Error} {Correction}},
	url = {http://arxiv.org/abs/2502.19971},
	doi = {10.48550/arXiv.2502.19971},
	urldate = {2026-02-26},
	publisher = {arXiv},
	author = {Hu, Gengyuan and Ouyang, Wanli and Lu, Chao-Yang and Lin, Chen and Zhong, Han-Sen},
	month = jun,
	year = {2025},
	note = {arXiv:2502.19971 [quant-ph]},
}

@article{torlaiNeuralDecoderTopological2017,
	title = {Neural {Decoder} for {Topological} {Codes}},
	volume = {119},
	url = {https://link.aps.org/doi/10.1103/PhysRevLett.119.030501},
	doi = {10.1103/PhysRevLett.119.030501},
	number = {3},
	urldate = {2026-02-26},
	journal = {Physical Review Letters},
	publisher = {American Physical Society},
	author = {Torlai, Giacomo and Melko, Roger G.},
	month = jul,
	year = {2017},
	pages = {030501},
}

@misc{poulinIterativeDecodingSparse2008,
	title = {On the iterative decoding of sparse quantum codes},
	url = {http://arxiv.org/abs/0801.1241},
	doi = {10.48550/arXiv.0801.1241},
	urldate = {2026-02-25},
	publisher = {arXiv},
	author = {Poulin, David and Chung, Yeojin},
	month = jul,
	year = {2008},
	note = {arXiv:0801.1241 [quant-ph]},
}

@article{berlekampInherentIntractabilityCertain1978,
	title = {On the inherent intractability of certain coding problems ({Corresp}.)},
	volume = {24},
	issn = {1557-9654},
	url = {https://ieeexplore.ieee.org/document/1055873/},
	doi = {10.1109/TIT.1978.1055873},
	number = {3},
	urldate = {2026-02-26},
	journal = {IEEE Transactions on Information Theory},
	author = {Berlekamp, E. and McEliece, R. and van Tilborg, H.},
	month = may,
	year = {1978},
	pages = {384--386},
}

@misc{iyerHardnessDecodingQuantum2013,
	title = {Hardness of decoding quantum stabilizer codes},
	url = {http://arxiv.org/abs/1310.3235},
	doi = {10.48550/arXiv.1310.3235},
	urldate = {2026-02-26},
	publisher = {arXiv},
	author = {Iyer, Pavithran and Poulin, David},
	month = oct,
	year = {2013},
	note = {arXiv:1310.3235 [quant-ph]},
}

@misc{gidney_magic_2024,
	title = {Magic state cultivation: growing {T} states as cheap as {CNOT} gates},
	shorttitle = {Magic state cultivation},
	url = {http://arxiv.org/abs/2409.17595},
	doi = {10.48550/arXiv.2409.17595},
	urldate = {2026-02-26},
	publisher = {arXiv},
	author = {Gidney, Craig and Shutty, Noah and Jones, Cody},
	month = sep,
	year = {2024},
	note = {arXiv:2409.17595 [quant-ph]},
}

@article{litinski_magic_2019,
	title = {Magic {State} {Distillation}: {Not} as {Costly} as {You} {Think}},
	volume = {3},
	issn = {2521-327X},
	shorttitle = {Magic {State} {Distillation}},
	url = {http://arxiv.org/abs/1905.06903},
	doi = {10.22331/q-2019-12-02-205},
	urldate = {2026-02-26},
	journal = {Quantum},
	author = {Litinski, Daniel},
	month = dec,
	year = {2019},
	note = {arXiv:1905.06903 [quant-ph]},
	pages = {205},
}

@misc{shorQuantumErrorCorrectingCodes1996,
	title = {Quantum {Error}-{Correcting} {Codes} {Need} {Not} {Completely} {Reveal} the {Error} {Syndrome}},
	url = {http://arxiv.org/abs/quant-ph/9604006},
	doi = {10.48550/arXiv.quant-ph/9604006},
	urldate = {2026-02-25},
	publisher = {arXiv},
	author = {Shor, Peter W. and Smolin, John A.},
	month = apr,
	year = {1996},
	note = {arXiv:quant-ph/9604006},
}

@misc{panteleevAsymptoticallyGoodQuantum2022,
	title = {Asymptotically {Good} {Quantum} and {Locally} {Testable} {Classical} {LDPC} {Codes}},
	url = {http://arxiv.org/abs/2111.03654},
	doi = {10.48550/arXiv.2111.03654},
	urldate = {2026-02-25},
	publisher = {arXiv},
	author = {Panteleev, Pavel and Kalachev, Gleb},
	month = jan,
	year = {2022},
	note = {arXiv:2111.03654 [cs]},
}

@article{calderbankQuantumErrorCorrection1997,
	title = {Quantum {Error} {Correction} and {Orthogonal} {Geometry}},
	volume = {78},
	url = {https://link.aps.org/doi/10.1103/PhysRevLett.78.405},
	doi = {10.1103/PhysRevLett.78.405},
	number = {3},
	urldate = {2026-02-25},
	journal = {Physical Review Letters},
	publisher = {American Physical Society},
	author = {Calderbank, A. R. and Rains, E. M. and Shor, P. W. and Sloane, N. J. A.},
	month = jan,
	year = {1997},
	pages = {405--408},
}

@article{fernHowSU2$_4$Anyons2017,
	title = {How {SU}(2)\$\_4\$ {Anyons} are {Z}\$\_3\$ {Parafermions}},
	volume = {3},
	issn = {2542-4653},
	url = {https://scipost.org/10.21468/SciPostPhys.3.6.037},
	doi = {10.21468/SciPostPhys.3.6.037},
	number = {6},
	urldate = {2026-02-24},
	journal = {SciPost Physics},
	author = {Fern, Richard and Kombe, Johannes and Simon, Steven},
	month = dec,
	year = {2017},
	pages = {037},
}

@article{jochym-oconnorDisjointnessStabilizerCodes2018,
	title = {Disjointness of {Stabilizer} {Codes} and {Limitations} on {Fault}-{Tolerant} {Logical} {Gates}},
	volume = {8},
	issn = {2160-3308},
	url = {https://link.aps.org/doi/10.1103/PhysRevX.8.021047},
	doi = {10.1103/PhysRevX.8.021047},
	number = {2},
	urldate = {2026-02-24},
	journal = {Physical Review X},
	author = {Jochym-O’Connor, Tomas and Kubica, Aleksander and Yoder, Theodore J.},
	month = may,
	year = {2018},
	pages = {021047},
}

@article{iqbalqutrit,
	title = {Qutrit toric code and parafermions in trapped ions},
	volume = {16},
	copyright = {2025 The Author(s)},
	issn = {2041-1723},
	url = {https://www.nature.com/articles/s41467-025-61391-z},
	doi = {10.1038/s41467-025-61391-z},
	number = {1},
	urldate = {2026-02-24},
	journal = {Nature Communications},
	publisher = {Nature Publishing Group},
	author = {Iqbal, Mohsin and Lyons, Anasuya and Lo, Chiu Fan Bowen and Tantivasadakarn, Nathanan and Dreiling, Joan and Foltz, Cameron and Gatterman, Thomas M. and Gresh, Dan and Hewitt, Nathan and Holliman, Craig A. and Johansen, Jacob and Neyenhuis, Brian and Matsuoka, Yohei and Mills, Michael and Moses, Steven A. and Siegfried, Peter and Vishwanath, Ashvin and Verresen, Ruben and Dreyer, Henrik},
	month = jul,
	year = {2025},
	pages = {6301},
}

@article{qcqutrit,
	title = {Quantum computing with parafermions},
	volume = {93},
	url = {https://link.aps.org/doi/10.1103/PhysRevB.93.125105},
	doi = {10.1103/PhysRevB.93.125105},
	number = {12},
	urldate = {2026-02-24},
	journal = {Physical Review B},
	publisher = {American Physical Society},
	author = {Hutter, Adrian and Loss, Daniel},
	month = mar,
	year = {2016},
	pages = {125105},
}
\end{document}